\documentclass[prb, aps, twocolumn, amsmath, amssymb, graphicx, longbibliography, nofootinbib]{revtex4-2}

\usepackage{bm}
\usepackage{graphicx}
\usepackage{braket}

\usepackage{color}
\usepackage[usenames, dvipsnames]{xcolor}
\usepackage[colorlinks=true]{hyperref}
\hypersetup{colorlinks=true, linkcolor=MidnightBlue, citecolor=MidnightBlue, urlcolor=OliveGreen}

\usepackage[]{cleveref}
\Crefname{section}{Sec.}{Secs.}
\Crefname{equation}{Eq.}{Eqs.}
\Crefname{figure}{Fig.}{Figs.}
\Crefname{tabular}{Tab.}{Tabs.}

\usepackage{siunitx}
\DeclareSIUnit{\ueV}{\micro\electronvolt}
\DeclareSIUnit{\meV}{\milli\electronvolt}

\usepackage{soul}
\usepackage{url}

\usepackage{tikz}

\usepackage{tabularx}
\usepackage{array}
\usepackage{multirow}
\usepackage{makecell}

\usepackage{comment}

\newcolumntype{C}{>{\centering\arraybackslash}X}

\newcommand{\cH}{\mathcal{H}}
\newcommand{\nket}[1]{|#1)}

\newcommand{\dd}{\text{d}}

\DeclareMathOperator{\tr}{tr}

\DeclareMathOperator{\Pf}{Pf}

\newcommand{\SM}{\mathrm{\scriptscriptstyle SM}}
\newcommand{\SC}{\mathrm{\scriptscriptstyle SC}}

\newcommand{\DeltaInd}{\Delta_\mathrm{ind}}

\renewcommand{\bra}[1]{\langle #1|}
\renewcommand{\ket}[1]{|#1 \rangle}

\newcommand{\Tr}{{\rm Tr}}

\newcommand{\EC}{E_\mathrm{\scriptscriptstyle C}}
\renewcommand{\ng}{n_\mathrm{g}}
\newcommand{\Ng}{N_\mathrm{g}}
\newcommand{\omegar}{\omega_\mathrm{r}}
\newcommand{\Vg}{V_\mathrm{g}}
\newcommand{\Vrf}{V_\mathrm{rf}}
\newcommand{\Cg}{C_\mathrm{g}}
\newcommand{\Q}{\mathrm{\scriptscriptstyle Q}}

\newcommand{\CQ}{C_\Q}
\newcommand{\Hqd}{H_\mathrm{qd}}
\newcommand{\tHqd}{\tilde{H}_\mathrm{qd}}
\newcommand{\Lwire}{L_{\rm wire}}
\newcommand{\Lqd}{L_{\rm qd}}
\newcommand{\muwire}{\mu_{\rm wire}}
\newcommand{\muqd}{\mu_{\rm qd}}

\newcommand{\fCQ}{\widetilde{C}_Q}

\begin{document}

\title{ 
Predictive simulations of the dynamical response of mesoscopic devices
}

\author{Samuel Boutin}
\affiliation{Station Q, Microsoft Corporation, Goleta, California, USA}
\author{Torsten Karzig}
\affiliation{Station Q, Microsoft Corporation, Goleta, California, USA}
\author{Tareq El Dandachi}
\affiliation{Station Q, Microsoft Corporation, Goleta, California, USA}
\author{Ryan~V.~Mishmash}
\affiliation{Station Q, Microsoft Corporation, Goleta, California, USA}
\author{Jan Gukelberger}
\affiliation{Station Q, Microsoft Corporation, Goleta, California, USA}
\author{Roman M. Lutchyn}
\affiliation{Station Q, Microsoft Corporation, Goleta, California, USA}
\author{Bela Bauer}
\affiliation{Station Q, Microsoft Corporation, Goleta, California, USA}

\date{\today{}}

\begin{abstract}
As the complexity of mesoscopic quantum devices increases, simulations are becoming an invaluable tool for understanding their behavior. This is especially true for the superconductor-semiconductor heterostructures used to build Majorana-based topological qubits, where quantitatively understanding the interplay of topological superconductivity, disorder, semiconductor quantum dots, Coulomb blockade and noise has been essential for progress on device design and interpretation of measurements~\cite{Aghaee23, Aghaee24}. In this paper, we describe a general framework to simulate the low-energy quantum dynamics of such complex systems. We illustrate our approach by computing the dispersive gate sensing (DGS) response of quantum dots coupled to topological superconductors. We start by formulating the DGS response as an open-system quantum dynamics problem, which allows a consistent treatment of drive backaction as well as quantum and classical noise.
For microscopic quantum problems subject to Coulomb-blockade, where a direct solution in the exponentially large many-body Hilbert space would be prohibitive, we introduce a series of controlled approximations that incorporate ideas from tensor network theory and quantum chemistry to reduce this Hilbert space to a few low-energy degrees of freedom that accurately capture the low-energy quantum dynamics.
We demonstrate the methods introduced in this paper on the example of a single quantum dot coupled to a topological superconductor and a microscopic realization of the fermion parity readout setup of Ref.~\cite{Aghaee24}.
\end{abstract}

\maketitle
\makeatletter
\def\l@subsection#1#2{}
\def\l@subsubsection#1#2{}
\def\l@f@section{}
\makeatother

\tableofcontents

\section{Introduction}
A standard tool to probe mesoscopic devices such as nanowires and quantum dots (QDs) are transport measurements (see, e.g. Refs.~\cite{datta_electronic_1997,nazarov_quantum_2009}), which are typically operated in the dc limit. Many of the interesting dynamical phenomena in these devices, however, require faster readout techniques, typically operating in the rf regime~\cite{vigneau_probing_2023}. This is particularly true for qubit devices such as spin qubits~\cite{burkard_semiconductor_2023} or topological qubits~\cite{karzig2017scalable}. While the dynamical response of mesoscopic devices can in some regimes be mapped to static quantities like the quantum capacitance, a quantitative description of the readout signal and corresponding signal-to-noise ratio (SNR) requires modeling the measurement backaction onto the device, which may require a non-perturbative treatment beyond linear response~\cite{maman_2020,peri_2024}.

Modeling of the dynamics of mesoscopic quantum devices can be complicated by the presence of many comparable energy scales. In the topological qubit device design of Ref.~\cite{Aghaee24}, the behavior is characterized by a competition between temperature ($k_{\rm B} \cdot \SI{50}{\milli\kelvin}  \approx \SI{4.3}{\ueV}$), couplings between QDs and Majorana zero modes (MZMs) ($\approx \SI{5}{\ueV}$), measurement drive amplitude at the QDs ($\approx \SI{5}{\ueV}$) and measurement frequency ($h \cdot \SI{500}{\mega\hertz} \approx \SI{2}{\ueV}$). Other nearby energy scales such as the (topological) gap ($\qtyrange[range-units=single]{10}{100}{\ueV}$), the charging energy of QDs ($\qtyrange[range-units = single]{40}{200}{\ueV}$), and the level spacing in the semiconductor ($\qtyrange[range-units = single]{10}{50}{\ueV}$) may also come into play. Moreover, the coupling to the environment through charge and other noise ($\approx \SI{1}{\ueV}$) is also crucial for a quantitative description. While qualitative intuition can be gained when ignoring some of these scales, a quantitative description requires a consistent treatment of the device dynamics, fast measurements and noise.

Here, we describe a framework that is able to capture all of these effects in a controlled approximation. We focus on dispersive gate sensing, which is a common technique to probe quantum devices through microwave resonators; however, the framework can easily be adapted to other settings. In \Cref{sec:rf-simulation}, we cast the problem as the evolution of an open quantum system, which can be tackled using established methods for open-system dynamics. We focus here on the recently proposed Universal Lindblad Equation (ULE)~\cite{Nathan20}, which treats the coupling to the environment perturbatively using the Born-Markov approximation but crucially remains valid in the limit of nearly degenerate ground states (as is typical for topological qubit devices) and can treat arbitrary time-dependence of the Hamiltonian (allowing us to include effects of the microwave drive). We demonstrate our framework for the example of a MZM coupled to a QD (or equivalently a double QD) in \Cref{sec:QD-MZM}.

While there are many competing scales at work, there are still well-defined approximations that can be applied to make this problem numerically tractable. For example, typically device dynamics happen at scales of the order of $\sim \SI{1}{\nano\second}$ and the response time of readout resonators typically is of the order of $\sim \SI{10}{\nano\second}$ for dispersive gate sensing measurements. These time scales correspond to energies $<\SI{1}{\micro \electronvolt}$. Combined with device temperatures and measurement frequencies of the order of a few $\SI{}{\micro\electronvolt}$ means that while the dynamics needs to take into account the complex structure of the device, it is typically sufficient to focus on moderately sized Hilbert spaces. This observation has been the basis of similar computational studies of semiconductor qubits, see e.g. Refs.~\cite{nielsen2010configuration,shehata2023modeling}

It can, however, be challenging to identify the relevant low-energy states from microscopic models in a systematic and controlled fashion. In \Cref{sec:methods}, we describe a combination of approaches that can be used to find the low-energy states of microscopic models involving many thousands of degrees of freedom, and that taken together can be used to put the numerical solution of the master equation for a microscopic model within reach. Our approach rests on ideas inspired by methods common in quantum chemistry. Specifically, we introduce the \emph{aggregated truncated configuration interaction} (ATCI) method where we first construct a low-energy Hilbert space at a given parameter point from few-particle excitations above the generalized Hartree-Fock (gHF)~\cite{bach1994generalized} approximation to the ground state. Taking the direct sum of these Hilbert spaces across several nearby parameter points, we define an \emph{aggregated low-energy Hilbert space}. By comparing with density-matrix renormalization group (DMRG) simulations~\cite{white1992density}, we demonstrate that variational energies within this aggregated low-energy Hilbert space are accurate within $\sim10\,\unit{\nano\electronvolt}$, while being much faster to obtain and the convergence of the simulation being independent of the geometry of the simulated model. We then show in \Cref{sec:projectionDynamics} how ATCI can be used to construct a sufficiently rich low-energy Hilbert space to simulate the low-energy dynamics of the system.

These simulations can be further accelerated by the \emph{natural orbitals} approximation introduced in \Cref{sec:natural-orbitals} which reduces the number of microscopic degrees of freedom by projecting onto the states with the largest contribution to the single-particle reduced density matrix.

Finally, in \Cref{sec:MPR}, we combine these methods to perform simulations of an interferometric parity readout. In Ref.~\cite{Aghaee24}, the corresponding response was simulated based on a simplified few-level model which could be fit to the experimental data. Similar open-systems approaches were employed in Ref.~\cite{munk2020parity,steiner2020readout} to illustrate decay mechanism and the readout using a quantum point contact. In other recent work~\cite{sau2024capacitance}, the quantum capacitance response was studied for a microscopic model in the linear response regime. Here, we go beyond these results by studying a microscopic Rashba model for the proximitized nanowire and including the effects of interactions, finite rf drive and noise. Unlike the few-level models, this allows us to access the full topological phase diagram. We start with a microscopic model of $\mathcal{O}(1000)$ degrees of freedom, which can be projected using natural orbitals to $\mathcal{O}(100)$ degrees of freedom. Using the aggregated truncated configuration interaction approach we can then construct many-body Hilbert spaces of $\mathcal{O}(10)$ states which we use to solve the dynamics of the problem. Following similar steps as in this example, these approximations enable studying the dynamics, including noise and fast readout, of a wide variety of complex mesoscopic devices by systematically reducing a system to the relevant degrees of freedom.

\section{Open-system description of rf response of quantum devices}
\label{sec:rf-simulation}

In this section, we outline the general framework we use to simulate the dynamics and rf response of quantum devices in the presence of non-idealities such as finite temperature, finite amplitude of the rf drive tone, and decoherence mechanisms such as charge noise. This framework can be applied to various quantum devices and we demonstrate some of the subtleties of the approach and make connections with analytic results in \Cref{sec:QD-MZM} where we discuss the example of a QD coupled to a MZM (which can be mapped to a double QD). 

\subsection{Theoretical background}

In general, a quantum device in the presence of noise can be described by the Hamiltonian
\begin{equation}
    H_{0} = \Hqd + H_{\rm noise}\,
    \label{eq:H0}
\end{equation}
where $\Hqd$ describes the quantum device while $H_{\rm noise}=\sum_i X_i \Phi_i +H_{\rm env}$ captures the coupling of the quantum device to a noisy environment. We use the convention where $X_i$ are dimensionless qubit device operators and $\Phi_i$ is an environment operator that contains any dimensionful coupling constants. In principle, $H_{\rm noise}$ includes a macroscopic number of degrees of freedom, many of which are not precisely controlled, and thus a direct solution of the time-dependent Schrödinger equation of $H_{\rm 0}$ is intractable. A more manageable approach is to describe the system in terms of the evolution of the reduced density matrix $\rho(t)$ of the quantum device, which can be thought of as capturing our knowledge about the quantum state of the device at any time, while the imperfectly controlled degrees of freedom in the environment are being traced out.

There exist various methods in the literature to obtain an approximate form of the equation of motion of $\rho(t)$,
\begin{equation}
    \dot{\rho}(t) = f[\rho]   \label{eq:EOM}
\end{equation}
Notable examples are the Nakajima–Zwanzig equation~\cite{nakajima_1958,zwanzig_1960}, Bloch-Redfield equation~\cite{redfield_1965}, coarse-grained master equation \cite{Mozgunov20}, and Universal Lindblad equation (ULE)~\cite{Nathan20}. While the general framework to describe the rf response below can be applied with any method that can map a time-dependent Hamiltonian and coupling to environmental noise to $\rho(t)$, the specific calculations in this paper will use the ULE formulation, which we review in Sec.~\ref{sec:ULE}.

\subsection{Dynamical readout response of the quantum device}
\label{sec:readout}
Using Eqs.~\eqref{eq:H0} and \eqref{eq:EOM} as a starting point, we now discuss how to augment the model to simulate the readout of a QD that forms part of the quantum device and is coupled to a readout resonator. We will focus on the case of a single QD coupled to a single readout resonator, but the formalism can be trivially extended to the readout of multiple QDs. When including the readout, the Hamiltonian takes the form
\begin{equation}
    H = H_{0} + H_{\rm qd-res} + H_{\rm res} +H_{\rm I/O} \label{eq:full_H}\,
\end{equation}
where $H_{\rm res}=\hbar \omegar \hat{a}^\dagger \hat{a}$ is the Hamiltonian of the resonator in terms of the resonator (angular) frequency $\omegar$ and the photon operators $\hat{a}$, $H_{\rm qd-res}$ captures its coupling to the quantum device, and $H_{\rm I/O}$ describes the input and output lines used to drive and read out the resonator.

We denote the gate voltage of the gate connected to the resonator as $\Vg$. This voltage will couple to the QD via a capacitance $\Cg$ and cause an offset charge $\ng=\alpha e\Vg/(2\EC)$ in the charging energy term of the QD, $\EC(\hat{n}-\ng)^2$, where $\alpha=2\Cg \EC/e^2$, $\EC$ denotes the charging energy, and $\hat{n}$ is the number operator of the QD. In the classical limit, photons in the resonator lead to an oscillating component $\ng\to \ng+\delta \ng  \cos(\omega t)$, with $\delta \ng =e\alpha \Vrf/(2\EC)$ where $\Vrf$ and $\omega$ are the amplitude and angular frequency of the voltage oscillations in the resonator supplied by the readout drive. Identifying $\cos(\omega t) \hat{=} (\hat{a}+\hat{a}^\dagger)/(2 |a|)$, where  $|a|=\sqrt{\langle \hat{a}^\dagger\hat{a}\rangle}$, and using $\hbar \omegar |a|^2=C\Vrf^2/2$, where $C$ is the full capacitance (including parasitics) of the resonator, motivates the quantum version of the coupling between the qubit device and resonator photons:
\begin{equation}
     H_{\rm qd-res} +H_{\rm res}=\hbar(\omegar-\omega) \hat{a}^\dagger \hat{a} + g \hat{n} (\hat{a}e^{-i\omega t}+\hat{a}^\dagger e^{i\omega t})\,
\end{equation}
where $g=e\alpha \sqrt{\hbar \omegar /2C}$, and we transformed into the rotating frame where the external drive becomes static.

The coupling to $H_{\rm I/O}$ is most conveniently described by standard input/output (I/O) theory, see, e.g. Ref.~\cite{clerk2010}. Here we assume a single input tone at angular frequency $\omega$ and corresponding input/output photon operators $\hat{b}_{\rm in,out}$. In the rotating frame the equation of motion of the resonator photons then takes the form
\begin{equation}
        \dot{\hat{a}} = i(\omega-\omegar)\hat{a} -i\hat{\Lambda}(t) -\frac{\kappa_{\rm e}+\kappa_{\rm i}}{2}\hat{a}-\sqrt{\kappa_{\rm e}}\hat{b}_{\rm in}\, 
        \label{eq:aEOM}
\end{equation}
where $\kappa_{\rm e}$, $\kappa_{\rm i}$ are, respectively, the coupling rate of the resonator to the I/O line and the internal resonator loss rate. The quantum device-resonator coupling gives rise to $\hat{\Lambda}(t)= g \hat{n}(t) e^{i\omega t}$, where $\hat{n}(t)=U(t)^\dagger \hat{n} U(t)$ is the number operator $\hat{n}$ in the Heisenberg picture as described by the time evolution operator $U(t)$ of the quantum system.

For small fermionic Hamiltonians $H_0$ and with resonator occupations of only a few photons it becomes possible to solve the (open) system dynamics of the coupled system of \Cref{eq:full_H} with an explicit treatment of the quantum character of the photons. For typical dispersive gate sensing measurements of QDs, however, $|a|\gg 1$ and the resonator photons couple weakly to the fermionic Hamiltonian. In this case, one can neglect entanglement of the photons and the fermionic system, which can be formally implemented by taking the expectation value of Eq.~\eqref{eq:aEOM} and assuming that the resonator photons are described by a classical complex field $\hat{a} \to a$. We will discuss this classical approximation and the comparison to the explicit treatment of the photons using the example of a double QD coupled to resonator photons in \Cref{sec:quantum_photons}. Staying within the classical approximation gives rise to \begin{equation} \label{eqn:Lambda}
  \hbar\Lambda(t) = g\langle \hat{n}\rangle(t)e^{i\omega t} \,, 
\end{equation}
where $\langle \hat{n}\rangle(t)=\tr(\rho(t) \hat{n})$ is the expectation value of $\hat{n}$ which acquires an oscillating component due to the classical drive.

We can thus describe the coupled evolution of the density matrix and the resonator photons via
\begin{eqnarray}
\tHqd &=& \Hqd+2g\hat{n} |a|\cos(\omega t) \label{eq:driven_H}\\
 \dot{\rho} &=& \tilde{f}[\rho] \\
 \dot{a} &=& i(\omega-\omegar-\chi)a -\frac{\kappa_{\rm e}+\kappa_{\rm i}+\kappa_{\rm qd}}{2}a-\sqrt{\kappa_{\rm e}}b_{\rm in}\, \label{eq:class_aEOM}
\end{eqnarray}
where $\tilde{f}$ describes the right hand side of the equation of motion \eqref{eq:EOM} modified due to using $\tHqd$ instead of $\Hqd$. We split up the real and imaginary part of $\Lambda$ via
\begin{equation}
    \Lambda(t)/|a|=\chi-i\kappa_{\rm qd}/2
\end{equation}
to emphasize that this method yields the dynamical response which can describe both a frequency shift $\chi$ of the resonator and a shift in the internal photon loss $\kappa_{\rm qd}$, where $\chi$ ($\kappa_{\rm qd}$) is due to in (out of) phase oscillations of $\langle \hat{n}\rangle(t)$ relative to the drive $\propto \cos(\omega t)$. Here we used that, without loss of generality, the phase offset of the drive can be set to zero for the purpose of calculating $\chi$ and $\kappa_{\rm qd}$ so that the equations of motions of $\rho$ and the resonator photons only couple via the absolute value $|a|$.

The classical treatment of the drive described here bears similarities to semiclassical master equations derived in the literature (see, e.g. Ref~\cite{vigneau_probing_2023}) which are typically combined with a quasi-adiabatic approximation of the drive. By keeping the full quantum evolution due to the drive and embedding it consistently in the Lindblad master equation we avoid the additional requirement of low measurement frequencies. Following the same goal, Peri et al.~\cite{peri_2024} recently developed a framework using a polaron-transformation of the Hamiltonian in a single QD system. A comparison between these approaches would be an interesting direction of further research.

The shifts $\chi,\kappa_{\rm qd}$ due to the quantum device lead to shifts in the quadratures of the resonator photons $a$ which can be detected by the change in amplitude and phase of the outgoing field~\cite{clerk2010}
\begin{equation}
b_{\rm out}=b_{\rm in}+\sqrt{\kappa_{\rm e}}a. \label{eq:bout}
\end{equation}
This allows to directly calculate the outcoming signal $b_{\rm out}$ of the measurement when the resonator is driven by an input tone of strength $b_{\rm in}$. The output power of the signal is given by $P_{\rm out}=\hbar\omega |b_{\rm out}|^2$. To capture inevitable noise of the measurement (either quantum noise or noise via the amplification chain) one can add a Gaussian random variable $b_\sigma $ via
\begin{equation}
    b_{\rm out} \to b_{\rm out} + b_{\sigma}, \label{eq:bnoise}
\end{equation} with a standard deviation $\sigma = \sqrt{P_{\rm N}/\hbar\omega}$. As an example, the noise power of an amplification chain with effective noise temperature $T_{\rm N}$ and integration time of $\tau_{\rm m}$ is given by $P_{\rm N}=k_{\rm B} T_{\rm N}/\tau_{\rm m}$, where $k_{\rm B}$ is the Boltzmann constant.

In summary, one can use Eqs.~\eqref{eq:driven_H}-\eqref{eq:class_aEOM} together with Eq.~\eqref{eq:bout} and the substitution~\eqref{eq:bnoise} to simulate the outcoming signal of the measurement that is directly affected by the dynamics of the quantum device. At the same time, one can obtain the time evolution of the quantum device that is driven due to the measurement resonator. Below we will address some examples and potential additions to the method described above.

\subsection{Dynamical quantum capacitance}\label{sec:dynCQ}

The photon dynamics of $|a|$ change on the scale given by $\kappa_{\rm e}+\kappa_{\rm i}=\omega/Q$  (where $Q\gg 1$ is the overall quality factor of the resonator) which is typically much slower than the dynamics of $\tHqd$. In some cases $|a|$ even becomes static where fast fluctuations in $\chi$ and $\kappa_{\rm qd}$ are averaged out on the scale of the resonator dynamics and potential slower state flips only affect the phase of $a$.

Such scenarios then allow for decoupling the equations of motion of the quantum device and the photons in which the response can be cast in form of a complex dynamical quantum capacitance
\begin{equation}
    \CQ = \frac{2C}{\hbar \omegar |a|}g \langle\overline{\hat{n}(t)e^{i\omega t}}\rangle \,.
    \label{eq:dynCQ}
\end{equation}
Here, the overline indicates time averaging over a period of the drive which eliminates higher frequency components of $\omega$ in the response which are averaged out at the level of the slowly changing photon response. The real part of $\CQ$ is related to the frequency shift and the imaginary part is related to the shift in loss. In the static limit of $\Vrf,\omega \to 0$, this dynamic form of the quantum capacitance coincides with the static quantum capacitance $\CQ^{(0)}=-(e\alpha/2\EC)^2 (\partial^2/\partial \ng^2) \langle \Hqd\rangle$. 

\subsection{Universal Lindblad Equation}
\label{sec:ULE}

The universal Lindblad equation (ULE) formalism~\cite{Nathan20} describes a set of approximations to derive an equation of motion of the form $\eqref{eq:EOM}$ from Eq.~\eqref{eq:H0}. In the Markovian setting, the most general equation that describes the noisy evolution of a quantum device coupled to its environment and preserves the physicality of $\rho$ is given by a Lindblad master equation,
\begin{eqnarray}
     \dot{\rho} = -\frac{i}{\hbar}[{\tilde{H}}_{\rm qd},\rho]  + \sum_i \mathcal{D}[L_i]\rho \label{eq:Lindblad-me} \\
    \mathcal{D}[L]\rho= L\rho L^\dagger -(L^\dagger L \rho + \rho L^\dagger L)/2.
\end{eqnarray}
Here, the $L_i$ are a set of operators often referred to as Lindblad or jump operators. The ULE framework describes how to systematically obtain approximate forms of the operators $L_i$ from Eq.~\eqref{eq:H0}.

To derive the operators $L_i$, the ULE assumes a weak coupling to a Gaussian environment with short memory time (Born-Markov approximation). The behavior of the noise is then fully captured by the noise spectral functions
\begin{equation}
S_i(\omega)=\int \dd t\langle \Phi_i(t) \Phi_i(0)\rangle_{\rm env}e^{i\omega t},
\end{equation}
where the expectation value $\langle \dots \rangle_{\rm env}$ is taken with respect to the density matrix of the environment.   
While $1/f$ charge noise can be non-Gaussian in general, in the limit of weak coupling to the environment the Gaussian contribution dominates; see Ref.~\cite{paladino2014} for a broader review of this subject.

With these assumptions, one can derive a time-local master equation for $\rho$ by directly applying second-order perturbation theory in the system-environment coupling. The resulting Bloch-Redfield equation, however, will in general not be of Lindblad form and can lead to unphysical results for $\rho$. For typical qubit systems this is dealt with by applying a rotating wave approximation (RWA) to bring the master equation into Lindblad form. This is well-justified for a two-level system with large level spacing, but the RWA breaks down for many mesoscopic quantum devices and in particular for topological qubits with a degenerate qubit subspace. Recently it was realized that, while convenient when valid, the RWA is not required to derive a Lindblad form of the dissipator. In fact, as argued by Nathan and Rudner~\cite{Nathan20}, the Born-Markov approximation of the environment introduces a family of approximate master equations that includes the Bloch-Redfield equation but can also be constructed to be of Lindblad form~\cite{Nathan20,Schaller08,Kirsanskas18,Mozgunov20} using the same assumptions. It can be rigorously shown~\cite{Nathan20,Mozgunov20} that these approximate master equations have small errors as long as the fastest dephasing times $1/\Gamma_{\rm sys}$ of the system are long compared to the correlation time of the environment $\tau_{\rm env}$.

Here we use the ULE expressions for deriving the Lindblad operators that enter into Eq.~\eqref{eq:Lindblad-me}. Specifically,
\begin{equation} 
    L_i = \int \dd s\, g_i(t-s) U(t,s)X_i U(s,t)\,,
    \label{eq:L_ULE}
\end{equation}
where $g(t)=\int \dd \omega\sqrt{S_i(\omega)}e^{-i\omega t}/2\pi$ and $U$ is the time evolution operator associated with $\tilde{H}_{\rm qd}$. The rigorous error bounds for the applicability of these Lindblad operators can be expressed in terms of the small parameter $\Gamma_{\rm sys}\tau_{\rm env}$ with
\begin{eqnarray}
\Gamma_{\rm sys} &=& 4 \left[ \int \dd t |g(t)| \right]^2 \\
\tau_{\rm env} &=& \frac{\int \dd t |g(t) t|}{\int \dd t |g(t)|}.
\end{eqnarray}
In addition to the dissipative part of the master equation $\mathcal{D}[L_i]$, the coupling to the environment will also introduce a Lamb shift. Here we will not explicitly incorporate the Lamb shift but instead assume that the quantum device Hamiltonian already contains parameters renormalized by the coupling to the noise.

\subsection{Treatment of classical and quantum noise}
\label{sec:cl_q_noise}

In principle there are two ways of including noise in the master equation framework of \Cref{eq:Lindblad-me}. Noise can either be added at the level of the Lindblad operators as discussed above or in the form of classical noise by adding time-dependent fluctuations of the Hamiltonian. The latter approach has been used previously to describe dephasing without a master equation framework, e.g. in Ref.~\cite{mishmash2020dephasing}, and one can follow the approach discussed there to generate an ensemble of noise trajectories. Both of these approaches have advantages and disadvantages. The Lindblad operators can capture quantum noise with spectral functions that are asymmetric, i.e. where $S(\omega)\neq S(-\omega)$, but this description is limited to weak Markovian noise in order for $\Gamma_{\rm sys}\tau_{\rm env}$ to remain a small parameter. The classical noise, on the other hand, is limited to capture frequency-symmetric spectral functions, but can capture noise of arbitrary strength and with non-Markovianity.

It can be useful to use a combination of these approaches for noise that is strong at low frequencies. A prime example are forms of pink noise which often occur in mesoscopic devices~\cite{paladino2014} and can lead to a divergence in $\Gamma_{\rm sys}$ which prohibits perturbative treatments of noise. Splitting up the noise into a strong classical part and a weak quantum part can then extend the regime of applicability of the ULE. Specifically, consider a noise spectral function $S$ that violates the condition of sufficiently small $\Gamma_{\rm sys}$.
One can attempt to decompose the noise into two parts, $S(\omega)=S_\mathrm{q}(\omega)+S_\mathrm{r}(\omega)$, such that the quantum part $S_\mathrm{q}(\omega)$ gives rise to a sufficiently small $\Gamma_{\rm sys}$ while the remainder is approximately symmetric and can be treated as classical noise with spectral function $S_\mathrm{cl}(\omega) = \left[ S_\mathrm{r}(\omega) + S_\mathrm{r}(-\omega)\right]/2$. Intuitively, such a decomposition can be obtained by choosing some cutoff frequency $\omega_0 < k_B T$ and defining $S_\mathrm{q}(\omega)$ to match $S(\omega)$ for $\omega > \omega_0$ while saturating to some finite value below $\omega_0$, while taking the classical part to be the difference between the desired $S(\omega)$ and $S_\mathrm{q}(\omega)$. Therefore, $\omega_0$ serves as high-frequency cutoff for the classical noise and low-frequency saturation point for the quantum noise. We provide a concrete example for the case of $S(\omega) \propto 1/\omega$ in \Cref{sec:charge_noise}.

\subsection{Dynamical $\CQ$ in the Floquet steady state}
If one is interested in calculating the dynamical $\CQ$ response of \Cref{eq:dynCQ} for a system that is time-independent except for the periodic drive of the measurement tone, it is sufficient to consider the driven steady state instead of solving the full equation of motion of $\rho$. Let's consider the Hamiltonian from \Cref{eq:driven_H} of the system driven by the measurement $\tilde{H}_{\rm qd}=H_{\rm qd}+ V_{\rm D} \hat{n}\cos(\omega t)$, where we introduced $V_{\rm D}=2g|a|$. 

We can then make use of the fact that $\tilde{H}_{\rm qd}$ is time periodic, which by construction leads to time-periodic Lindblad operators $L(t)$. In the driven steady state also $\rho(t)$ will become time periodic and we can thus expand $\rho(t)=\sum_n \rho_n e^{-i\omega n t}$ and $L(t)=\sum_n L_n e^{-i\omega n t}$ to obtain an algebraic equation for the Fourier components of the steady state density matrix
\begin{multline} \label{eq:steady_state}
    i\omega n \rho_n = i[H_{\rm qd},\rho_n] +i\frac{V_{\rm D}}{2}[\hat{n},\rho_{n+1}+\rho_{n-1}] \\
    -\!\sum_{l,m}\! \Big(\! L_l \rho_{n+m-l} L_m^\dagger - \frac{1}{2}\big\{L^\dagger_m L_l,\rho_{n+m-l} \big\}\!\Big). 
\end{multline}
In practice, it is often sufficient to keep a small number of Fourier components because the occupation of components with $|n|\omega \gg k_{\rm B} T$ is small. This allows for a fast solution of \Cref{eq:steady_state}, which can then be used to directly calculate the dynamical $\CQ$ via
\begin{equation}
    \CQ = \frac{2e^2\alpha^2}{V_{\rm D}}\tr\left\{\hat{n}\rho_1\right\}\,.
\end{equation}

\subsection{Quantum trajectories}

The Lindblad form of the master equation of the quantum device lends itself to an alternative description in terms of quantum trajectories of pure states, with the Lindblad operators taking the role of "jump operators" that induce discontinuous transitions in the state (see, e.g., Ref.~\cite{wiseman2010measuerment}). This description is equivalent in the sense that averaging over such trajectories will recover the same result as obtained from evolving under the master equation (up to statistical uncertainty due to sampling a finite number of trajectories).

In this framework, a single pure-state trajectory of the non-normalized wavefunction $\ket{\psi}$ evolves via
\begin{equation}
    \frac{\partial}{\partial t} \ket{\psi} = f_{\rm qt}[\psi]
\end{equation}
where $f_{\rm qt}[\psi]=-[(i/\hbar)\tHqd+\sum_i L^\dagger_i L_i/2]\ket{\psi}$. At the beginning of the evolution, a number $r$ is chosen uniformly in the interval $[0,1]$. At the time $t$ such that $\braket{\psi(t)|\psi(t)} = r$, a jump operator $L_i$ is chosen with probability $w_i \propto \bra{\psi}L^\dagger_i L_i\ket{\psi}$, the state jumps to $\ket{\psi}\to L_i\ket{\psi}/\sqrt{w_i}$, and a new random number $r$ is chosen\footnote{When computing time-dependent observables in this framework, it is important to take into account that the wavefunction is not normalized.}.

This approach is particularly useful when applied to quantum measurements, where the jump operators can be used to capture the (potentially partial) projection of the system, and different trajectories are used to capture the statistics of measurement outcomes. Such trajectories can then be used to estimate fidelities of qubit measurements and measurement-based operations, as done in the context of measuring Majorana qubits using quantum point contacts in Ref.~\cite{schulenborg2023detecting}.

\section{Rf readout of a QD coupled to a MZM}
\label{sec:QD-MZM}

We will now demonstrate how to apply the formalism introduced in \Cref{sec:rf-simulation} and discuss some possible subtleties for the example of a QD coupled to a single MZM. The Hamiltonian takes the form
\begin{equation}
    H_{\rm QD-MZM} = E_C(\hat{n}_d - \ng)^2 + t_{\rm C} (d\gamma + \text{h.c.}) \,
\end{equation}
where $\hat{n}_d=d^\dagger d$ in terms of the  QD operator $d$, $E_C$ is the charging energy of the QD and $t_{\rm C}$ the strength of the tunnel coupling to the MZM $\gamma$. For a fixed total parity sector and close to charge degeneracy ($n_{\rm g}=0.5$) the Hamiltonian can be recast as a simple 2-level system 
\begin{equation}
    H_{\rm TLS} = \frac{\Delta}{2}\sigma_Z + t_{\rm C} \sigma_X\,,
    \label{eq:TLS}
\end{equation}
where $\Delta = \EC(1-2\ng)$ is the detuning and we use the basis states corresponding to an empty/occupied QD (up to an arbitrary offset charge) such that the QD number operator becomes $\hat{n}_d=\sigma_Z/2 + \openone/2$ (for notational simplicity, we will drop the irrelevant constant term $\openone/2$). From the general form \Cref{eq:TLS} it is clear that this setup equally describes a simple double QD system.

We consider a measurement where the rf readout is coupled to the gate that controls the detuning of the QD. Using the results of \Cref{sec:rf-simulation}, the Hamiltonian in the presence of the readout drive takes the form
\begin{equation}
    \tilde{H}_{\rm TLS} = \left[\Delta+V_{\rm D} \cos(\omega t)\right]\frac{\sigma_Z}{2} + t_{\rm C} \sigma_X\,,
    \label{eq:H_tilde}
\end{equation}
where $V_{\rm D}=e\alpha V_{\rm rf}$ is the drive amplitude of the detuning oscillations due to the measurement tone at frequency $\omega$.

Following \Cref{eq:dynCQ}, the dynamical capacitive response becomes
\begin{equation}
    \CQ = \frac{e^2 \alpha^2}{V_{\rm D}}\braket{\overline{\sigma_Z(t) e^{i\omega t}}}\,.
    \label{eq:CQ_DQD}
\end{equation}

\subsection{Noise channels}

The various noise channels of the environment can be specified via noise spectral functions and corresponding Lindblad operators. A key coupling mechanism of noise is via the occupation of the QD, i.e. $X_{\rm g}=\hat{n}_d=\sigma_Z$/2. Examples of underlying mechanism that couple in such a way are charge noise and phonons. Additionally, it is possible for the coupling term to fluctuate, leading to an incoherent tunneling contribution described by the noise operator $X_{\rm t}=\sigma_X$.

\subsubsection{Phonon noise}

A microscopic derivation of $S_{\rm ph}$ is beyond the scope of this paper. Instead one can use the approximate form
\begin{equation}
S_{\rm ph}= \frac{\hbar \omega}{2\pi \rho v^3}\left(D^2\omega^2/v^2 + e^2h_{14}^2\right) \frac{1}{1-e^{-{\hbar \omega}/{k_{\rm B} T}}}
\label{eq:S_ph}
\end{equation}
which abstracts away details of various form factors and is motivated in \Cref{app:phonons}. A typical material for the QDs coupled to MZMs is InAs and one can thus estimate the relevant parameters from the values of bulk InAs with deformation potential $D=\SI{5.1}{\electronvolt}$, the piezoelectric coupling $h_{14}=\SI{3.5e6}{\volt/\cm}$, the density $\rho=\SI{5.7}{\gram/\cm^3}$ and phonon velocity $v=\SI{4}{\km/\second}$~\cite{vurgaftman01,madelung04}. For the InAs parameters mentioned above one finds typical $S_{\rm ph}\sim 1/\SI{10}{\nano\second}$ at energy scales $\hbar \omega\sim \SI{5}{\ueV}$ which are typical scales for QD-MZM devices \cite{Aghaee24}.

\subsubsection{Charge noise}
\label{sec:charge_noise}
Charge noise typically dominates QD devices at low frequencies~\cite{paladino2014} and is often treated via a classical $1/f$ spectral function. In order to capture the right interplay between dephasing and relaxation/excitation due to the noise, a description in terms of classical noise breaks down once $\hbar \omega$ approaches the temperature $k_{\rm B} T$, as a classical (infinite-temperature) noise source would heat up the steady state of the quantum system. In order to deal with this issue we use a quantum version of the $1/f$ noise,
\begin{equation}
    S_{C}(\omega) =\frac{\alpha_C^2}{|\omega|}\frac{2}{1+e^{-\frac{\hbar \omega}{k_{\rm B} T}}}\,,
    \label{eq:S_C}
\end{equation}
where the latter term assumes that the charge noise is compatible with a noise environment at thermal equilibrium\footnote{This may require sufficient filtering of the instrument noise at high frequencies.} so that $S_{C}(\omega)/S_{C}(-\omega)=\exp(\hbar \omega/k_{\rm B} T)$ but otherwise recovers standard expressions for the $1/f$ noise at low frequencies. For InAs QDs  $\alpha_C \sim \SI{1}{\ueV}$ is a typical value~\cite{Aghaee24}.

Since a spectral function of the form \eqref{eq:S_C} would lead to divergent rates $\Gamma_{\rm sys}$ we use the method described in \Cref{sec:cl_q_noise} to split up the noise into a quantum version $S_{\rm q}=2\alpha_C^2/\sqrt{\omega^2+\omega_0^2}/[1+\exp(-\hbar\omega/k_{\rm B}T)]$ with cutoff $\omega_0$ that enters the Lindblad operators while capturing the remaining noise classically. We show in \Cref{app:cutoff} that for moderate charge noise a cutoff $\omega_0<k_{\rm B} T/\hbar$ can be chosen such that $\Gamma_{\rm sys} \tau_{\rm env} \sim (\alpha_C/k_{\rm B} T)^2\ll 1$.

While in principle, the classical noise requires repeating the time evolution for many different noise realizations, we find that in many practical situations the classical charge noise can be added via a convolution of the calculated response with a Gaussian broadening $\sigma_\Delta$ of the detuning parameter given by
\begin{equation}
    \sigma_\Delta^2 = \int_{-\infty}^{\infty} \frac{d\omega}{2\pi} S_{\rm cl}(\omega)\approx \frac{\alpha_C^2}{\pi}\log(\omega_0 \tau_{\rm m}),
    \label{eq:sigmaDelta}
\end{equation} where the measurement time of a collected data point $\tau_{\rm m}$ provides an effective low-frequency cutoff of the classical noise and we evaluated the integral to leading logarithmic order.

\subsection{Simplified noise model}
For qualitative discussions it is often useful to parameterize the noise by a simple single-parameter spectral function of the form
\begin{equation}
    S_{\rm eff} = \frac{\hbar^2 \gamma_{\rm eff}}{1+e^{\hbar \omega/k_{\rm B} T}}
    \label{eq:S_eff}
\end{equation}
which can be thought of as a phenomenological description of the noise described by relaxation/excitation processes which act at a rate $\gamma_{\rm eff}$ while also fulfilling detailed balance.  The relevant scale for detuning noise ($\gamma_{\rm eff}=\gamma_{\rm g}$) is determined by combining phonon and $1/f$ charge noise at the typical energy scales that are important for relaxation/excitation processes. Extrapolating the charge noise measurements of Ref.~\cite{Aghaee24} yields typical values of $\gamma_{\rm g}\sim \SI{1}{\giga\hertz}$ for the detuning noise of InAs QDs. In the following examples we will use this value of $\gamma_{\rm g}$ together with the coupling operator $X_{\rm g}=\sigma_Z/2$ and work with the simpler detuning noise $S_{\rm eff}$. We demonstrate the effect of incoherent tunneling terms by also including a noise term with effective strength $\gamma_{\rm t}$ and corresponding operator $X_{\rm t}=\sigma_X$. A showcase of the full treatment of $1/f$ noise including the slow classical noise is discussed in \Cref{sec:MPR}.

\subsection{Quantum treatment of resonator photons}
\label{sec:quantum_photons}

We will now compare the formalism of a classical drive of \Cref{eq:H_tilde,eq:CQ_DQD} to the explicit treatment of resonator photons. The latter is implemented via the Hamiltonian
\begin{multline}\label{eq:H_ph} 
    H = H_{\rm TLS}+(\omega_{\rm r}-\omega)\hat{a}^\dagger\hat{a}+\epsilon^* \hat{a}+\epsilon\hat{a}^\dagger \\
    + \frac{g}{2}\sigma_Z(\hat{a}e^{-i\omega t}+\hat{a}^\dagger e^{i\omega t})\,
\end{multline}
where the drive parameter is related to \Cref{eq:aEOM} via $i\epsilon=\sqrt{\kappa_{\rm e}}b_{\rm in}$. The dynamics of \Cref{eq:H_ph} can be solved within the ULE formalism using the same Lindblad operators of the fermionic system $H_{\rm TLS}$ as in the case of the classical drive together with the photon Lindblad operators $L_\uparrow=\sqrt{\kappa n_B}a^\dagger$ and $L_\downarrow=\sqrt{\kappa (1+n_B)}a$ where $\kappa=\kappa_{\rm e}+\kappa_{\rm i}$ and $n_B=(\exp(\omega_{\rm r}/k_{\rm B}T)-1)^{-1}$ is the thermal occupation of the undriven resonator \cite{wiseman2010measuerment}.

\begin{figure}
    \centering
    \includegraphics[width=\columnwidth]{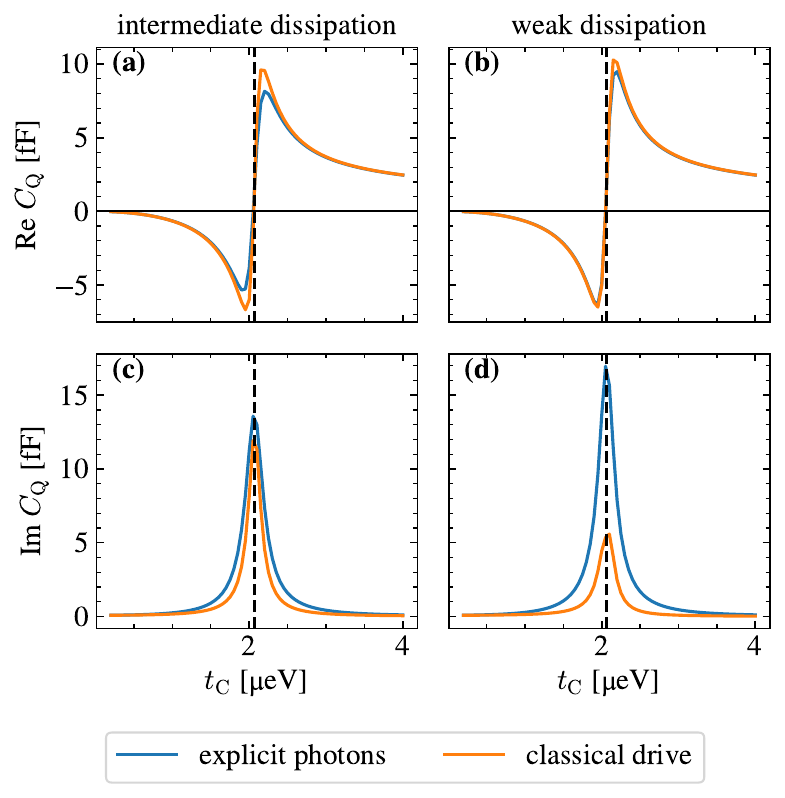}
    \caption{Comparison of explicit quantum treatment of the resonator photons to the classical drive approximation for different values of dissipation. Panels (a) and (c) show the real and imaginary $\CQ$ response for intermediate dissipation. Parameters: $E_C=\SI{50}{\micro\electronvolt}$, $\omega_{\rm r}/2\pi=\omega/2\pi=\SI{1}{\giga\hertz}$, $V_{\rm D}=\SI{1}{\micro\electronvolt}$, $Q_e=Q_i=100$, $C=\SI{600}{\femto\farad}$, $\alpha=0.5$, $\gamma_{\rm g}=\SI{1}{\giga\hertz}$, $\gamma_{\rm t}=\SI{0.2}{\giga\hertz}$, $T=\SI{50}{\milli\kelvin}$ corresponding to $g/\hbar\approx \SI{0.56}{\giga\hertz}$. Vertical dashed lines mark the resonance condition $\hbar\omega=2t_{\rm C}$. Panels (b) and (d) represent an example with weaker dissipation. Parameters are same as in the left column except $\gamma_{\rm g}=\gamma_{\rm t}=\SI{0.2}{\giga\hertz}$. }
    \label{fig:explicit_photons}
\end{figure}

A comparison between the explicit quantum treatment of photons and the classical treatment of the drive described by \Cref{eq:H_tilde,eq:CQ_DQD} is shown in \Cref{fig:explicit_photons}. In order to map the photon response to a complex quantum capacitance we use the $\langle \hat{a}\rangle$ of the the steady state solution and then identify frequency shift and loss by comparing to the steady state solution of \Cref{eq:class_aEOM}. We observe that the $\CQ$ response is generally described well by the classical treatment of the drive. Sizable deviations can occur at resonance of the fermionic and photonic system $2t_{\rm C}\approx \omega_{\rm r}$, especially when driven resonantly $\omega=\omega_r$ and in the limit of weak dissipation $\gamma_{\rm g},\gamma_{\rm t} \ll g/\hbar$. In this regime the combination of entanglement between the fermionic and photon degrees of freedom together with the explicit treatment of photon noise via  $L_\uparrow$ and $L_\downarrow$ leads to much stronger contribution to $\mathrm{Im}\CQ$ compared to the classical treatment of the drive. The phase response described by $\mathrm{Re}\CQ$ is more robust and is generally still captured well by the classical drive model. For large dissipation in the fermionic system $\gamma_{\rm g},\gamma_{\rm t} \gg g/\hbar$ the entanglement to the photons is quenched and the two models agree exactly.

The typical charge noise in QD systems coupled to MZMs \cite{Aghaee24} is sizable and the corresponding dispersive gate sensing measurements are performed away from resonance. With an eye towards this target system, we will thus in the following use the classical treatment of the drive, which applies very well in this regime.

\subsection{Comparison to linear response theory}
To benchmark the formalism and build intuition about the various response regimes it is helpful to compare the simulated dynamical $\CQ$ response to analytical expressions that can be derived in a linear response framework. In \Cref{sec:lin_resp}, we extend the approach of Ref.~\cite{Aghaee24}, treating the noise that can cause transitions between states in a simplified way via a scalar imaginary component of the Green's function.

\begin{figure}
    \centering
    \includegraphics[width=\columnwidth]{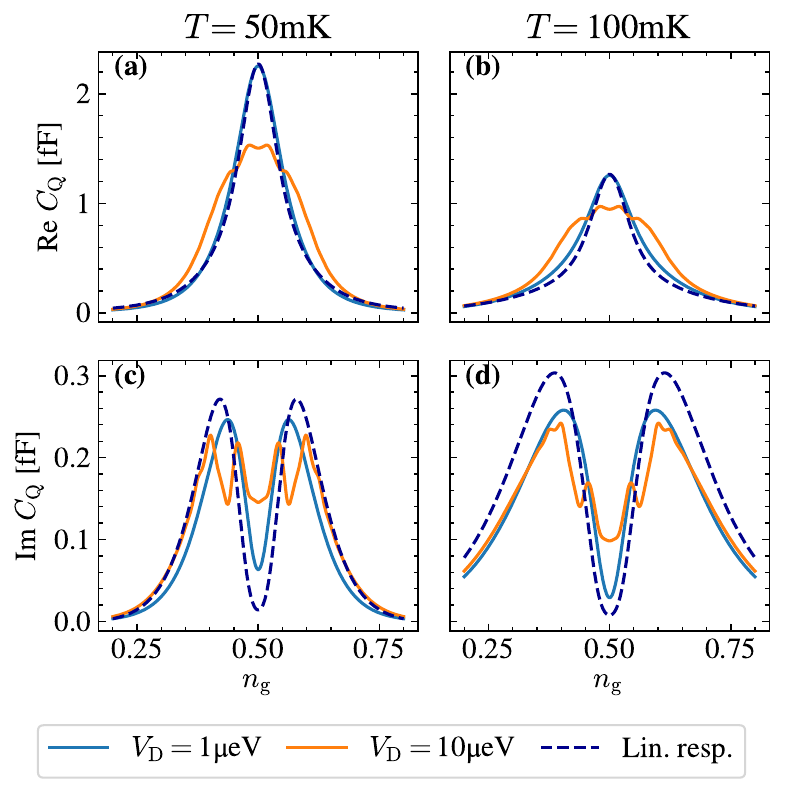}
    \caption{Comparison of dynamical $\CQ$ simulation based on \Cref{eq:CQ_DQD} and linear response theory of \Cref{sec:lin_resp} in the presence of temperature and power broadening. Panels (a) [(b)] and (c) [(d)] show ${\rm Re}~\CQ$ and ${\rm Im}~\CQ$ for $\SI{50}{\milli\kelvin}$ [$\SI{100}{\milli\kelvin}$]. Blue (orange) lines mark the result from the dynamical $\CQ$ simulation with $V_{\rm D}=\SI{1}{\micro\electronvolt}(\SI{10}{\micro\electronvolt})$. Dashed lines are the linear response theory results. Other parameters: $t_{\rm C}=\SI{3}{\micro\electronvolt}$, $E_C=\SI{50}{\micro\electronvolt}$, $\omega/2\pi=\SI{0.5}{\giga\hertz}$, $\gamma_{\rm g}=\SI{1}{\giga\hertz}$, $\gamma_{\rm t}=\SI{2}{\giga\hertz}$.
    }
    \label{fig:comparison_keldysh}
\end{figure}

Figure~\ref{fig:comparison_keldysh} shows a comparison of the simulated dynamical $\CQ$ to the linear response prediction for moderately small $t_{\rm C}=\SI{3}{\micro\electronvolt}$ and for sizable noise terms corresponding to charge noise $\gamma_{\rm g}$ and incoherent couplings $\gamma_{\rm t}$. The real part of $\CQ$ agrees well between the two approaches in the weak drive regime. The imaginary part of $\CQ$ shows qualitative agreement capturing, e.g., the dip in the loss at charge resonance as expected for a Sisyphus resistance contribution, see e.g. Ref.~\cite{vigneau_probing_2023}. We attribute the remaining quantitative differences in the imaginary part to the simplified treatment of noise in our linear response framework. In the regime of strong measurement drive, the dynamical $\CQ$ simulations show a broadening of the $\CQ$ response but also resonance features (particularly visible in ${\rm Im}~\CQ$) that correspond to Landau-Zener-Stückelberg interferometry patterns~\cite{shevchenko_2010} of a strongly driven two-level system. These effects are not captured in the lowest order perturbation treatment of the linear response framework but could in principle be included in a Floquet setting~\cite{Dong17}, which goes beyond the scope of this paper.

The effect of increasing temperature can be observed by comparing the left and right columns of \Cref{fig:comparison_keldysh}. The agreement between both frameworks remains similar as for low temperatures. The key features are a suppression of the $\CQ$ response and a temperature broadening of the $\CQ$ peak. The latter can be understood in terms of contributions from the incoherent tunneling capacitance~\cite{vigneau_probing_2023}, which adds a $\mathrm{Re}~\CQ$ response $\propto\cosh(\Delta/2k_{\rm B}T)^{-2}$. To highlight this effect we chose a sizable incoherent coupling term $\gamma_{\rm t}$ in \Cref{fig:comparison_keldysh} which manifests as an additional broadening of the $\CQ$ peak in the larger temperature regime. This becomes especially pronounced in the imaginary part which shows a significantly wider response than in the low temperature regime.

\subsection{Drive backaction}
A key strength of the formalism introduced above is that it captures the backaction effects of finite drive strength. This interplay is crucial to predict the maximal signal that can be obtained in a given measurement time and thus the SNR. 
In the limit of small capacitive shifts $\CQ \ll C/Q$, the measured signal is proportional to $\CQ$ and $V_{\rm D}$, see, e.g. Ref.~\cite{Aghaee24}.\footnote{Note that here we are defining the signal relative to the response in the Coulomb valley where $\CQ\approx 0$.} We thus define the relevant signal as $S=V_{\rm D}\CQ$, which would diverge for large $V_{\rm D}$ in the absence of backaction effects. However, in the presence of backaction, there is an optimal value of the drive amplitude that maximizes the signal.

To build intuition about backaction effects it is useful to consider the limiting case of a quasistatic drive which can be treated analytically. We thus start with the regime where the drive is slow and the avoided crossing $2t_{\rm C}$ is much larger than temperature such that the system follows the ground state of $\tilde{H}_{\rm TLS}$. The regime of applicability of the drive being sufficiently slow is given by the Landau-Zener condition (see, e.g. \cite{bauer_dynamics_2018} or the original Refs.~\cite{majorana1932atomi,zener1932non,landau1932theorie}) 
\begin{equation}
    P_{\rm LZ}=\exp\left(- \frac{2\pi t_{\rm C}^2}{V_{\rm D} \hbar\omega}\right) \ll 1\,
    \label{eq:LZ}
\end{equation}
which defines a bound on the drive $V_{\rm D} \ll 2\pi t_{\rm C}^2/\hbar \omega$ for the applicability of the quasi-static regime. The expectation value of $\sigma_Z(t)$ can be obtained by calculating the instantaneous ground state of $\tilde{H}_{\rm TLS}$ at each point in time yielding
\begin{equation}
    \langle \sigma_Z(t)\rangle = \frac{V_{\rm D}\cos(\omega t)}{\sqrt{4t_{\rm C}^2+V_{\rm D}^2 \cos^2(\omega t)}}\,.
\label{eq:sigma_Z_qs}
\end{equation}

The capacitive response and signal can then be calculated by averaging \Cref{eq:sigma_Z_qs} over a period of the drive. The general expressions of the corresponding integrals can be evaluated in terms of elliptical functions  \cite{maman_2020}. One can obtain some intuition by considering the limiting regimes of $V_{\rm D}$. For $V_{\rm D} \to 0$ the signal becomes $S/(e^2\alpha^2)=V_{\rm D}/4t_{\rm C}$ and is thus limited by $V_{\rm D}$ while for large $V_{\rm D}$ the signal saturates. In this adiabatic modeling the backaction thus leads to a suppression of $\CQ\propto 1/V_{\rm D}$ due to the measurement apparatus increasingly probing the system away from the charge resonance where the capacitive response is weak. This suppression alone, however, is not strong enough to explain a finite value of optimal drive amplitude, as within this model the signal would saturate (rather than decrease) as $V_{\rm D} \to \infty$. This is resolved by noting that for large drive amplitudes the Landau-Zener constraint \eqref{eq:LZ} will eventually be violated.\footnote{In principle there is also another constraint on $V_{\rm D} \ll E_C$ to stay within the limits of applicability of the single transition model since for $V_{\rm D}\sim E_C$ other charge states need to be taken into account.} The latter leads to an additional contribution to backaction by allowing the drive to induce transitions between the charge ground and excited state. We can incorporate this effect by estimating that in the presence of Landau-Zener transitions the signal will be suppressed by a factor $(1-4P_{\rm LZ})$ as the ground state and excited state give opposite signals and the avoided crossing is traversed twice in a period.

\begin{figure}
    \centering
    \includegraphics[width=1.0\columnwidth]{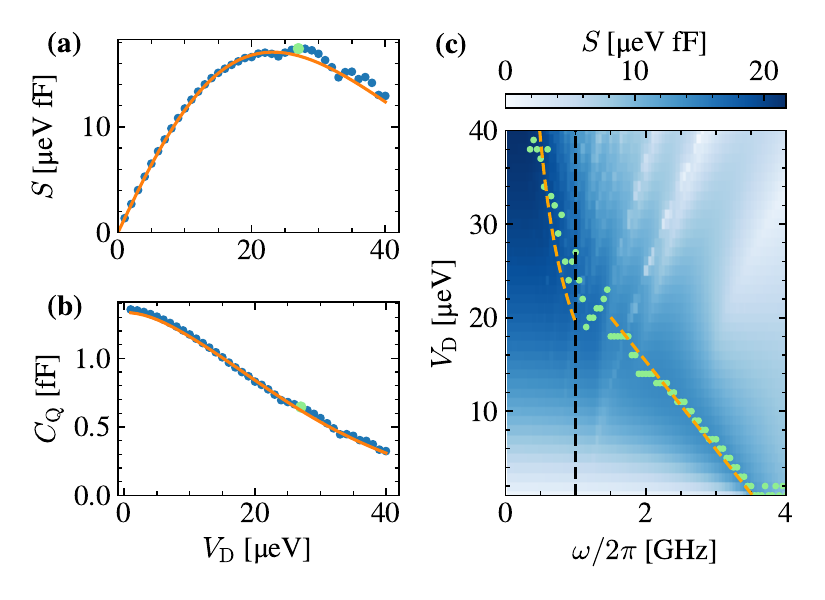}
    \caption{Measurement drive backaction. Panels (a) and (b) show the signal $S$ and $\CQ$ response in the low frequency backaction regime (blue datapoints) as a function of the measurement drive amplitude $V_{\rm D}$. The $\CQ$ response declines due to backaction effects which leads to an optimal choice of $V_{\rm D}=V_{\rm D}^{\rm opt}$ (marked as light green dot) to maximize the signal $S$. The solid orange line is a comparison to the backaction theory in the quasistatic regime discussed in the text. (c) Best signal $S=V_{\rm D} \CQ$ (maximized over $n_{\rm g}$) vs measurement drive amplitude $V_{\rm D}$ and frequency $\omega$. Simulation parameters: $t_{\rm C}=\SI{7.5}{\micro\electronvolt}$ (corresponding to resonance at $2t_{\rm C}/2\pi\approx \SI{3.6}{\giga\hertz}$), $E_C=\SI{50}{\micro\electronvolt}$ , $\gamma_{\rm g}=\SI{1}{\giga\hertz}$, $T=\SI{50}{\milli\kelvin}$. The vertical dashed line indicates the $\omega/2\pi=\SI{1}{\giga\hertz}$ linecut for panels (a) and (b). Light green dots mark the drive amplitude $V_{\rm D}^{\rm opt}$ that optimizes the signal for given $\omega$. The orange dashed lines are guides to the eye using $V_{\rm D}^{\rm opt}\propto t_{\rm C}^2/\omega$ in the low frequency regime and $V_{\rm D}^{\rm opt}\propto (2t_{\rm C}-\hbar \omega)$ close to resonance.}
    \label{fig:backaction}
\end{figure}

It is this additional suppression due to Landau-Zener transitions that leads to a finite optimal drive amplitude.
This effect is indeed reproduced by our simulation framework and illustrated in \Cref{fig:backaction}. Focusing first on panel (a), we identify the optimal drive amplitude as the maximum of the signal. It is interesting to note that the optimal regime corresponds to a quantum capacitance that is suppressed by approximately $1/2$ relative to the weak drive limit, see \Cref{fig:backaction}(b), thus indicating substantial backaction. We also observe that for sizable couplings $t_{\rm C}=\SI{7.5}{\ueV}$ and moderate frequencies $f=\SI{1}{\giga\hertz}$ the simulation results are close to the quasi-static estimate together with the simple Landau-Zener backaction which validates the intuition from the corresponding backaction described above. Finally, panel (c) highlights the frequency and drive amplitude dependence of the signal. We find that the optimal drive amplitude scales with $V_{\rm D}^{\rm opt}\propto t_{\rm C}^2/\omega$ in the low frequency regime, in line with the scaling of \eqref{eq:LZ}, while close to resonance $V_{\rm D}^{\rm opt}\propto (2t_{\rm C}-\hbar \omega)$. We also observe that the achievable signal at optimal drive increases for lower frequencies. In practice, this increase may not translate into SNR enhancements, though, as the performance of the microwave readout chain is typically improved at higher frequencies.

\section{Low-energy many-body spectrum of a Coulomb-blockade problem}
\label{sec:methods}
The numerical examples of \Cref{sec:QD-MZM} consider a model with few degrees of freedom which allows us to directly study the quantum dynamics by solving the ULE.
However, a detailed microscopic model of a quantum device can include hundreds or thousands of degrees of freedom. A direct numerical solution of equations of motion of the form of \Cref{eq:EOM} can then rapidly become numerically prohibitive given that the size of the density matrix $\rho$ scales exponentially with the number of degrees of freedom. Fortunately, in many cases of interests, the dynamics of the device is confined to a small number of low-energy states in the system. In this section, we introduce numerical methods to efficiently project a, potentially large, fermionic model to this low-energy subspace. In Sec.~\ref{sec:projectionDynamics}, we will then connect the methods introduced in this section back to the quantum dynamics problem discussed in previous sections.

\subsection{Overview}

Key to projecting an interacting fermionic model to its low-energy subspace is identifying a subspace that spans the low-energy many-body eigenstates of the Hamiltonian. This is generally an exponentially difficult problem, but physically motivated approximations tailored to a given class of physical models can reduce the computational complexity.
Focusing again on mesoscopic quantum devices similar to the topological qubit device of Ref.~\cite{Aghaee24}, this work is concerned with models where electron-electron interactions are due to Coulomb-blockade effects in multi-electron QDs. This family of models, in the limit where QDs have finite level-spacing, can be mapped to quantum impurity models~\cite{pustilnik2004kondo}. As such, motivated by both formal and numerical results~\cite{Bravyi2017, boutin2021quantum}, we consider methods where the low-energy subspace is described as a coherent superposition of non-orthogonal fermionic Gaussian states.

While many common methods, in particular all variants of configuration interaction popular in quantum chemistry, can be viewed as constructing the ground state as such a superposition, it is important to note that considering non-orthogonal Gaussian states is an important generalization. This is because the linear combination of two Gaussian states is not a Gaussian state, and therefore a set of Gaussian states cannot be orthogonalized while preserving their Gaussian nature and the cardinality of the set. Put differently, the space of states that can be constructed from $N$ non-orthogonal Gaussian states is strictly larger than that constructed from $N$ orthogonal Gaussian states (for $N > 1$), but smaller than the full Hilbert space dimension.\footnote{More formally, the space of states $\ket{\psi}$ that can be written as $\ket{\psi}=\sum_{i=1}^N c_i \ket{\phi_i}$ with $\ket{\phi_i}$ Gaussian is a strict superset of the space of states that can be written in such a way with the additional constraint $\braket{\phi_i | \phi_j}=\delta_{ij}$. Considering general states $\ket{\phi_i}$ instead of Gaussian states, these two spaces are trivially equivalent since any set of $\ket{\phi_i}$ can be orthogonalized without increasing $N$.}

\begin{figure}
    \centering
    \includegraphics[width=\columnwidth]{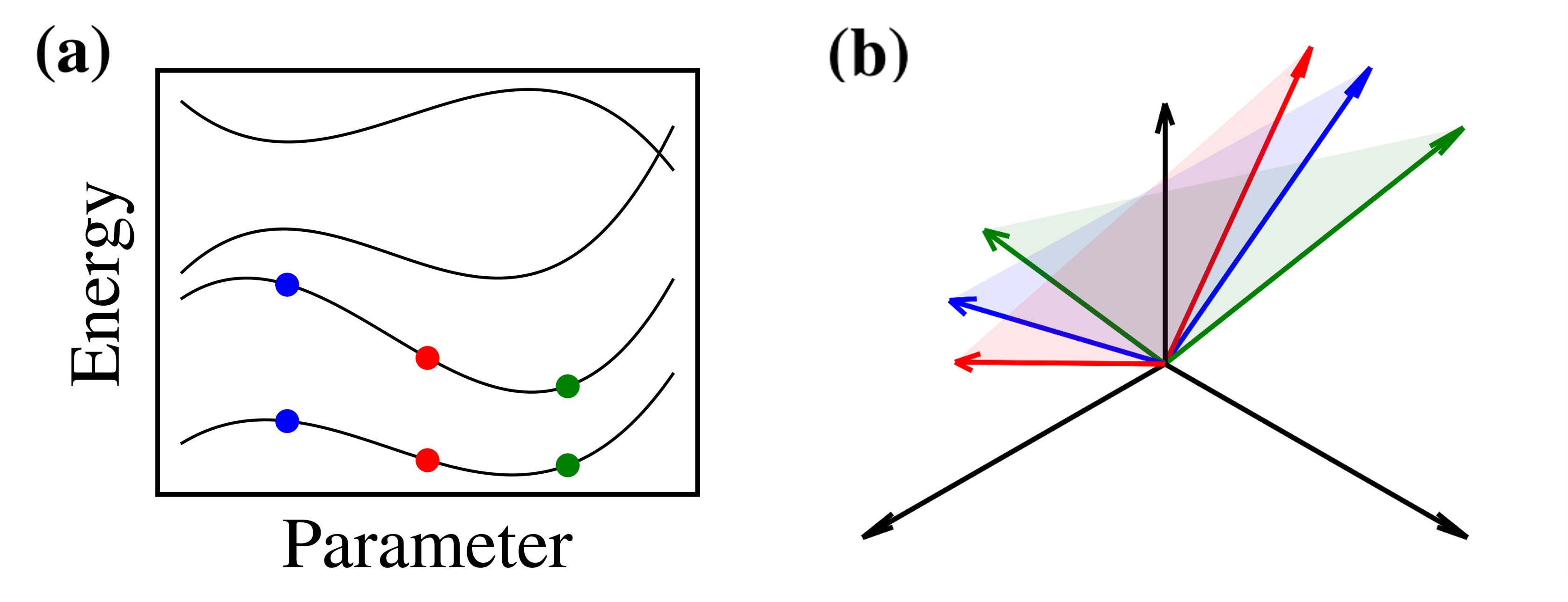}
    \caption{Simplified illustration of the ATCI method. (a) Given a parametrized model, we approximate the low-energy eigenstates at a set of parameter points, illustrated here by colored disks for the case where the two lowest energy eigenstates at three different points in parameter space are calculated. (b) The eigenstates at each parameter point, visualized here as colored vectors, form an orthogonal basis for a subspace of the full Hilbert space. While the eigenstates from different parameter points are not orthogonal, the span of all eigenvectors still forms a larger subspace of the full Hilbert space. It is important to note that the dimension of the full Hilbert space is exponentially large, and therefore--unlike in the three-dimensional cartoon--the basis vectors are with high probability linearly independent.
    }
    \label{fig:ATCI-sketch}
\end{figure}

The key challenge is then to efficiently identify a set of such non-orthogonal states. In the following, we introduce one such approach, which is based on ideas put forward in Ref.~\cite{boutin2021quantum}, but is much faster and, as we will show, very competitive in accuracy with state-of-the-art methods. The gain in performance compared to Ref.~\cite{boutin2021quantum} comes from tailoring the introduced methods to the class of models of interest in this work. Our approach to obtain the low-energy subspace relies on a combination of several methods: First, for a given set of parameters of the system, we perform a generalized Hartree-Fock (gHF)~\cite{bach1994generalized} simulation of the ground state, expressed in a convenient formalism of Gaussian covariance matrices (Sec.~\ref{sec:gHF}). Second, to increase the accuracy of the solution and approximate excited states, and similar to the well-known configuration interaction method in quantum chemistry, we form a basis of low-energy states from two-particle excitations above the gHF ground state and diagonalize the Hamiltonian in this space (Sec.~\ref{sec:truncatedCI}). Finally, we  combine the low-energy subspaces obtained from several parameter points within the parameter space of interest and form an Hilbert space as the span of these (non-orthogonal but typically linearly independent) basis vectors (\ref{sec:aggregateCI}). This not only increases the accuracy, but also the speed of simulations as the resulting \emph{aggregated} Hilbert space can be reused for efficient computations at any points within a continuous interval of parameter space.
This is of particular interests in situations where one is interested in computing the properties of the system over a fine grid of some tuning parameter such as the gate charge of a quantum dot.
Taken together, we refer to this combination of methods as the \emph{Aggregated-subspace Truncated Configuration Interaction} (ATCI) method. The general idea of the method is summarized in Fig.~\ref{fig:ATCI-sketch} and the associated caption. 

To benchmark the convergence and accuracy of the methods introduced in this work, we compare our results to reference density-matrix renormalization group (DMRG) calculations~\cite{white1992density}. To this end, we introduce below an extension of the QD-MZM model introduced in \Cref{sec:QD-MZM} to a spinless 1D model.
We demonstrate that the energy spectrum and the expectation value of the QD occupation calculated using ATCI converge rapidly to the results obtained from DMRG with the number of excited states included in the truncated configuration interaction step. Using 10 excited states, the calculated ground state energy and QD occupation are respectively within $\sim10 \unit{\nano\electronvolt}$ and within $10^{-4}$ electrons of the DMRG calculations. See \Cref{App:dmrg} for additional details on the reference DMRG simulations used in this section.

We find that for reasonable parameters, ATCI can be orders of magnitude faster than conventional methods, specially when scanning over large, high-dimensional parameter spaces. We illustrate this in \Cref{sec:MPR}, where we study the quantum capacitance response of an interferometer throughout the topological phase diagram of a spinful Rasbha wire. This computation would have been computationally prohibitive using methods such as DMRG.

\subsection{Example system: Multi-electron QD coupled to a spinless topological wire}
\label{sec:spinlessModel}

We consider the example physical system of a multi-electron QD coupled to a topological superconducting wire. Within this model, we will demonstrate the computation of the QD-MZM coupling from the many-body spectrum and the rf response using the open system dynamics formalism introduced in \Cref{sec:rf-simulation}. To illustrate the relevant methods, we focus first on the simplified case of a spinless model; the more realistic case of a spinful system will be discussed in \Cref{sec:MPR}.

The spinless system can be described by the Hamiltonian  
\begin{equation}
    H = H_0 + H_{\rm C},\label{eq:hqdmzm}
\end{equation}
with $H_0$ containing all the non-interacting (bilinear) terms of the Hamiltonian and 
\begin{equation}
    H_{\rm C} = \EC\left(\hat n_{\rm qd} - \ng\right)^2,
    \label{eq:HCspinless}
\end{equation}
the Coulomb blockade interaction in the QD, where $\EC$ is the charging energy and $\ng$ is the dimensionless gate charge.

The non-interacting part of the model can be written more explicitly as 
\begin{equation}
    H_0= \Hqd^{(0)} + H_{\rm sc} + H_t,
    \label{eq:H0spinless}
\end{equation}
where the three terms correspond to the non-interacting part of the QD Hamiltonian, the topological superconductor Hamiltonian and a tunnelling Hamiltonian between the two parts, respectively.
Introducing the fermionic annihilation (creation) operators $ c_j^{(\dag)}$, with $j \in [-N_{\rm qd}, -1]$ corresponding to fermionic modes in the QD and $j \in [0, N_{sc}-1]$ corresponding to fermionic modes in the wire,
the terms of Eq.~\eqref{eq:H0spinless} can be written as 
\begin{align}
    \Hqd^{(0)} &= -\mu_{\rm qd}  \hat n_{\rm qd}+ t_{\rm qd} \sum_{j=-N_{\rm qd}}^{-2} \left( c_j^\dag c_{j+1} + h.c.\right)
    \\
    H_{\rm sc} &=-\mu_{\rm sc}  \hat n_{\rm sc} + \sum_{j=0}^{N_{\rm sc} -2} \left( t_{\rm sc} c_j^\dag c_{j+1} + \Delta c_j c_{j+1}  + h.c.\right),
    \\
    H_t &=\tau_{12} c_{-1}^\dag c_0 + h.c.,
\end{align}
where $t_\alpha$ and $\mu_\alpha$ are the hopping amplitude and chemical potential in the QD ($\alpha=\rm qd$) and the superconductor ($\alpha = \rm sc$). The $p$-wave pairing amplitude is set by $\Delta$ and $\tau_{12}$ is the hopping amplitude between the last site of the QD and the first site of the superconductor. We denote by $N=N_{\rm qd} + N_{\rm sc}$ the total number of fermionic degrees of freedom.

Motivated by near-term topological devices, we consider throughout this section a parameter regime where the spinless superconductor is in a topological phase, but near the phase transition. In this regime, there is no clear hierarchy of energy scales, with the topological gap $\Delta_T$ being of a similar scale as the charging energy $\EC$ and the mean level spacing of the QD. 

\subsection{Generalized Hartree-Fock}
\label{sec:gHF}

As an initial step to treating interactions, we use a generalized Hartree-Fock (gHF) method~\cite{bach1994generalized} for a given point of the parameter space of the model to estimate the ground state wavefunction and energy. Generalized Hartree-Fock should be viewed as the most general mean-field method for a system of fermions, equivalent to a variational search over all possible Gaussian states, i.e. states that are fully characterized by their two-point correlation functions. As such, it also generalizes self-consistent BCS mean-field theory. For convenience, we formulate gHF entirely through these two-point correlation functions (see, e.g., Ref.~\cite{lowdin1955}) using the Gaussian covariance matrix formalism of Ref.~\cite{bravyi2004lagrangian}, and apply the imaginary-time evolution method described in Ref.~\cite{kraus2010generalized}.

The gHF formalism is most naturally described in the language of Majorana operators. Decomposing the fermionic annihilation operators as $c_j = \gamma_{2j-1} + i \gamma_{2j}$, where the hermitian Majorana operators obey the anti-commutation relation
$\left\{\gamma_j\,,\, \gamma_k\right\} = 2 \delta_{j,k}$, one can describe any fermionic Gaussian state $\ket{\Gamma}$ in terms of a covariance matrix with matrix elements 
\begin{equation}
    \Gamma_{kl} = \frac{-i}{2}\bra{\Gamma} [\gamma_k \,,\, \gamma_l] \ket{\Gamma},
    \label{eqn:CMdef}
\end{equation}
with $\Gamma$ a $2N \times 2N$ real skew-symmetric matrix. A fermionic Gaussian state $\ket{\Gamma}$ obeys a Wick theorem such that any physical expectation value can be expressed as a polynomial of the matrix elements of $\Gamma$; see Appendix~\ref{App:matrix-elements} for an introduction to the covariance matrix formalism.

The system described in Sec.~\ref{sec:spinlessModel} is part of a broader class of Hamiltonians formed from a sum of terms either quadratic or quartic in the number of fermionic operators. Any Hamiltonian in this class takes the form  
\begin{equation}
    \mathcal{H} = i \sum_{k,l=1}^{2N} A_{kl}\gamma_k \gamma_l + 
    \sum_{k,l,p,q}^{2N} U_{klpq} \gamma_k \gamma_l \gamma_p \gamma_q,
    \label{eqn:HMajoranaOps}
\end{equation}
where $A$ is a real skew-symmetric matrix and $U$ a real rank-4 tensor skew-symmetric with respect to the exchange of any neighboring indices.
For a Gaussian state with covariance matrix $\Gamma$, the expectation value of $\mathcal{H}$ can be expressed as~\cite{kraus2010generalized}
\begin{equation}
    \bra{\Gamma} \mathcal{H} \ket{\Gamma} = \Tr \left[\Gamma \overline{A}(\Gamma)\right],
\end{equation}
with $\overline{A}(\Gamma)_{kl} = A_{kl} + 3\sum_{m,n} U_{klmn} \Gamma_{mn}$.

The gHF method can be formulated as a variational method for finding the approximate ground state within the class of Gaussian states. 
As shown in Ref.~\cite{kraus2010generalized}, one of the ways this optimization can be performed in practice is through imaginary time evolution. The covariance matrix follows the equation of motion
\begin{align}
    \partial_\tau \Gamma = 4\left[F \Gamma F + F\right],
    \label{eqn:imtimeEOM}
\end{align}
where we introduced the Fock Matrix
\begin{equation}
    F_{kl}(\Gamma) = A_{kl} + 6\sum_{m,n} U_{klmn} \Gamma_{mn}.
    \label{eqn:FockMatrix}
\end{equation}
 The steady-state solution of \Cref{eqn:imtimeEOM} is the gHF solution of $\mathcal{H}$ and can be obtained by integrating the differential equation over a sufficiently long time interval. See \Cref{App:ImaginaryEvolution} for additional details on the implementation of the imaginary-time evolution.

\begin{figure}
    \centering
    \includegraphics[width=\columnwidth]{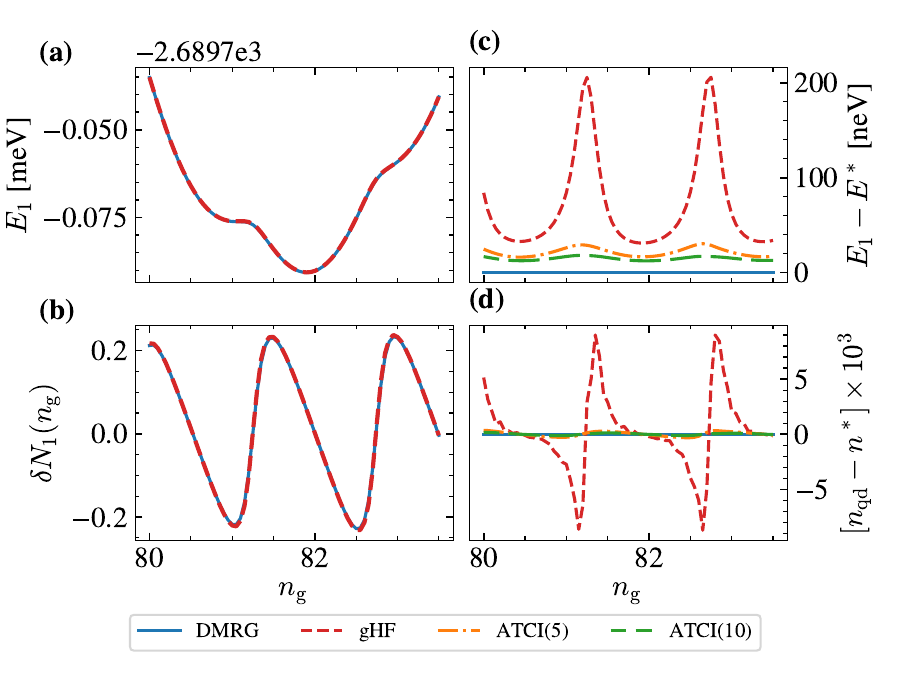}
    \caption{
    Variational ground-state of the spinless QD-MZM model. Comparison of (a) the ground-state energy and (b) the QD occupation obtained using standard DMRG (solid blue curve) and the gHF solution (dashed red curve) obtained by a long time evolution of Eq.~\eqref{eq:EOM}. The linear background ($\propto n_{\rm g}$) is subtracted.  (c) Ground-state energy difference between the DMRG solution ($E^*$) and the different methods introduced in this section. (d) Difference in the occupation of the QD in the ground state $\langle n_{\rm qd}\rangle$ and the DMRG solution ($n^*$). The orange dot-dashed (green dashed) curve correspond to the difference between the DRMG solution and the ATCI solution where 5 (10) excited states were included in the excitation subspace (see Sec.~\ref{sec:aggregateCI}).
    \emph{Parameters:}
    $N_{\rm qd}=400$, $N_{\rm sc}=200$, $\Delta=$\SI{0.1}{\meV}, $t_{\rm sc}=t_{\rm qd} = \SI{5}{\meV}$, $\tau_{12}=\SI{4}{\meV}$, $E_C=\SI{0.05}{\meV}$, $\mu_{\rm qd}=\SI{-8.0}{\meV}$, $\mu_{\rm sc}=\SI{-9.8}{\meV}$. 
    }
    \label{fig:wfn-convergence1}
\end{figure}

As a first benchmarking of the method, Fig.~\ref{fig:wfn-convergence1}(a) compares the ground state energy ($E_1$) of the spinless QD-MZM model introduced in Sec.~\ref{sec:spinlessModel} obtained using gHF (dashed red curve) and DMRG (solid blue curve). The parameters of the model are chosen to be of a similar scale as those of Ref.~\cite{Aghaee24} (see \Cref{sec:spinless-results} for more details). The metaparameters of the DMRG simulation were chosen to ensure high accuracy at the cost of increased computational cost such that DMRG can act as a benchmark of the methods introduced in this work. 
Both methods yield quantitatively comparable results for the dispersion of the ground state energy as a function of gate charge. To better resolve the difference in convergence of the methods, panels (c,d) of Fig.~\ref{fig:wfn-convergence1} show the difference between the two curves shown in panels (a,b) as the dashed red curves.  Denoting by $E^*$ the ground-state energy obtained by DRMG, the difference in ground state energy between the two methods is illustrated in Fig.~\ref{fig:wfn-convergence1}(c). The difference is maximal near charge steps and minimal in Coulomb valleys with a maximal difference of approximately $\SI{200}{\nano\electronvolt}$.

Fig.~\ref{fig:wfn-convergence1}(b) plots the ground state occupation of the QD obtained using both gHF and DMRG. To verify that the Coulomb blockade physics is correctly reproduced by the gHF ground state and better highlight possible difference in convergence between methods, we subtract a linear offset from the QD occupation and plot
\begin{equation}
    \delta N_1(n_{\rm g}) = \langle \hat n_{\rm qd}\rangle - (\beta n_{\rm g} + n_0),
    \label{eq:deltaN}
\end{equation}
where the same slope coefficient $\beta$ and offset $n_0$ is used for both methods. With this subtraction, the negative slope region of the curve corresponds to Coulomb blockade plateaus, while the positive slope regions corresponds to the charge steps. Again we observe good agreement between the methods; for a more quantitative comparison, in Fig.~\ref{fig:wfn-convergence1}(d) we subtract the QD occupation in the ground state calculated using DMRG ($n^*$) from the one calculated using gHF. The difference is at most $\pm 0.01$ with the maximal difference observed again near the charge steps.

\subsection{Truncated basis configuration interaction}
\label{sec:truncatedCI}
In order to calculate dynamical properties, one must not only estimate the ground state of the interacting problem, but a set of low-energy many-body eigenstates. 
To this end, we use a truncated configuration interaction (TCI) method~\cite{TCI_dmft_2012}, where the interacting Hamiltonian is projected on a subspace formed by the span of $N_{T}$ states expected to form a good approximation of the low-energy subspace of the Hamiltonian. The accuracy of the method can be controlled by the number of states $N_{T}$ included in the subspace, with the limit $N_{T} \rightarrow 2^N$ corresponding to exact diagonalization of the Hamiltonian (also known in quantum chemistry as ''full configuration interaction'').

Building on the gHF methods and results of the previous section, we follow a common strategy for building the projection subspace by considering low-energy excitations of the mean-field Hamiltonian\footnote{For a review of how to construct low-energy excited states of HF solutions, see Ref.~\cite{lowdin1955}.}
\begin{equation}
    \mathcal{F}(\overline{\Gamma}) = i\sum_{k,l=1}^{2N} F_{kl}(\overline{\Gamma}) \gamma_k \gamma_l,
\end{equation}
with $\overline{\Gamma}$ the gHF variational ground state obtained following the approach described in Sec.~\ref{sec:gHF}. To build the low-energy subspace, the real and skew-symmetric Fock matrix $F(\overline{\Gamma})$ can be transformed into a normal form using an orthogonal transformation $S$. Introducing a new Majorana operator basis $b_j = \sum_k S_{kj}\gamma_k$ (also known as normal modes), this normal form can be written as
\begin{equation}
    \mathcal{F}(\overline{\Gamma}) = \frac{i}{2}\sum_{n=1}^N \epsilon_n b_{2n-1}b_{2n},
    \label{eqn:HeffGHF}
\end{equation}
with $\epsilon_n$ positive single-particle energies labeled in ascending order ($\epsilon_1 < \epsilon_2 < \dots < \epsilon_N$). The ground-state of $\mathcal{F}(\overline{\Gamma})$, $\ket{\Gamma_1}$ is then the state where $ib_{2n-1}b_{2n}\ket{\Gamma_1} = -\ket{\Gamma_1}$ for all $n=1,\dots,N$.\footnote{
    In principle the ground state of $\mathcal{F}(\overline{\Gamma})$ is $\ket{\overline{\Gamma}}$, however, given practical numerical considerations, such as a finite imaginary-time evolution,$\ket{\overline{\Gamma}}$ can differ from $\ket{\Gamma_1}$.
}
We are interested in the low-energy many-body states in the same parity sector as $\ket{\Gamma_1}$. The first few of those can be constructed by applying two quasi-particle operators to the ground state. For example, the first excited state in the same parity sector as the ground state is given by
\begin{align}
    \ket{\Gamma_2} = \frac{1}{4}(b_1 - i b_2)(b_3-ib_4)\ket{\Gamma_1}.
    \label{eqn:gamma1}
\end{align}
One can then build a truncated basis of low-energy doubly-excited states $\left\{\ket{\Gamma_n} \right\}$  where $n=1,\dots N_T$. 

Using this approximate low-energy subspace, we can obtain an improved approximation of the ground state (compared to gHF) and the low-energy excited states by using TCI. We project the interacting Hamiltonian $\mathcal{H}$ onto the subspace forming a $N_T\times N_T$ matrix $h$ with elements ($n,m=1, \dots N_T$)
\begin{align}
    h_{nm} = \bra{\Gamma_n} \mathcal{H} \ket{\Gamma_m},
    \label{eq:matElemH}
\end{align}
which can easily be diagonalized, with the new eigenstates linear superposition of the states forming the truncated subspace.
In the case of number-conserving problems, this approach is known in the quantum chemistry literature as configuration interaction singles (CIS) method~\cite{foresman1992toward}.

Different approaches of varying efficiency can be taken to evaluate the matrix elements of Eq.~\eqref{eq:matElemH}.
While it is straightforward to evaluate covariance matrices representing the excited states $\ket{\Gamma_n}$, the formalism is inefficient when working with orthogonal states~\cite{Bravyi2017}. Instead, we explicitly compute matrix elements using a representation of the excited states similar to Eq.~\eqref{eqn:gamma1}.
We introduce quadratic excitation operators such that $\ket{\Gamma_n} = \hat \Lambda_n\ket{\Gamma_1}$, with 
\begin{equation}
    \hat \Lambda_n = \sum_{k,l} [\Lambda_n]_{kl} \gamma_k \gamma_l
\end{equation}
a quadratic operator with $\Lambda_n$ a skew-symmetric complex matrix formed from columns of the orthogonal matrix $S$ defined above. The required Hamiltonian matrix elements can then be computed as $h_{nm}=\bra{\Gamma_1} \hat \Lambda_n^\dagger \mathcal{H} \hat \Lambda_m \ket{\Gamma_1}$, which after some simple but tedious algebra can be evaluated as a series of simple matrix operations. 

\subsection{Aggregated subspace TCI}
\label{sec:aggregateCI}
The TCI method discussed in the previous section allows us to estimate the low-energy many-body eigenstates of the interacting Hamiltonian $\mathcal{H}$. However, in many cases of interest, the Hamiltonian of Eq.~\eqref{eqn:HMajoranaOps} depends on one or more scalar parameters and one is interested in evaluating some quantity as a function of these parameters. To make this explicit, we denote as $\mathcal{H}[\mathbf{p}]$ a parameterized Hamiltonian with $\mathbf{p} = (p_1, \dots p_K)$ a vector of $K$ parameters. For example, the gate charge $\ng$ was swept in Fig.~\ref{fig:wfn-convergence1} corresponding to a parameterized Hamiltonian where $K=1$ and $p_1=\ng$.

Naively, when considering a calculation over a range of parameters, the TCI calculation must be repeated independently at each point of interest in parameter space. 
However, for physical systems, the spectrum and eigenstates of $\mathcal{H}[\mathbf{p}]$ are generally either smooth or piecewise smooth as a function of physical parameters $\mathbf{p}$. Taking advantage of this smoothness, we can further improve on TCI by creating a larger subspace formed by the union of the low-energy subspaces obtained from TCI at different points within the parameter space of interest.

This aggregation method, which we dub aggregate subspace truncated configuration interaction (ATCI), leads to two main improvements over TCI. First, by enriching the projection subspace compared to TCI, ATCI can further improve the accuracy of the low-energy many-body eigenstates and spectrum. Secondly, the aggregated subspace can be reused to efficiently evaluate the spectrum and eigenstates at additional parameter points. For example, considering again the case of sweeping a gate charge parameter $\ng$, the aggregated subspace can be constructed from a set of TCI calculations using a coarse sampling of the $\ng$ parameter space, while the same aggregated subspace can then be used at each point of a fine grid of $\ng$ points to solve for eigenstates and evaluate observables. In practice, we have found that this can lead to an acceleration by orders of magnitude, especially for high-resolution multi-dimensional sweeps such as gate-gate maps.

Concretely, the ATCI method can be described as follows. Denoting a set of points in parameter space used to build the aggregated subspace $\left\{\mathbf{p}_j\right\}$ with $j=1,\dots N_P$, we update the notation of Sec.~\ref{sec:truncatedCI}, by noting $\ket{\Gamma_n(j)}$ the $n^{\rm th }$ state obtained from TCI with Hamiltonian $\mathcal{H}[\mathbf{p}_j]$. The Hamiltonian at a different point of parameter space $\mathbf{p}'$ can then be projected onto the aggregated subspace, with the resulting matrix having the block structure
\begin{equation}
    h(\mathbf{p}') = \left( \begin{array}{cccc}
h^{(1,1)} &h^{(1,2)} &h^{(1,3)} &\ldots \\
\left[h^{(1,2)}\right]^\dagger &h^{(2,2)} &h^{(2,3)} &\ldots \\
\left[h^{(1,3)}\right]^\dagger &\left[h^{(2,3)}\right]^\dagger &h^{(3,3)} &\ldots \\
\vdots &\vdots &\vdots &
\end{array} \right),
\label{eq:hblocks}
\end{equation}
where the matrix blocks take the form
\begin{equation}
    h_{nm}^{(j,k)}(\mathbf{p}) = \bra{\Gamma_n(j)} \mathcal{H}[\mathbf{p}] \ket{\Gamma_m(k)},
    \label{eq:hblockElem}
\end{equation}
and where for brevity we omitted the explicit dependency of each matrix block $h^{(i,j)}$ on $\mathbf{p}'$ in Eq.~\eqref{eq:hblocks}.

The truncated basis at different points in parameter space will generically be non-orthogonal. We thus introduce the overlap matrix $G$ with the same block structure as in Eq.~\eqref{eq:hblocks}, with diagonal blocks $G^{(j,j)} = \openone_{N_T}$ and off-diagonal blocks
\begin{equation}
    G_{nm}^{(j,k)} = \Braket{\Gamma_n(j) | \Gamma_m(k)}.
    \label{eq:GblockElem}
\end{equation}
Contrary to the diagonal blocks, the overlaps and matrix elements of the off-diagonal blocks of Eqs.~\eqref{eq:hblockElem} and~\eqref{eq:GblockElem} can be computed efficiently using the covariance matrix formalism for non-orthogonal states of Ref.~\cite{Bravyi2017}. See Appendix~\ref{App:matrix-elements} for a short introduction to the formalism.

The many-body eigenstates and the spectrum can finally be obtained by solving the generalized eigenvalue problem
\begin{equation}
    h(\mathbf{p}) \phi_n(\mathbf{p}) = E_n(\mathbf{p}) G \phi_n(\mathbf{p}),
    \label{eq:generalizedEigen}
\end{equation}
with $E_n$ the energy and $\phi_n(\mathbf{p})$ a complex vector of length $N_T\times N_P$ listing the coefficients of the coherent linear superposition of the states forming the aggregated subspace. More explicitly, following the block indexing of Eqs.~\eqref{eq:hblockElem} and~\eqref{eq:GblockElem}, the $n^{th}$ eigenstate of $\mathcal{H}[\mathbf{p}]$ is approximated as
\begin{equation}
    \ket{\phi_n(\mathbf{p})} = \sum_{m=1}^{N_T}\sum_{j=1}^{N_P} [\phi_{n}(\mathbf{p})]_m^{(j)} \ket{\Gamma_m (j)},
\end{equation}
with the accuracy of that approximation controlled by $N_T$ the number of states in the TCI basis and the number ($N_P$)  and position of the points in parameter space used to build the aggregated subspace.

\begin{figure*}
    \centering
    \includegraphics[width=\textwidth]{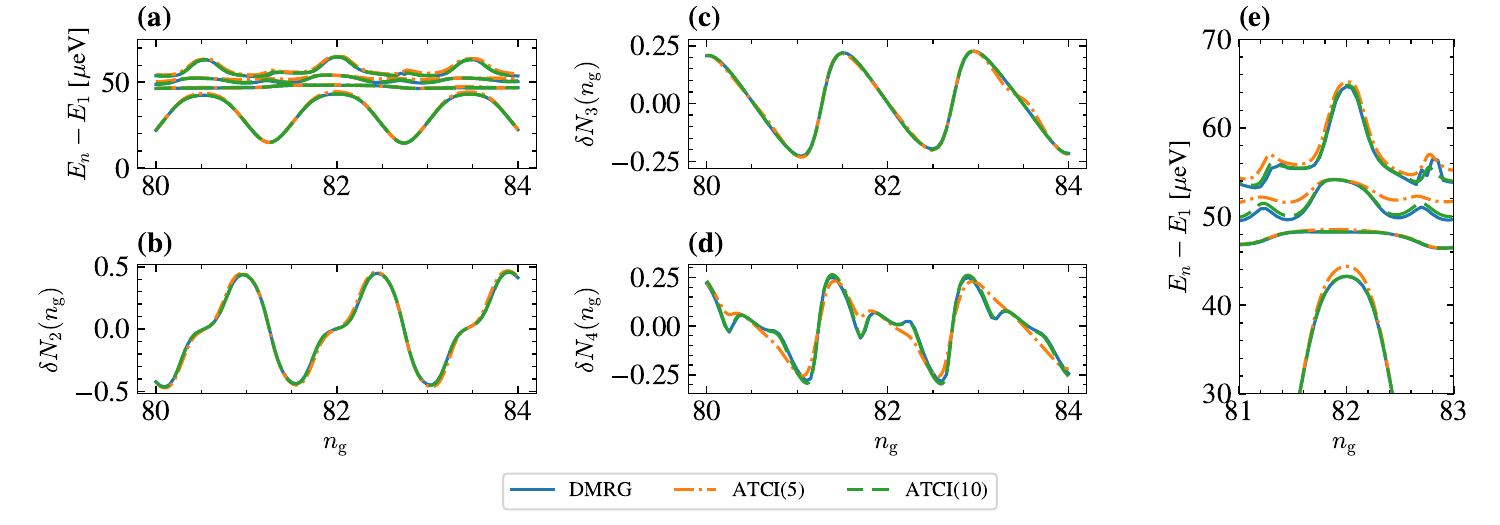}
    \caption{
        Spectrum and QD occupation of the low-energy many-body excited states of the spinless QD-MZM model obtained from DRMG (solid blue curve) and using ATCI($n$), where $n=5,10$ indicates the number of excited states included in the TCI basis. 
        (a) Difference in energy between the ground state ($E_1$) and excited states ($E_{n>1}$). (b,c,d) QD occupation of the first three excited states ($n=2,3,4$), with the linear background subtracted. [see Eq.~\eqref{eq:deltaN}]
        (e) Zoom-in of the excited states spectrum [panel (a)] near the gap edge. See caption of Fig.~\ref{fig:wfn-convergence1} and main text for a description of parameters.
    }
    \label{fig:wfn-convergence2}
\end{figure*}

\subsection{Example system: Convergence of low-energy eigenstates}
\label{sec:spinless-results}

To illustrate the convergence of the proposed ATCI method, we consider again the spinless QD-MZM model described in Sec.~\ref{sec:spinlessModel}. 
We first compare the accuracy of ground state properties calculated using ATCI to the previously discussed results for gHF and DMRG in Fig.~\ref{fig:wfn-convergence1}. Panel (c) shows the ground state energy difference between high-accuracy DMRG calculations and ATCI($n$) where $n=5,10$ denotes the number of excited states included in the TCI basis.
While the previous results for gHF showed a maximal energy difference of up to 200 neV near charge steps, ATCI increases the accuracy by an order of magnitude, with maximal energy difference of 30 neV for ATCI(5) and 20 neV for ATCI(10). Similarly, the convergence of the QD occupation is increased by more than one order of magnitudes, with maximal difference of $3\times 10^{-4}$ for ATCI(5) and $1\times 10^{-4}$ for ATCI(10). 

In addition to an increased accuracy of the variational ground state solution, the advantage of ATCI over gHF is the ability to accurately compute excited states. Fig.~\ref{fig:wfn-convergence2}(a) compares the excited-state spectrum calculated using high-accuracy DMRG, ATCI(5), and ATCI(10) (note that we have omitted showing gHF excited states, since they qualitatively misrepresent the behavior of the system). Up to small differences in convergence that we discuss below, all three methods reproduce the same features: (i) The first excited state exhibits avoided crossings at each charge step and (ii) higher-energy excited states correspond to a weakly dispersing set of states above the topological gap $\Delta_T$. Due to the finite size of the system, the above-gap states are separated by a finite level spacing.

To better compare the convergence of ATCI to DMRG, Fig.~\ref{fig:wfn-convergence2}(e) shows a zoomed-in portion of the spectrum of panel (a) where the first excited state approaches the topological gap edge. Notably, one observes a level repulsion between the first and second excited states. While the lower-precision ATCI(5) underestimates that repulsion by \SI{1}{\micro\electronvolt}, it is equally well captured by DMRG and ATCI(10). Not surprisingly, while qualitatively correct, details of the dispersion of the higher-energy states plotted ($n=4,5$) are missed by ATCI(5), but correctly captured by ATCI(10). This is in line with expectations that the number of states included in the truncated basis of TCI should be larger than the number of states of interest.

Fig.~\ref{fig:wfn-convergence2}(b-d) show the QD occupation in eigenstates $n=2,3,4$ with the linear contribution subtracted (see Eq.~\ref{eq:deltaN}). The same subtraction is done for all methods and all eigenstates. As expected from the convergence of the spectrum, ATCI(10) (dashed green curve) shows good quantitative agreement to DMRG (blue curve) for all three panels. However, ATCI(5) (orange dot-dashed curve) show small deviations for states $n=2,3$ and larger deviations for $n=4$ (panel d). Taken together, these results indicate that the convergence of ATCI is well-controlled by the choice of the number of states in the TCI basis.

An additional metaparameter of ATCI is the number of points used to build the aggregated subspace. The density of points should be large enough to sample all relevant features of the model such as all Coulomb valleys and charge steps in the example of Fig.~\ref{fig:wfn-convergence2}. There is however a point of diminishing return as a large density of points can lead to linearly-dependent states in the aggregated subspace with limited additional information.
The ATCI results shown in Figs.~\ref{fig:wfn-convergence1} and~\ref{fig:wfn-convergence2} were obtained using TCI results on a coarse grid of 21 points in the intervals $\ng \in (77,87)$.

\section{Quantum dynamics in low-energy subspaces}
\label{sec:projectionDynamics}
The ATCI method introduced in Sec.~\ref{sec:methods} allows us to efficiently calculate the low-energy many-body spectrum and eigenstates of a Coulomb-blockaded system. In this section, we now explain how one can use these eigenstates to solve the original problem of simulating the quantum dynamics of an open system and calculating the dynamical response. 

We first introduce in Sec.~\ref{sec:general-projection} a generalized projection (GP) method  which allows us to reduce the dimensionality of the quantum dynamics problem by projecting it to a low-energy subspace.
While GP bears many similarities to ATCI, both methods can be used together or separately. In particular, GP can be used in conjunction with other many-body solvers such as DMRG. In Sec.~\ref{sec:example-spinless-dynamics}, we summarize how GP and ATCI can be used together to efficiently compute the dynamical response of mesoscopic devices and, as an example, provide results for the dynamical response of the spinless QD-MZM model considered in the previous section.

\subsection{Generalized projection method for quantum dynamics simulation}\label{sec:general-projection}
A key component of the ATCI method introduced in Sec.~\ref{sec:methods} is that models of interest are parameterized by one or more scalar parameters. While we previously considered these parameters to be static, we now extend to the case where one or more of these parameters are time-dependent.
Building on the notation of Sec.~\ref{sec:methods}, we focus on systems where parameters are varied over time and can be described by a many-body time-dependent Hamiltonian taking the form 
\begin{equation}
    \mathcal{H}[\mathbf{p}(t)] = \mathcal{H}_0 + \sum_{n=1}^K p_n(t) \mathcal{H}_n,
    \label{eq:Ht}
\end{equation}
where $\mathcal{H}_0$ is the time-independent part of the Hamiltonian and the other term of the equation describes $K$ additional contributions that are time-dependent through the coefficients $\mathbf{p}(t) = [p_1(t), p_2(t), \dots, p_K(t)]$.
For each time-dependent function $p_n(t)$ of Eq.~\eqref{eq:Ht} one can define a lower bound $l_n$ and an upper bound $u_n$ for the function during the evolution from an initial time $t_i$ to a final time $t_f$. These bounds create a finite $K$-dimensional parameter-space with continuous time-dependent parameters $p_n(t)$ forming a path within this space.

As previously discussed, in most cases of physical interest, the eigenstates of $\mathcal{H}[\mathbf{p}]$ are smooth (up to possible permutations due to level crossings) over this parameter space. Similarly to ATCI we can approximate the low-energy eigenstates along the evolution path by calculating a finite number ($M$) of low-energy eigenstates at a finite number of points ($S$) within the parameter-space of interest. In this work, we focus on the case where these $M$ low-energy eigenstates are obtained using ATCI.

If the quantum dynamics of interest is slow compared to the energy scale of the high-energy degrees of freedom that have been discarded, the GP method as described below can be used to efficiently simulate the quantum evolution of the system. This low-energy subspace quantum evolution method can be summarized as:
\begin{enumerate}
    \item Select a set of $S$ points sampling the $K$-dimensional parameter space of time-dependent parameters: $\mathbf{S} = \left\{\mathbf{q}_1, \dots, \mathbf{q}_S\right\}$, where $\mathbf{q}_j=(q_{j,1}, q_{j,2}, \dots, q_{j,K})$ is a vector of parameters where $q_{j,n} \in [l_n, u_n]$
    \item For each element $\mathbf{q}_i$ of $\mathbf{S}$, find the $M$ lowest-energy eigenvectors $\left\{ \ket{\phi_n(\mathbf{q}_j)} \right\}$ of the many-body Hamiltonian $\mathcal{H}[\mathbf{q}_j]$.
    \item Build an orthonormalized basis out of the $M\times S$ eigenvectors calculated in the previous step.
    \item Project the time-dependent problem on the orthogonalized basis.
\end{enumerate}
The outcome of these four steps is a time-dependent Hamiltonian $\tilde{h}(t)$ defined on a (time-independent) orthonormal basis of size $M\times S$. One can then  solve the quantum dynamics within that approximate low-energy subspace.

The orthogonalization procedure of step 3 can be achieved by evaluating an overlap matrix which, similar to Eq.~\eqref{eq:GblockElem}, will have a block structure with $S\times S$ blocks of size $M\times M$ with elements
\begin{equation}
    G(\mathbf{S})_{n,m}^{(j,k)} = \braket{\phi_{n}(\mathbf{q}_j) | \phi_{m}(\mathbf{q}_k)}.
\end{equation}
The matrix $G(\mathbf{S})$ is hermitian and invertible in the case where all states are linearly independent.\footnote{In the case where the states are linearly dependent, an additional step is required. One can first diagonalize $G(\mathbf{S})$ and then project on the subspace of eigenvectors with positive nonzero eigenvalues to obtain a reduced linearly independent basis.}
This allows us to define the orthogonalized basis 
\begin{equation}
    \ket{\tilde{\phi}_\alpha(\mathbf{S})} =\sum_{\beta = 1}^{M\times S} \ket{\phi_\beta(\mathbf{S})} G^{-1/2}_{\beta, \alpha}(\mathbf{S}),
\end{equation}
where the Greek indices $\alpha, \beta$ are joint indices over the eigenstate index and the projection point index such that there is a one-to-one mapping between the sets of states $\left\{\ket{\phi_\beta(\mathbf{S})}\right\}$ and $\left\{ \ket{\phi_{n}(\mathbf{q}_j)}\right\}$.

Finally, step 4 of the above procedure corresponds to projecting the time-dependent Hamiltonian defined in Eq.~\eqref{eq:Ht} on the orthogonalized basis. The resulting projected Hamiltonian is 
\begin{equation}
    \tilde{h}(t) = \tilde{h}_0 + \sum_{n=1}^K p_n(t) \tilde{h}_n,
\end{equation}
with the projected terms of the Hamiltonian defined as 
\begin{equation}
[\tilde{h}_n]_{\alpha,\beta} = \bra{\tilde{\phi}_\alpha} \mathcal{H} \ket{\tilde{\phi}_\beta}.    
\end{equation}
A similar projection can be applied to other observables of interest and to the jump operators of the master equation.

Finally, the GP method allows for two parameters to verify and control convergence: i) the choice and number of points in $\mathbf{S}$ sampling the parameter space, and ii) the number $M$ of eigenstates calculated at each point. Assuming a smooth low-energy subspace over the time-evolution path, only  a small number of projection points are generally needed to obtain convergence of the quantum state evolution. The number of eigenstates $M$ needed to obtain convergence of the method will depend on i) the initial state of the evolution, and ii) the rate of change of the time-dependent parameters compared to the energy level spacing in the spectrum.

\begin{figure}
    \centering
    \includegraphics[width=\columnwidth]{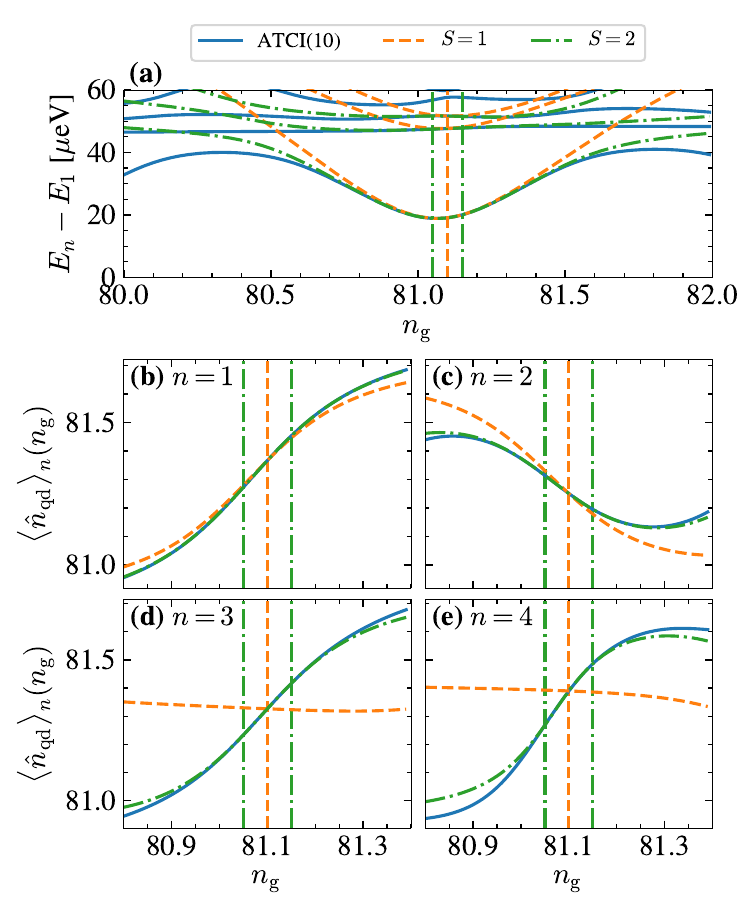}
    \caption{
        Application of the GP method near charge degeneracy point ($\ng^{(0)} = 81.1$) in the spinless QD-MZM model.
        (a) Excited states spectrum obtained using ATCI(10) (solid blue curve), one projection point ($S=1$, orange dashed curve), and two projection points ($S=2$, dot-dashed green curve).
        Vertical lines indicate corresponding projections with $q_1=\ng^{(0)}$, and $q_{1/2} = \ng^{(0)} \pm 0.05$ respectively.
        (b-e) Expectation value of the QD occupation $\langle \hat n_{\rm qd}\rangle_n$ in the first four eigenstates ($n=1,2,3,4$).
        Numerical results are for $M=4$ and $\tau_{12}=4.9$. See Fig.~\ref{fig:wfn-convergence1} for additional model parameters. 
    }
    \label{fig:dynamics-projection}
\end{figure}
\subsection{Example: Dynamical response of the spinless QD-MZM model}\label{sec:example-spinless-dynamics}
The GP strategy above is well-suited to the simulation of the readout of quantum devices as described in Sec.~\ref{sec:readout}. In the simplest case of the readout of a single QD, the time-dependent parameter is a harmonic modulation of the gate charge as described in Eq.~\eqref{eq:driven_H}, i.e. $K=1$ and 
\begin{equation}
    p_1(t) = \ng^{(0)} + \delta \ng \cos\omega t,
\end{equation}
with $\ng^{(0)}$ the average gate charge, $\delta\ng = V_{\rm D}/2\EC$ the drive amplitude in dimensionless units, and $\omega$ the drive frequency.
The trajectory in parameter space then repeatedly covers the interval $[\ng^{(0)} - \delta\ng, \ng^{(0)} + \delta\ng]$. As the modulation $\delta\ng$ is generally small compared to the distance between charge steps ($\delta\ng \lesssim 0.1$),
 calculating the low-energy eigenstates of the system at two gate charge values was observed to be sufficient to correctly reproduce the dynamics of the readout. 

To make things more concrete, Fig.~\ref{fig:dynamics-projection} illustrates the application of the GP procedure to the multi-electron spinless QD-MZM model considered in Sec.~\ref{sec:methods} for a dynamical response simulation. As an example, we consider a gate charge value near charge degeneracy and a dimensionless drive amplitude $\delta\ng \lesssim 0.1$, and compare two choices of projection points $\mathbf{S}$ ($S=1$ and $S=2$)
to a reference calculation that uses ATCI(10) at each point. While both projections correctly reproduce the avoided crossing of the excited states spectrum in panel (a), the slope of the QD occupation as a function of $\ng$ is better reproduced by the $S=2$ projection in panels (b-e), with the dispersion of the QD occupation for the states 3 and 4 missed by the $S=1$ projection. As the QD occupation slope directly relates to the dynamical response in the weak drive limit (see Sec.~\ref{sec:dynCQ}), its convergence is an important test for the projected model. Assuming a temperature $k_{\rm B} T \lesssim \Delta_T$, the dynamical response of the system should be well approximated by solving the quantum dynamics in the 8-dimensional subspace of the $S=2$ projection of Fig.~\ref{fig:dynamics-projection}. 

\begin{figure}
    \centering
    \includegraphics[width=0.9\columnwidth]{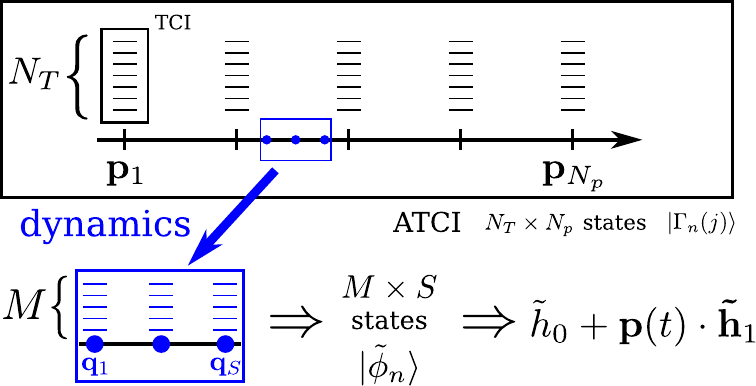}
    \caption{Overview of the flow of projections for obtaining the low energy Hamiltonian $\tilde{h}_0+\mathbf{p}(t)\cdot\mathbf{\tilde{h}}_1$ for calculating the dynamics of the system in a small region in parameter space (blue box). The larger space given by the ATCI may be used as a starting point to calculate the low-energy Hamiltonian around a series of projection points, e.g., for simulating the dynamical measurement result for a series of $n_{\rm g}$ points.}
    \label{fig:dynamics_flow_overview}
\end{figure}

While Fig.~\ref{fig:dynamics-projection} illustrates the application of the GP procedure for a given gate charge $\ng^{(0)}$, one is often interested in evaluating the dynamical response over a range of gate charges  $\ng^{(0)} \in (\ng^{\rm min}, \ng^{\rm max})$. This requires repeatedly applying the GP procedure and solving the quantum dynamics. Fig.~\ref{fig:dynamics_flow_overview} summarizes how the ATCI and GP methods can be used together to efficiently sample the desired range of gate charges and minimize computations. First, the black-framed box illustrate the steps needed to build the aggregated subspace of the ATCI method:
\begin{enumerate}
    \item Select a coarse grid of $N_p$ points 
    $\left\{ \mathbf{p}_j \right\}$
    sampling the $\ng$ axis over the interval $(\ng^{\rm min}, \ng^{\rm max})$.
    \item At each point do a TCI calculation to obtain $N_T$ low-energy approximate eigenstates $\ket{\Gamma_n(j)}$.
    \item Build the $N_T \times N_p$ aggregate subspace by calculating states overlaps Eq.~\eqref{eq:GblockElem} and project on the aggregated subspace each term $\mathcal{H}_n$ of Eq.~\eqref{eq:Ht}.
\end{enumerate}
After these steps, the GP procedure can be applied repeatedly within the interval $(\ng^{\rm min}, \ng^{\rm max})$. The blue-framed box illustrates one such application, where $M$ low-energy eigenstates are computed at projection points $\mathbf{q_j}$, where $j=1,\dots S$. For each projection point, the eigenstates are obtained by solving the generalized eigenvalue problem of Eq.~\eqref{eq:generalizedEigen}. Finally, these eigenstates can be orthogonalized and used to project the time-dependent Hamiltonian and solve the quantum dynamics problem with the final low-energy subspace.

\begin{figure}
    \centering
    \includegraphics[width=0.9\columnwidth]{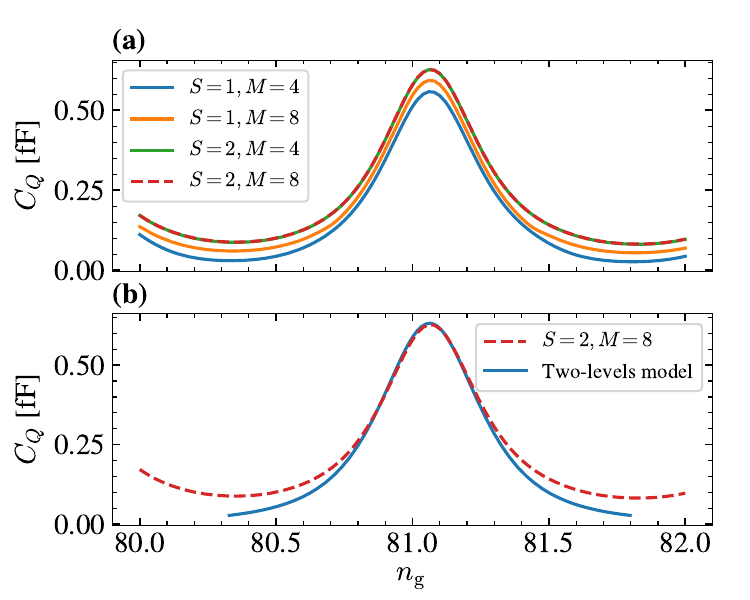}
    \caption{(a) Convergence of the dynamical response (quantum capacitance $\CQ$) computed using the GP-ATCI procedure for the spinless multi-electron QD-MZM model. In the $S=1$ case, the projection point $q_1=\ng$ is used, and for $S=2$ the points $q_{1/2} = \ng \pm \delta\ng$ are used, with $\delta\ng = 0.05$ the unitless drive amplitude. For all calculations we use drive frequency $\omega/2\pi = 0.5$~GHz, lever arm $\alpha=0.5$, temperature $T=100$~mK and $\gamma_{\rm eff}=1$~GHz using the noise spectral function of Eq.~\eqref{eq:S_eff}.
    See Fig.~\ref{fig:dynamics-projection} for other model parameters.
    (b) Comparison of the converged dynamical response of panel (a) (red dashed curve) to a calculation using the effective two-levels model of \Cref{sec:QD-MZM} (solid blue curve). The effective charging energy, lever arm and QD-MZM coupling are obtained from the first two levels of the many-body energy spectrum calculated using ATCI(10). 
}
    \label{fig:GP_convergence}
\end{figure}
We apply the above procedure to evaluate the dynamical capacitive response $\CQ$ in the spinless QD-MZM model. Figure~\ref{fig:GP_convergence} shows the results for different choice of parameters within the GP procedure. Similarly to the simplified model considered in Sec.~\ref{sec:QD-MZM}, the dynamical response is peaked at the avoided crossing. While all curves are qualitatively similar, the cases where a single projection point was used ($S=1$, blue and orange solid curve) present a smaller quantum capacitance compared to the calculation where two projection points were used. The difference is reduced as the number of eigenstates $M$ is increased in the $S=1$ case. On the contrary, the $S=2$ appears to be converged, with the same response observed for $M=4$ and $M = 8$.

As illustrated by \Cref{fig:GP_convergence}(a), the combined ATCI and GP methods allows us to efficiently and systematically project a microscopic interacting model of a mesoscopic device to a low-energy model. The quantum dynamics can then be solved within that subspace using the same methods and with a similar numerical complexity to simplified and more phenomenological few level models.
Finally, \Cref{fig:GP_convergence}(b) compares the converged dynamical response of panel (a) to the dynamical response of an effective QD-MZM model (see \Cref{sec:QD-MZM}). A quantitative agreement between the models can be obtained near charge resonance when accounting for both renormalization of parameters due to quantum charge fluctuations~\cite{Flensberg1993, Lutchyn16} and due to the finite-level spacing in the microscopic model. These effects result in a renormalization of the charging energy and the lever arm in the effective model that can be estimated from the curvature of the first two-levels of the many-body spectrum of the microscopic model. Near the Coulomb valleys, corrections to the spectrum curvature due to the presence of additional states increases the quantum capacitance in the microscopic model compared to the effective model. These discrepancies between the compared models disappear in the small QD-MZM coupling limit.

\section{Reduced basis using Natural orbitals projection}
\label{sec:natural-orbitals}
The ATCI method introduced in \Cref{sec:methods} allows us to approximate, in a controlled manner, the low-energy eigenstates of a mesoscopic device in presence of Coulomb blockade interactions. While the method reduces the exponential scaling of exact solutions to a cubic scaling, the computational cost can still become significant for large systems and when scanning high-dimensional parameter spaces.
To further accelerate the simulations of large systems, prior to solving the Coulomb blockade problem using ATCI, we now describe a projection method that can reduce the number of degrees of freedom at the non-interacting level.
This projection preserves the structure of the system, such that effective degrees of freedom are localized to a part of the system such as a QD or topological wire. We refer to this approach as \emph{natural orbitals projection}.

Our approach is based on a projection onto eigenvectors of the reduced single-particle density matrix in each part of the system; therefore, borrowing terminology from widely used methods in quantum chemistry, we refer to these states as \emph{natural orbitals}. However, an important distinction to standard chemistry methods is that we use eigenvectors of the \emph{reduced} density matrices. This is motivated by ideas from renormalization group and tensor network methods, in particular the insight that low-energy eigenstates of a part of the system do not form a good basis to describe the entire system~\cite{white1992real} while eigenstates of the reduced density matrix do~\cite{white1992density}. As such, the ansatz described here could also be viewed as a Gaussian Matrix Product State (GMPS)~\cite{schuch2012gaussian,fishman2015compression,schuch2019matrix} where each part of the quantum device forms a single site in the GMPS; however, since the goal here is not to find the ground state of a non-interacting system in a more efficient way but rather determine the relevant degrees of freedom for the treatment of interactions, the numerical approach taken is quite different from previous publications.

For sake of clarity of the presentation, we first describe the natural orbital folding approach for a normal-state system, i.e. in the absence of superconductivity. We defer to \Cref{app:natOrb} the description of the extension of the method to the superconducting case.

\subsection{Normal-state case}

We start from a microscopic description of a quantum device, which is comprised of several parts $p=1, \ldots, P$, corresponding for example to different QDs and superconducting wires. We consider a system of fermions with corresponding creation and annihilation operators $c_i^\dagger$, $c_i$; grouping the fermion modes in some part $p$ together, we define the vector of fermion creation operators for that part, $\mathbf{c}_p^\dagger$. By $\mathbf{c}$, we denote the vector of all fermionic modes. The fermionic degrees of freedom are, at the non-interacting or mean-field level, described by a Hamiltonian of the form
\begin{align}
\cH_f &= \mathbf{c}^\dagger H \mathbf{c}  \nonumber\\ &= 
    \sum_p \mathbf{c}_p^\dagger H^{(p)} \mathbf{c}_p
    + \sum_{p_1,p_2} \left[ \mathbf{c}_{p_1}^\dagger H^{(p_1,p_2)} \mathbf{c}_{p_2} + {\rm h.c.} \right].
    \label{eq:HfNatOrb}
\end{align}
Here, $H^{(p)}$ describes the coupling between fermionic modes within the part $p$ and $H^{(p_1,p_2)}$ describes the tunneling of fermions between parts $p_1$ and $p_2$. 
The structure of Eq.~\eqref{eq:HfNatOrb} can also be understood as $H$ having a block structure, with $H^{(p)}$ and $H^{(p_1,p_2)}$ forming respectively diagonal and off-diagonal blocks.
We can find the single-particle eigenenergies $\left\{ \epsilon_n\right\}$ and corresponding eigenstates $\left\{\psi_n\right\}$ by diagonalizing the hermitian matrix $H$ in a time that scales at worst cubically with the number of fermionic modes in the system; a few states in a desired energy window can typically be obtained using iterative methods in a time that depends on the sparsity of the Hamiltonian, but is often linear in its size.

In order to build an effective model that preserves the block structure of Eq.~\eqref{eq:HfNatOrb} and is valid in an energy window bounded by $\epsilon_l$  and $\epsilon_h$, we build a many-body state $\ket{\psi(\epsilon_l, \epsilon_h)}$ where all single-particles states within the energy-window are occupied. We can then compute the reduced density matrix for each part as
\begin{equation}
D_p(\epsilon_l, \epsilon_h) = \bra{\psi(\epsilon_l, \epsilon_h)} \mathbf{c}_p^\dagger \otimes \mathbf{c}_p \ket{\psi(\epsilon_l, \epsilon_h)},
\label{eq:Dreduced}
\end{equation}
where the product $\otimes$ is to be understood as outer product of the vector of fermion operators. Note that the $D_p$ are diagonal blocks of the full single-particle Green's function $D_{ij} = \bra{\psi(\epsilon_l, \epsilon_h)} c_i^\dagger c_j \ket{\psi(\epsilon_l, \epsilon_h)}$. They satisfy $D_p = D_p^\dagger$ with eigenvalues in the interval $[0,1]$.
The eigenvalues of $D_p$ can be related to the occupation of the corresponding eigenmodes and its entanglement to modes in other parts of the system~\cite{schuch2019matrix}, with $\lambda_n=1$ corresponding to an occupied mode, $\lambda_n=0$ an empty mode and a value in between those bounds to a mode entangled with another part of the quantum system.

To obtain the reduced model, we define the natural orbitals $\nket{n}_p$ for each part (using the notation $\nket{\cdot}_p$ for normalized vectors with support only in the part $p$) as the eigenvectors of the density matrix $D_p$:
\begin{equation}
D_p \nket{n}_p = \lambda_n^{(p)} \nket{n}_p,
\end{equation}
where we arrange eigenvalues as $\lambda_n^{(p)} \geq \lambda_{n+1}^{(p)} \geq \lambda_{\rm min}$ with $\lambda_{\rm min}$ a cutoff with vectors with eigenvalues below the cutoff considered \emph{frozen}. For each part, we define the $N_p' \times N_p$ (where $N_p$ is the number of fermionic modes in part $p$) isometry $Q_p$ that projects the part $p$ onto the $N_p'\leq N_p$ natural orbitals with the largest corresponding eigenvalues $\lambda_n^{(p)}$. Its elements are given by
\begin{equation}
\left( Q_p \right)_{nm} = \left[ \nket{n}_p \right]_m,
\end{equation}
where the right-hand side is the $m$'th entry in the $n$'th natural orbital for part $p$. Being an isometry, this matrix obeys $Q_p Q_p^\dagger = \openone_{N_p'}$ and $(Q_p^\dagger Q_p)^2 = Q_p^\dagger Q_p$. For $N_p'= N_p$, it is unitary.

We can define an operator acting on the entire system that performs the transformation to natural orbitals by constructing a projector that follows the block structure of $H$ in Eq.~\eqref{eq:HfNatOrb}. For example in the spinless QD-MZM model considered in previous sections with $P=2$ blocks, the projector takes the form
\begin{equation}
Q = \left( \begin{array}{ccc}
Q_1 &0 \\
0 &Q_2
\end{array} \right).
\end{equation}
The transformed Hamiltonian in the natural orbital basis is given by $H'= Q H Q^\dagger$, where $Q$ can in principle contain non-trivial transformations for one or several parts.

\subsection{Treatment of charging energy}\label{sec:natOrbBlockade}

Crucially, the transformation onto the natural orbitals preserves the block-structure of the Hamiltonian. Therefore, it is easy to introduce a charging energy term for a given part. In the original fermionic basis, this term is given by
\begin{equation}
\mathcal{H}_C = \EC^{(p)} \left( \mathbf{c}_p^\dagger \cdot \mathbf{c}_p - \Ng^{(p)} \right)^2,
\end{equation}
where $(\cdot)$ denotes the vector inner product, and $\Ng^{(p)}$ is the dimensionless gate voltage applied to part $p$. If $p$ is a part to which no non-trivial transformation was applied (i.e., $Q_p = \openone$), then the form of the charging energy is manifestly unchanged under the transformation $Q$.

If, on the other hand, $Q_p \neq \openone$, we can still approximately implement the charging energy term. If we denote by $\mathbf{d}_p = Q_p \mathbf{c}_p$ the ($N_p'$) new fermionic modes that describe part $p$ after the transformation, then we can approximate
\begin{equation}
\mathbf{c}_p^\dagger \cdot \mathbf{c}_p \approx \mathbf{d}_p^\dagger \cdot \mathbf{d}_p + {\rm const}.
\end{equation}
Here, we have essentially assumed that the $N_p' - N_p$ natural orbitals that are not being included are inert at the energy scales relevant to the interacting calculation, and have included them in the constant term.
Therefore, the new charging energy term takes the form
\begin{equation}
\mathcal{H}_C = \EC^{(p)} \left( \mathbf{d}_p^\dagger \cdot \mathbf{d}_p - \tilde{N}_g^{(p)} \right)^2,\label{eqn:EcNatOrb}
\end{equation}
where $\tilde{N}_g^{(p)}$ is related to $\Ng^{(p)}$ by a shift. If the energy window used to compute the natural orbitals is sufficiently large such that the inert orbitals assumption is valid, the shift in gate charge can be estimated by $\Ng^{(p)} - \tilde{N}_g^{(p)}\approx \langle \mathbf{c}_p^\dagger \cdot \mathbf{c}_p  \rangle_0 - \langle \mathbf{d}_p^\dagger \cdot \mathbf{d}_p \rangle_0$ where $\langle \cdot \rangle_0$ indicates the expectation value of the non-interacting ground state.

\begin{figure}
    \centering
    \includegraphics[width=\columnwidth]{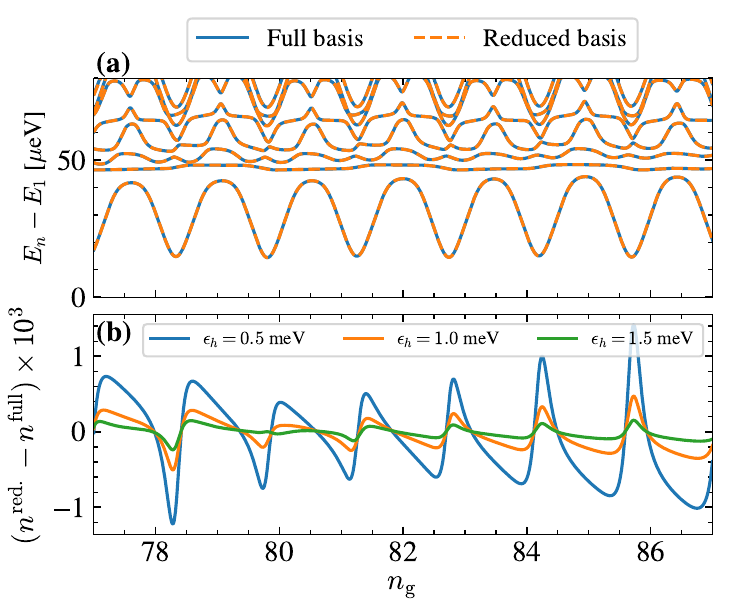}
    \caption{
    (a) Comparison of the many-body excited states spectrum obtained using ATCI(10) starting from the lattice basis (solid blue curves) and the reduced natural orbitals basis (dashed orange curves, $\epsilon_h = -\epsilon_l = \SI{0.5}{\milli\electronvolt}$).
    (b) Convergence of the ground-state QD occupation calculated in the reduced basis ($n^{\rm red}$) compared to the full basis ($n^{\rm full}$) as a function of the bound $\epsilon_h = -\epsilon_l$ of the energy window of interest defining the natural orbitals. The corresponding reduced basis sizes are $N' =56,85,118$ compared to a full basis of $N=600$ sites. All results are obtained for $\lambda_{\rm min} = 10^{-6}$.
    See \Cref{fig:wfn-convergence1} for model parameters.
    }
    \label{fig:nat_orb_convergence}
\end{figure}

\subsection{Example application}

As an example application of the natural orbitals method, \Cref{fig:nat_orb_convergence}(a) compares the low-energy excited states spectrum obtained in \Cref{sec:methods} for the spinless QD-MZM model to a calculation performed using a reduced natural orbitals basis. The reduced basis was obtained considering an energy range of $\pm\SI{0.5}{\milli\electronvolt}$ around the Fermi level leading to a reduced basis size $N'=\sum_p N_p' =56$.
Strikingly, despite a reduction of the basis size by more than an order of magnitude, the results for ATCI(10) performed in the natural orbitals basis (dashed orange curves) are, at this resolution, indistinguishable from the results obtained in the full lattice basis (solid blue curves). 
Given the cubic scaling of ATCI, 
the reduction of the basis size by a factor $N/N'$ translates to an asymptotic speed up of order $(N/N')^3$. For this specific example, we thus achieve a speedup by more than three orders of magnitudes.

To better understand the convergence of quantities of interest with the energy window of states included in the construction of $\ket{\psi(\epsilon_l, \epsilon_h)}$, \Cref{fig:nat_orb_convergence}(b) shows the difference of the ground-state QD occupation in the reduced natural orbitals basis ($n^{\rm red}$) and in the initial lattice basis ($n^{\rm full}$) for three different energy windows $\epsilon_h = -\epsilon_l = 0.5, 1.0,\SI{1.5}{\milli\electronvolt}$,  which controls the accuracy of the natural orbitals projection. As $\epsilon_h$ increases, the accuracy of $n^{\rm red}$ improves with the results converging towards the full basis calculation. Even for the smallest basis size used in panel (a), the difference in QD occupation compared to the full basis calculation is at most of the order of one thousandth of an electron.

\section{Application: Majorana parity interferometer}
\label{sec:MPR}
Revisiting the initial motivation for this work, we now apply the methods outlined in the previous sections to the study of the dynamical response of a Majorana parity interferometer. In particular, the methods developed in \Cref{sec:methods,sec:projectionDynamics,sec:natural-orbitals} allow us to go beyond the phenomenological models considered in Ref.~\cite{Aghaee24} and consider a microscopic spinful model of the topological superconductor and multi-electron QD~\cite{Lutchyn10,Oreg10}. This more detailed model allows us to study the dynamical response of the interferometer throughout the topological phase diagram.

\begin{figure}
    \centering
    \includegraphics[width=1.0\columnwidth]{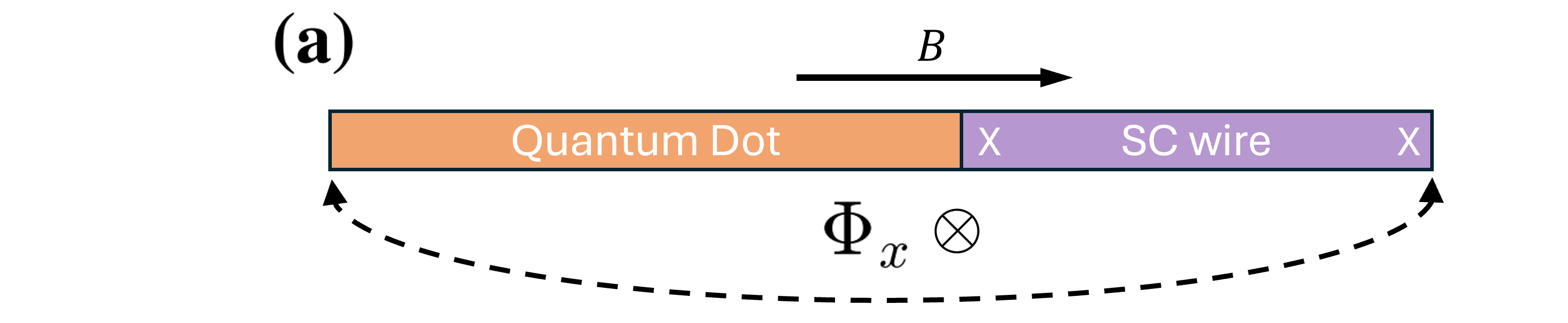}
    \includegraphics[width=\columnwidth]{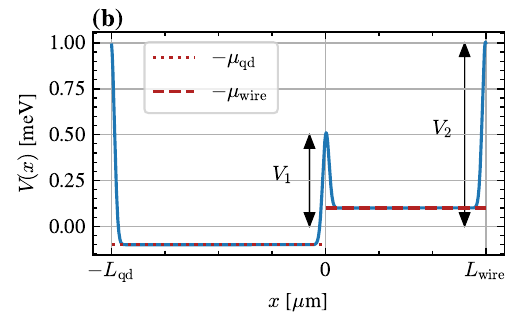}
    \caption{
        (a) Sketch of the Majorana parity interferometer simulated in \Cref{sec:MPR}. A QD of length $\Lqd$ is coupled to a superconducting (SC) wire of length $\Lwire$ to form a loop with a magnetic flux $\Phi_x$ threading the loop. (b) Example of the potential profile where Gaussian barriers of height $V_1=\,$\SI{0.5}{meV} and $V_2=2V_1$ and width $w_1=w_2=\,$\SI{60}{nm} are used to control the coupling between the QD and the wire for $\muqd=\,$\SI{0.1}{meV} and $\muwire=\,$\SI{-0.1}{meV}.
    }
    \label{fig:sketchMPR}
\end{figure}
\subsection{Microscopic model}\label{sec:rashba-model}
The Majorana parity interferometer studied in this work is sketched in \Cref{fig:sketchMPR}(a). We consider a superconducting nanowire of length $\Lwire$ coupled at both ends to a large one-dimensional multi-electron QD of length $\Lqd$. The resulting loop formed by the wire and the QD is threaded by a magnetic flux $\Phi_x$ and a magnetic field $B$ is applied parallel to the wire.

We consider a system where both the QD and the wire are formed of the same semiconductor (SM) structure, but where the wire is also covered by a superconductor (SC). Both the QD and wire are taken in the single sub-band regime and can hence be described by a one-dimensional model.  Integrating out the parent superconductor covering the wire and linearizing the resulting Hamiltonian~\cite{Stanescu11, Aghaee23}, we obtain the standard Rashba nanowire model
\begin{align}
    H_0 =&\, H_\SM + \DeltaInd O_\SC, \label{eqn:Rashba}
\end{align}
where $H_\SM$ is the Hamiltonian of the semiconductor, $\DeltaInd = \Delta_0 \Gamma/(\Delta_0 + \Gamma)$ is the induced superconducting gap in the wire, with $\Delta_0$ the parent superconductor gap and $\Gamma$ the SM-SC coupling. Finally, the pairing term is defined in the wire segment as
\begin{align}
    O_\SC =&\! \int\limits_0^{L_{\rm wire}} \!dx \, \bigl(
        \psi_\uparrow^\dag(x) \psi^\dag_\downarrow(x)
        + \mathrm{h.c.}
    \bigr),
\end{align}
where we take the wire to be defined on the interval $x \in (0, \Lwire)$, and the QD the interval $x \in (-\Lqd, 0)$.
The fermionic field operator $\psi_\sigma (x)$ (annihilating a fermion of spin $\sigma$ at position $x$) is defined on a ring with twisted boundary conditions
\begin{equation}
    \psi_\sigma(-\Lqd) =  \psi_\sigma(\Lwire)
    e^{ie \Phi_x /\hbar}.
    \label{eq:bc}
\end{equation}
to account for the magnetic flux $\Phi_x$ threading the loop.

The semiconductor part of the Hamiltonian takes the form
\begin{align}
\begin{split}
    \label{eq:Hsm}
    H_\SM =& \int\limits_{-L_{\rm qd}}^{\Lwire} \!dx \,
    Z(x) \psi_{\sigma}^\dag(x) \left(
        \!-\frac{\partial_x^2}{2m^*}\!
         - \!V(x)\! \right.\\
        &  \qquad+\!i\alpha_{\rm SO} \hat{\sigma}_y \partial_x\!
        + g_\mathrm{SM} \mu_\mathrm{B} B/2\, \hat{\sigma}_x\!
    \bigg)_{\!\sigma\sigma'}
    \!\psi_{\sigma'}(x),
    \end{split}
\end{align}
where $m^*$ is the effective mass, $V(x)$ an effective potential, $\alpha_{\rm SO}$ the Rashba spin-orbit coupling, $g_\mathrm{SM}$ the g-factor of the semiconductor, and $B$ the applied magnetic field parallel to the wire. The renormalization factor
\begin{equation}
    Z(x) = \Theta(x) \frac{\Delta_0}{\Delta_0 + \Gamma} + \Theta(-x)
\end{equation}
accounts for the renormalization of the SM parameters by the integrated-out parent superconductor in the wire ($x>0$) but not in the QD ($x<0$) (here, $\Theta(x)$ is the Heaviside step function).

The parameters simulated are motivated by the material stack used in Ref.~\cite{Aghaee24} and chosen to accurately represent the relevant regime of low density, where the most robust topological phases are found. The precise parameters are listed in \Cref{tab:rashba-params}. For numerical calculations, the continuous model of \Cref{eqn:Rashba} is discretized on a lattice with spacing $a$ using finite difference methods.

An example of the effective potential $V(x)$ introduced in Eq.~\eqref{eq:Hsm} is shown in Fig.~\ref{fig:sketchMPR}(b). 
The QD and wire are coupled to each other through tunnel junctions that are implemented as Gaussian potential barriers of width $w_{1,2}$ and height $V_{1,2}$. In addition, the chemical potential of the QD, $\muqd$, and of the wire, $\muwire$, is subtracted from the potential. In general, $V(x)$ could also include a disorder potential~\cite{Aghaee23}, but this is beyond the scope of this work and we focus throughout on the clean case.

As in the spinless model of \Cref{sec:spinlessModel}, the full Hamiltonian is 
$H = H_0 + H_{\rm C}$, where we add to the non-interacting Hamiltonian $H_0$ the charging energy term 
\begin{equation}
    H_{\rm C} = \EC \left(\hat n_\mathrm{qd} - \ng\right)^2,
\end{equation}
where, in the continuum limit, the QD number operator takes the form
\begin{equation}
    \hat n_\mathrm{qd} =   \sum_\sigma \int_{-L_{\rm qd}}^{0} \!dx \, \psi^\dag_\sigma (x) \psi_\sigma(x).
\end{equation}
In the following simulations, we use $\EC=\,$\SI{75}{\micro\electronvolt}.
Finally, following the approach described in \Cref{sec:QD-MZM}, we separate the effect of charge noise on the QD gate charge, $\ng$, in two parts. First, we include a quantum contribution by adding a Lindblad operator with the spectral function $S_{\rm eff}$ as defined in \Cref{eq:S_eff} and taking $T=\SI{50}{mK}$ and $\gamma_{\rm eff}=\SI{1}{GHz}$. Second, we account for the low-frequency part of the $1/f$ charge noise by a Gaussian broadening of the simulation results along the $\ng$ parameter; see \Cref{eq:sigmaDelta} and related discussion.

\begin{table}[t]
    \centering
    \begin{tabularx}{\columnwidth}{|*{1}{C|}c|*{6}{C|}} \hline
        $m^*$ & $\alpha_{\rm SO}$ & $g_\mathrm{SM}$ & $\Gamma$ & $\Delta_0$ &
        $a$ & $L_{\rm wire}$ & $L_{\rm qd}$
        \\ \hline\hline
        -- & meV$\cdot$nm & -- & meV & meV & nm& $\SI{}{\micro\meter}$& $\SI{}{\micro\meter}$
        \\ \hline
        0.0275 & 7.6  & -8.8  & 0.2 & 0.3 & 5 & 3 & 4
        \\ \hline
    \end{tabularx}
    \caption{Parameters for the effective Rashba model of Eqn.~\eqref{eqn:Rashba}. 
    }
    \label{tab:rashba-params}
\end{table}

\subsection{Dynamical quantum capacitance of Majorana parity readout}\label{sec:MPR-clean}
\begin{figure}
    \centering
    \includegraphics[width=\columnwidth]{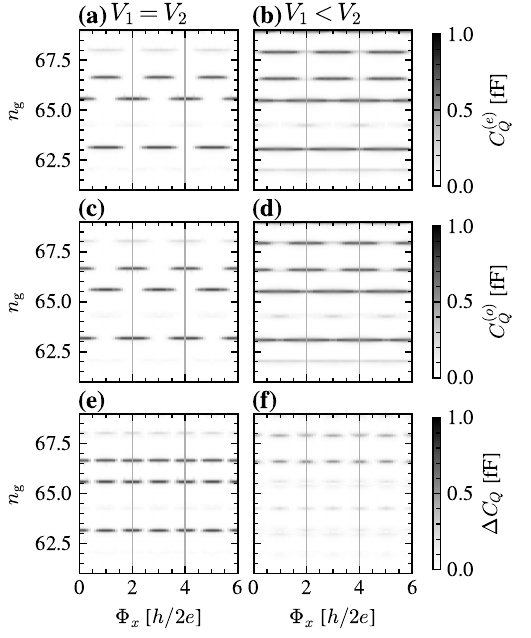}
    \caption{
        Dynamical quantum capacitance of the interferometer in the topological phase ($B=$\SI{1.5}{\tesla}, $\muwire=$\SI{0}{meV}) in the even parity sector [$\CQ^{(e)}$, panels (a,b)], in the odd parity sector [$\CQ^{(o)}$ panels (c,d)] and their difference $\Delta \CQ = |\CQ^{(e)} - \CQ^{(o)}|$ (e,f). The first column (a,c,e) corresponds to a balanced interferometer where $V_1=V_2=$\SI{1.2}{meV}, while the second column (b,d,f) corresponds to an unbalanced interferometer where $V_1=$\SI{0.9}{meV} and $V_2=$\SI{1.2}{meV}.
        In all panels, we consider a unitless drive amplitude $\delta\ng = 0.05$, drive frequency $\omega/2\pi = 0.5$~GHz, lever arm $\alpha=0.4$, and charge noise broadening $\sigma_\Delta=0.04$. The simulation is performed on the interval $\Phi_x \in (0,1)$ and extended to the $(0,6)$ interval using the Hamiltonian symmetries. See main text for other parameters.
    }
    \label{fig:MPR-topo-phase}
\end{figure}
We first consider the dynamical response of a Majorana parity interferometer where the wire is in the topological phase. In Fig.~\ref{fig:MPR-topo-phase}, we show the dynamical response as a function of gate charge $\ng$ and external flux $\Phi_x$ for $B=\,$\SI{1.5}{\tesla} and $\muwire =\SI{0}{meV}$, where the topological gap is $\Delta_T \approx \ \SI{40}{\micro\electronvolt}$. The chemical potential of the QD is chosen as $\mu_{\rm qd}=\SI{1}{\meV}$ to form a multi-electron quantum dot containing approximately 65 electrons in the non-interacting limit. We consider in the first column (panels a,c,e) a well-balanced interferometer where $V_1 \sim V_2$. Our quantum dynamics model does not include quasi-particle poisoning which is expected to be slow compared to other relevant scales of the model. This allows us to plot separately the dynamical responses $\CQ^{(e,o)}$ where the total fermionic parity of the system is even (panel a) or odd (panel c).
As expected from analytical expressions of simplified models~\cite{Aghaee24}, the dynamical response within a fixed parity sector is periodic with period $h/e$, with the response shifted by half a period ($h/2e$) between the two parity sectors.

The more microscopic model of the interferometer allows us to observe additional features that are not as apparent in simpler models. First, even for the fixed barrier parameters and the clean model considered in Fig.~\ref{fig:MPR-topo-phase}(a,c), the simulated quantum capacitance varies for different charge transitions of the multi-electron QD. This variation in the visibility of the interferometric signal can be understood by different QD charge states having different matrix elements coupling to the MZM due to both variations in the spin and the orbital components of the wavefunctions. Similarly, the phase of the effective coupling between the QD and MZMs can change for different charge transitions. For example, the interferometric signal of the charge transition at $\ng\sim65.5$ is shifted by $h/2e$ from the response at $\ng\sim66.5$ in panels (a,c).
To better relate the simulation results to experimentally measurable quantities, Fig.~\ref{fig:MPR-topo-phase}(e) shows the difference in signal between the parity sectors
\begin{equation}
    \Delta \CQ  = \left|\CQ ^{(e)} - \CQ^{(o)} \right|,
\end{equation}
which is, in the topological phase, $h/2e$ periodic. We note that this difference in quantum capacitance between the two parity sectors allows the fast readout of fermion parity demonstrated in Ref.~\cite{Aghaee24} and underlies the measurement-based qubit architecture of Ref.~\cite{karzig2017scalable}.

The second column of Fig.~\ref{fig:MPR-topo-phase} (panels b,d,f) shows the dynamical response of the topological interferometer in the regime where the tunnel barriers are unbalanced with $V_1 = 0.75~V_2$. The different barrier heights lead to different matrix elements coupling the QD states to the left and right MZMs.
As expected, unbalancing the interferometer reduces the flux-dependence of the dynamical response. Indeed, while the maximal dynamical response in panels (b,d) is only reduced by \SI{50}{aF} compared to the balanced case of panels (a,c), the amplitude of the maximal interferometric signal $\Delta \CQ$ is reduced by \SI{200}{aF} (see \Cref{fig:MPR-topo-phase}f). 

\begin{figure}
    \centering
    \includegraphics[width=\columnwidth]{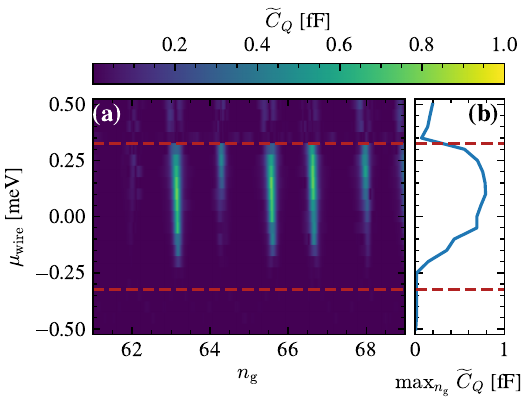}
    \includegraphics[width=\columnwidth]{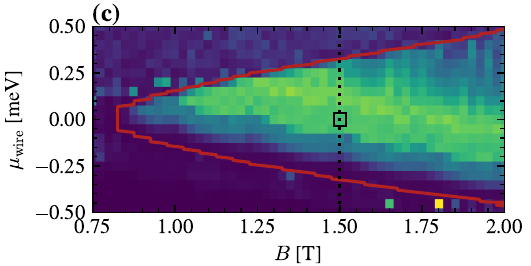}
    \caption{
        (a) $h/2e$ oscillation amplitude of the dynamical quantum capacitance difference $\Delta \CQ$ ($\fCQ$) at $B=\,$\SI{1.5}{\tesla} for a balanced interferometer with $V_1=V_2 =\,$\SI{1.2}{meV}. Dashed red lines indicate the change of sign of the Pfaffian topological invariant (see main text)
        (b) Maximum of $\fCQ$ over the simulated $\ng$ axis.
        (c) Maximum of $\fCQ$ over the $\ng$ axis in the wire phase diagram. See panel (a) for colorbar. The solid red line indicates the topological phase transition based on the Pfaffian invariant. The dotted black line corresponds to the cut shown in panels (a) and (b). The black-outlined square corresponds to the data of Fig.~\ref{fig:MPR-topo-phase}(e).
        Other than $\mu_{\rm wire}$ and $B$, the model parameters follow those of \Cref{fig:MPR-topo-phase}. 
    }
    \label{fig:MPR-phasediagram}
\end{figure}
Now that we have a baseline of the expected interferometric signal in the topological phase, we consider how this signal changes as one explores the topological phase diagram of the superconducting wire. We focus on the balanced regime where $V_1=V_2=\,$\SI{1.2}{\milli\electronvolt}. In order to visualize the interferometric signal in the $(B,\muwire)$ phase-space, one must reduce over the $(\ng, \Phi_x)$ parameter space previously plotted for a single point of phase space in  Fig.~\ref{fig:MPR-topo-phase}. To this end, we focus on the peak-to-peak $h/2e$ oscillation amplitude of $\Delta \CQ$ which we denote as $\fCQ$. We focus on $\fCQ$ as an $h/2e$ periodic signal is expected in the topological phase allowing us to discard trivial flux-independent signal and possible $h/e$ periodic signals of non-topological nature~\cite{Hell2018}.

Figure~\ref{fig:MPR-phasediagram}(a) shows the simulated $h/2e$ oscillation amplitude for a cut of the phase diagram at $B=\,$\SI{1.5}{\tesla}. To relate the simulated signal to the topological phase diagram, we calculate the Pfaffian topological invariant~\cite{Kitaev01} for the bulk of the superconducting wire and indicate the change of the invariant by dashed red horizontal lines, with the center region of the plot, $\muwire \in (-0.3, 0.3)$ corresponding to the topological phase. As observed in Fig.~\ref{fig:MPR-topo-phase}, the amplitude of the $h/2e$ response varies for each QD charge transition due to variations of matrix elements, with the maximal periodic response observed near the center of the topological phase. No signal is observed in the trivial  phase where $\muwire \lesssim \SI{-0.3}{\milli\electronvolt}$ as the chemical potential is below the bottom of the band and the semiconductor is depleted. In the trivial phase where $\muwire \gtrsim \SI{+0.3}{\milli\electronvolt}$, a small ($\lesssim \SI{0.2}{\femto\farad}$) residual $h/2e$ periodic signal is observed on each side of the charge transition. Overall, we observe a good correlation between a large magnitude of the $h/2e$ periodic signal $\fCQ$ and the topological invariant.

To further reduce the dimensionality of the dataset, Fig.~\ref{fig:MPR-phasediagram}(b) takes the maximum over the $\ng$ axis of panel (a). Repeating this analysis as a function of the magnetic field $B$ leads to \Cref{fig:MPR-phasediagram}(c) which shows the maximal $h/2e$-periodic dynamical response throughout the $(B, \muwire)$ phase space. The red curve indicates the topological phase transition as calculated with the Pfaffian invariant. The $h/2e$-periodic response is maximal in the topological phase. We note that a few points near the phase transitions appear noisy. The numerical noise level could be further reduced by adjusting some of the parameters controlling the convergence of the ATCI method and the GP procedure near the topological phase transition.

To emphasize the importance of highly-efficient numerical methods to compute the dynamical response, we note that each horizontal linecut of \Cref{fig:MPR-phasediagram}(a) is the result of 22 separate calculations using the ATCI method sampling 11 points in the $\Phi_x \in (0,1)$ interval, as well as the two parity sectors. Each of those calculations allows us to efficiently build a low-energy subspace within the plotted gate charge interval $n_{\rm g} \in (61,69)$. These subspaces are then the starting point of the open system quantum dynamics simulation used to calculate the dynamical quantum capacitance. 
Effectively sampling that large gate charge interval over multiple charge transitions using independent many-body calculations would require two to three orders of magnitude more simulations. In that case, computing the phase diagram of \Cref{fig:MPR-phasediagram}(c),  which samples the four-dimensional subspace ($\mu_{\rm wire}, B, \ng, \Phi_x$)
and required more than 20,000 aggregate subspace simulations, would necessitate millions of independent simulations which would be prohibitively costly.

\section{Conclusions and Outlook}

In this paper we introduced a general framework for simulating the dynamical response of mesoscopic devices, which is particularly suitable to simulations of rf measurements. Our approach starts from a formulation of dispersive gate sensing as an open-system quantum dynamics problem, and we discuss various advantages of that approach, in particular its ability to model finite-drive effects such as measurement backaction. By introducing a series of controlled approximations, we show that predicting the rf response within microscopic models is feasible.

As an example application, we model the interferometric parity readout of a pair of MZMs in a model that includes microscopic degrees of freedom and goes beyond the description in Ref.~\onlinecite{Aghaee24}. This allows us to explore the full topological phase diagram and demonstrate a good correlation between the topological invariant and a significant flux-$h/2e$-periodic contribution to the quantum capacitance signal. Such a periodicity was observed in the measurements of Ref.~\onlinecite{Aghaee24}.

The ability to treat quasi-one-dimensional Hamiltonians for a hybrid superconductor-semiconductor nanowire together with the charging energy and readout dynamics of nearby QDs opens the possibility of many interesting further research directions. First and foremost are the quantitative description of the readout of single and multi-qubit measurements in topological qubit devices taking into account the effects of disorder and other non-idealities. In addition to topological systems these tools may also be useful for the description of other quantum systems, such as spin qubits and other multi-quantum-dot devices.

\acknowledgements

We acknowledge useful discussions with Andrey E. Antipov, William S. Cole, Christina Knapp, Chetan Nayak, Marcus P. da Silva, and Mark Rudner. We also acknowledge compute infrastructure work and technical help from Joseph Weston and Jeffrey Jones.

\appendix
\addtocontents{toc}{\string\tocdepth@munge}
\section{Phonon spectral function}
\label{app:phonons}

A simple model for electron-phonon coupling is given by scalar deformation potential theory, see, e.g., Ref.~\cite{cleland2013} where longitudinally polarized phonons lead to lattice dilatations that couple to the electronic potential. The corresponding coupling Hamiltonian is given by
\begin{equation}
    \hspace{-1.8mm} H_{D} = -iD \sum_{\bold{k},\bold{q}}\!\left(\frac{\hbar}{2N_{\rm a} M \omega_q}\right)^{1/2}\hspace{-1mm} q \left(a_\bold{q} c^\dagger_{\bold{k}+\bold{q}}c_{\bold{k}}-\text{h.c.}\right)\!,
\end{equation}
where $D$ is the deformation potential, $N_{\rm a}$ and $M$ are the number and mass of atoms, $\omega_q$ is the phonon dispersion, and $c_{\bold{k}},a_{\bold{q}}$ are electron and phonon operators. Here we are mainly interested in the coupling of the phonons to the number operator of the QD, $N=\sum_\bold{k} c^\dagger_\bold{k}c_\bold{k}$. In the energy regime of interest the phonon wavelength in InAs exceeds the typical size of the QDs (hundreds of nm). This allows to approximate
\begin{equation}
    H_D= N \Phi_D
\end{equation}
with $\Phi_D=-iD \sum_{\bold{q}}\left(\hbar/2N_{\rm a} M \omega_q\right)^{1/2}\hspace{-1mm} q \left(a_\bold{q} -\text{h.c.}\right)$. The corresponding spectral function $S_D(\omega)=\int \dd t \langle \Phi_D(t)\Phi_D \rangle e^{i\omega t}$ can then be evaluated yielding
\begin{equation}
    S_D(\omega) =\int \frac{\dd q^3}{(2\pi)^2}\frac{\hbar}{2\rho vq}D^2q^2  \frac{\delta(|\omega|-\omega_q)}{1-e^{-\frac{\hbar \omega}{k_{\rm B} T}}}\,
\end{equation}
where we used $\langle a^\dagger_\bold{q} a_\bold{q}\rangle=(\exp(\hbar \omega_q/k_{\rm B} T)-1)^{-1}$ and assumed a linear phonon dispersion $\omega_q=v|q|$. Performing the momentum integral then leads to the first contribution of $S_{\rm ph}$ in Eq.~\eqref{eq:S_ph}.

In materials with broken inversion symmetry (as in InAs) there is also a significant piezoelectric component of the electron-phonon coupling. The coupling to the QD charge acts via the phonon-induced electric fields that couple to the charge dipole associated with changing the QD occupation. In zincblende materials piezoelectric fields emerge from the shear component of strain which are associated with transverse polarized phonons. This allows to add the piezoelectric component as an independent contribution to $S_{\rm ph}$. In the simplest form we can then estimate the strength of the piezoelectric contribution by replacing $Dq \to eh_{14}$~\cite{cleland2013} which yields Eq.~\eqref{eq:S_ph}.

\section{Low frequency cutoff of $1/f$ noise in ULE framework}
\label{app:cutoff}
In this section we will study the dependence of the small parameter of the ULE framework on the choice of the low frequency cutoff $\omega_0$. For the convergence of the integrals it is helpful to also add a high frequency cutoff $\omega_{\rm c}$ e.g. via $S_{\rm q} \to S_{\rm q}e^{-(\omega/\omega_{\rm c})^2}$. The results will only weakly depend on this cutoff and it's naturally given in terms of the highest energy states that are involved in the evolution of the system. In order to derive the scaling of $\Gamma_{\rm sys}$ and $\tau_{\rm env}$ we will measure the cutoffs relative to temperature via $\hbar\omega_0=x_0 k_{\rm B}T$ and $\hbar\omega_{\rm c}=x_{\rm c} k_{\rm B}T$. The regime of applicability of the classical noise and the occupation of the high energy states then set boundary conditions $x_0 \ll 1$ and $x_{\rm c} \gg 1$. Note that since the temperature dependence is typically exponential the relations $\ll$ and $\gg$ are quickly justified once the cutoffs are below/above their respective bounds. In addition it becomes useful to introduce the time scale $\tau_T=\hbar/(k_{\rm B}T)$.

\begin{figure}
    \centering
    \includegraphics[width=\columnwidth]{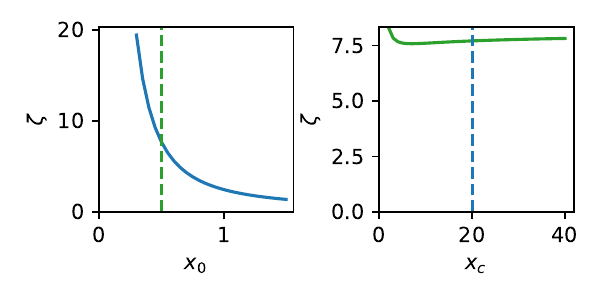}
    \caption{Dependence on the numerical integral \Cref{eq:zeta_integral} on the cutoffs $x_0$ (left) and $x_{\rm c}$ (right). The green (blue) dashed lines indicate the choice of $x_0$ ($x_{\rm c}$) for the plot on the right (left).}
    \label{fig:cutoffs}
\end{figure}

With these definitions we find
\begin{eqnarray}
    g(t) &=& \alpha_C \sqrt{T} \tilde{g}(t/\tau_T) \\
    \tilde{g}(y) &=& \int \frac{\dd x}{2\pi} \frac{e^{-(x/x_{\rm c})^2/2}}{(x^2+x_0^2)^{1/4}}e^{-ixy}\,.
\end{eqnarray}
The perturbative parameter in the ULE framework then takes the form $\Gamma_{\rm sys}\tau_{\rm env} = (\alpha_C^2/T^2) \zeta$, with
\begin{equation}
    \zeta = 4\left(\int \dd y |\tilde{g}(y)|\right)\left(\int \dd y_1|y_1\tilde{g}(y_1)|\right)\,.
    \label{eq:zeta_integral}
\end{equation}
The corresponding integrals can be evaluated numerically as shown in \Cref{fig:cutoffs}. We observe a very weak dependence on $x_{\rm c}$ and chose $x_{\rm c}=20$ going forward which is well inside the regime of applicability. The cutoff $x_0$ represents a tradeoff between the applicability of the ULE (large $x_0$ significantly suppress $\zeta$) and the applicability of the splitting procedure of the $1/f$ noise into a classical and quantum part. Here we will use $x_0=0.5$ which yields $\zeta \approx 7.7$. For typical values of $\alpha_C=\SI{1}{\micro\electronvolt}$ and $T=50{\rm mK}$ the corresponding ULE parameter $\Gamma_{\rm sys}\tau_{\rm env}\approx 0.4$ which is also in the regime of applicability. Note that while the error bounds of the ULE derived by Nathan and Rudner \cite{Nathan20} are rigorous they are not necessarily tight and the ULE may still perform well for $\Gamma_{\rm sys}\tau_{\rm env}\gtrsim 1$.

\section{Linear response treatment of dynamical $\CQ$}
\label{sec:lin_resp}
Ref.~\cite{Aghaee24} derived an expression for dynamical $\CQ$ in the limit of infinitesimal dissipation. Here we extend this to finite dissipation to better compare with the simulated results. As a starting point we use that the dynamical $\CQ$ can be expressed to lowest order in the coupling to the readout resonator in terms of the following response function \cite{Aghaee24} 
\begin{eqnarray}
    \CQ(\omega)^{\rm (lr)} &=& -(\text{e}\alpha)^2 \int \frac{d \Omega_1}{2\pi}\Tr\Big[V G^K(\Omega_1)V G^R(\Omega_1+\hbar\omega) \nonumber\\
    & & \hspace{1.5cm} + V G^A(\Omega_1-\hbar\omega)V G^K(\Omega_1)\Big]. \label{eq:dyn_CQ_int}
\end{eqnarray}
Here, ${\rm e}$ is the elementary charge, $V=\partial H_{\rm TLS}/\partial \Delta=\sigma_Z/2$, while $G^R$, $G^A$, $G^K$ are, respectively, the retarded, advanced and Keldysh Green's functions of the system.

It is convenient to work in the eigenbasis where $H_{\rm TLS}$ is diagonal and shift the energy spectrum so that the ground state is at zero energy while the excited state is at energy $E=\sqrt{\Delta^2+4t_C^2}$. Then,
\begin{equation}
    G^R(\Omega)=\left(\begin{matrix}
        (i\eta/2 +\Omega)^{-1} & 0 \\
        0 & (i\eta/2 +\Omega - E)^{-1}
    \end{matrix} \right)\,
\end{equation}
while $G^A(\Omega)=G^R(\Omega)^*$ and $G^K(\Omega)=[G^R(\Omega)-G^A(\Omega)]\tanh(\Omega/2k_BT)$. The latter relation assumes that to lowest order in the coupling to the resonator the occupation of the system remains thermal. 

Performing the integral in \eqref{eq:dyn_CQ_int} yields
\begin{equation}
    \CQ^{\rm (lr)}=\frac{2\text{e}^2 \alpha^2 t_{\rm C}^2}{E(E^2-(\hbar\omega+i\eta/2)^2)}\tanh\left(\frac{E}{2k_{\rm B}T}\right)\,, \label{eq:CQ_infi}
\end{equation}
in the limit of infinitesimal $\eta$ \cite{Aghaee24} which describes the real part switching sign while the imaginary part takes the form of a delta function around $\omega=E$. Away from this resonance condition and for small but finite $\eta$, the real part is typically well-described by the $\eta=0$ limit while the imaginary part scales proportional to $\eta$ but has additional contributions compared to \eqref{eq:CQ_infi}. The integral can be evaluated either numerically or approximately analytically (for not too small temperatures) by summing over the poles in \eqref{eq:dyn_CQ_int} and taking the first few poles of $\tanh$ into account. The resulting analytic expressions are lengthy and not explicitly reported here but used in the plots of Fig.~\ref{fig:comparison_keldysh}.

In order to make a connection with the simulations based on Lindblad master equations, we interpret $\eta$ as the decay rate of the eigenstates due to transitions between ground and excited states induced by noise coupling to the system. Specifically, we use
\begin{eqnarray}
    \eta &=& \frac{S_{\rm g}(E)+S_{\rm g}(-E)}{2} |\langle 1|\hat{n}_d|0\rangle|^2 \\
    &&+\frac{S_{\rm t}(E)+S_{\rm t}(-E)}{2} |\langle 1|\sigma_X|0\rangle|^2\,,
\end{eqnarray}
where $S_{\rm g}$ and $S_{\rm t}$ are the noise spectral functions described by the effective rates $\gamma_{\rm g}$ and $\gamma_{\rm t}$ and the matrix elements capture the efficiency of the noise to transition the system between the ground to excited state. The treatment of capturing the effect of noise via a single scalar is a simplification and we thus do not expect a 1:1 correspondence with the simulation results but rather a qualitative agreement. A more rigorous treatment of the linear response regime is beyond the scope of this paper but could be obtained by replacing $\eta$ with the (generally frequency-dependent and matrix-valued) self energy introduced after integrating out the noisy environmental degrees of freedom.

\section{Details of DMRG simulations}\label{App:dmrg}
The DMRG simulations presented in Sec.~\ref{sec:methods} are obtained using the implementation of Ref.~\cite{ITensor}. In order to implement the charging energy interaction term of \Cref{eq:HCspinless}, we use a matrix product operator (MPO) representation of the charging energy part of the Hamiltonian similar to Ref.~\cite{Pavesic2022A}. Following the site indexing introduced in \Cref{sec:spinlessModel}, the tensors $W_j$ forming the MPO representation of the charging energy are only nontrivial for sites $j \in [-N_{\rm qd}, -1]$, corresponding to the QD sites.
Away from the QD boundaries ($j \in [-N_{\rm qd}+1, -2]$), the tensors are
\begin{equation}
    W_j = \left(\begin{matrix} 1& \EC (1- 2\ng)\hat n_j & 2E_c \hat n_j \\ 0 & 1 & 0 \\ 0  &\hat n_j & I \end{matrix} \right),
\end{equation}
where the left boundary tensor is
\begin{equation}
    W_{-N_{\rm qd}} = \left(\begin{matrix} 1& \EC (1- 2\ng)\hat n_{-N_{\rm qd}} & 2E_c \hat n_{-N_{\rm qd}} \end{matrix} \right),
\end{equation}
while the right boundary tensor reads
\begin{equation}
    W_{-1} = \left(\begin{matrix}
    \EC(1- 2\ng)\hat n_{-1}
    \\
    1
    \\
    \hat n_{-1}
    \end{matrix} \right).
\end{equation}
This representation allows for a compact MPO representation of \Cref{eq:HCspinless} despite the long-range nature of the charging energy interaction term.
 
For the simulation results of \Cref{fig:wfn-convergence2}, which requires the calculation of excited states, we use an iterative procedure where we introduce penalty terms proportional to the overlap with lower-energy states into the Hamiltonian~\cite{Stoudenmire12}.
This method is known to occasionally get stuck in local minima and miss one or more eigenstates. To reduce this risk, we use the following strategy. First, for a fixed maximum bond dimension we iteratively find the $n_s=5$ first eigenstates. Second, we sort these eigenstates based on their respective energy. Finally, we repeat the process using as an initial state the respective sorted states. We repeat this loop with a gradually increasing bond dimension up to convergence for bond dimensions $\chi \sim 300$. The final variational eigenstates are then orthogonalized.

\section{Matrix elements of non-orthogonal Gaussian states} \label{App:matrix-elements}
The implementation of the numerical methods of \Cref{sec:methods} relies on the covariance matrix representation of fermionic Gaussian states.
While we refer the reader to Ref.~\cite{Bravyi2017} for more details and derivations, we give here a summary of the main ideas and equations needed for calculating matrix elements using the covariance matrix formalism.

The mapping between a Gaussian state $\ket{\Gamma}$ and its covariance matrix $\Gamma$ is incomplete as an additional phase parameter is needed to fully describe the state. Indeed, the covariance matrix $\Gamma$ is invariant under gauge transformations of the form $\ket{\Gamma} \rightarrow e^{i\phi}\ket{\Gamma}$.
A key element of the covariance matrix formalism with non-orthogonal states is then the introduction of a reference state $\ket{\Gamma_{\rm ref}}$ which is non-orthogonal to all states in the aggregated subspace. One can use the gauge freedom of each state to define $\braket{\Gamma_{\rm ref} | \Gamma_n^{(k)}}  = \braket{\Gamma_n^{(k)} | \Gamma_{\rm ref}} = g_n^{(k)} >0$, with $g_n^{(k)}$ a real number. This gauge choice fixes a phase reference and allows to compute matrix elements in terms only of gauge-invariant quantities. For example, the overlap between two Gaussian states $\ket{\Gamma_1}$ and $\ket{\Gamma_2}$ can be written as 
\begin{equation}
    \braket{\Gamma_1 | \Gamma_2} = \frac{
    \braket{\Gamma_{\rm ref} | \Gamma_1}
    \braket{\Gamma_1 | \Gamma_2}
    \braket{\Gamma_2 | \Gamma_{\rm ref}}
}{
    \braket{\Gamma_{\rm ref} | \Gamma_1}
    \braket{\Gamma_2 | \Gamma_{\rm ref}}
}, \label{eq:overlapGaugeInvariant}
\end{equation}
with the numerator a gauge-invariant quantity that can be computed using a covariance matrix representation of the states~\cite{Bravyi2017}
\begin{align}
\begin{split}
    &\braket{\Gamma_{\rm ref} | \Gamma_1}
    \braket{\Gamma_1 | \Gamma_2}
    \braket{\Gamma_2 | \Gamma_{\rm ref}} = \\
    &
    \qquad \Pf\left[\frac{1}{2}(\Gamma_1 +\Gamma_2)\right] \Pf\left[\frac{1}{2}(\Delta^{1,2}+\Gamma_{\rm ref})\right],
    \label{eq:overlap}
\end{split}
\end{align}
with $\Pf[A]$ the Pfaffian of matrix $A$ and where we introduced the complex matrix 
\begin{equation}
    \Delta^{1, 2} = \left[i(\Gamma_1 - \Gamma_2)-2\openone\right](\Gamma_1 + \Gamma_2)^{-1}.
    \label{eq:delta12}
\end{equation}
Similarly, to Eq.~\eqref{eq:overlap} the denominator of Eq.~\eqref{eq:overlapGaugeInvariant} can also be computed in terms of covariance matrices given the gauge choice above using
\begin{align}
    \braket{\Gamma_{\rm ref} | \Gamma_n}
    = \left|\braket{\Gamma_{\rm ref} | \Gamma_n}\right|
    = \sqrt{\left|\Pf \left[\frac{\Gamma_{\rm ref} + \Gamma_n}{2}\right]\right|},
\end{align}
where the first equality follows from our gauge choice. In practice, while the reference state needs to be non-orthogonal to all states of interests, the overlap to these states can be very small as efficient numerical routines avoiding underflow are available~\cite{Wimmer12}.

Similarly, one can derive expressions for the matrix elements of operators. Defining a  monomial of Majorana operators with distinct indices $\mathcal{I}=i_1, i_2,\dots i_n$, its matrix element can be computed as
\begin{align}
    \bra{\Gamma_1}c_{i_1} c_{i_2} \dots c_{i_n} \ket{\Gamma_2} &=  \braket{\Gamma_1 | \Gamma_2} \Pf  \left[i \Delta^{1,2}(\mathcal{I})\right]^*,
    \label{eq:mat_elemDelta}
\end{align}
where $\Delta^{1,2}(\mathcal{I})$ is a $n\times n$ submatrix of $\Delta^{1,2}$ obtained by taking the rows and columns corresponding to the indices in $\mathcal{I}$.

In Eq.~\eqref{eq:delta12}, the matrix $\Gamma_1 + \Gamma_2$ is singular if the states $\ket{\Gamma_1}$ and $\ket{\Gamma_2}$ are orthogonal. In that limit, the above expressions involving the complex matrix $\Delta^{1, 2}$ are not valid. While alternate expressions valid in the limit of orthogonal states are derived in Ref.~\cite{Bravyi2017}, those are significantly more expensive to use numerically. Hence, in this work, we only apply the covariance matrix formalism in cases where the Gaussian states are non-orthogonal.

Finally, in order to relate the operator formalism used in Sec.~\ref{sec:truncatedCI} for orthogonal states and the covariance matrix formalism of this section, we choose the arbitrary phase factor in the definition of $\hat \Lambda_n^{(j)}$ such that $\braket{\Gamma_{\rm ref} | \Gamma_n^{(j)}} = \braket{\Gamma_{\rm ref} | \hat \Lambda_n^{(j)} | \Gamma_0^{(j)}}$.

\section{Generalized Hartree-Fock implementation}\label{App:ImaginaryEvolution}
In this Appendix, we expand on the implementation of the imaginary time evolution used to obtain the gHF approximate ground state of the interacting problem introduced in \Cref{sec:gHF}.

For a pure quantum state, normalization leads to the constraint $\Gamma^2 = -\openone$.
While the \Cref{eqn:imtimeEOM} can be solved using an ordinary differential equation solver, care must be taken to ensure that the normalization of the covariance matrix is preserved. In this work, this is achieved by using an orthogonal transformation to implement time-evolution. As introduced in Ref.~\cite{kraus2010generalized}, evolving from imaginary time $\tau$ to $\tau + {\rm d}\tau$ to first order in ${\rm d}\tau$ can be computed by applying the similarity transformation
 \begin{equation}
     \Gamma(\tau +{\rm d}\tau) = R(\tau)\Gamma(\tau)R(\tau)^{\rm T},
 \end{equation}
 with the orthogonal matrix
     $R(\tau) = \exp\left\{\frac{{\rm d}\tau}{2} \left[\Gamma\,,\, \partial_\tau \Gamma \right]\right\}$  ($RR^{\rm T} = \openone$).
In addition, the normalization of the covariance matrix can be corrected by decomposing the real skew-symmetric matrix in its canonical form
 \begin{equation}
    \Gamma = S \bigoplus_{j=1}^N 
    \left(\begin{matrix}
    0 & \lambda_j \\ -\lambda_j &0
    \end{matrix}\right)
    S^T,
    \label{eqn:CM-normal-form}
\end{equation}
and rescaling the coefficients $\lambda_j$ to unity.

\section{Natural orbitals for superconducting systems}\label{app:natOrb}
The natural orbital folding procedure introduced in \Cref{sec:natural-orbitals} can also be applied to the case of systems with (mean-field) superconductivity. We consider an extension of the generic Hamiltonian Eq.~\eqref{eq:HfNatOrb} to a Bogoliubov-de Gennes (BdG) basis. Introducing the Nambu spinors, $\Psi_i = (c_i, c_i^\dagger)$, and the vector of spinors in part $p$, $\boldsymbol{\Psi}_p$, the Hamiltonian now reads
\begin{multline}
\cH_{\rm BdG} = \boldsymbol{\Psi}^\dagger H \boldsymbol{\Psi} = 
    \sum_p \boldsymbol{\Psi}_p^\dagger H^{(p)} \boldsymbol{\Psi}_p +\\
    \sum_{p_1,p_2} \left[ \boldsymbol{\Psi}_{p_1}^\dagger H^{(p_1,p_2)} \boldsymbol{\Psi}_{p_2} + {\rm h.c.} \right].
\end{multline}
The folding procedure follows in the same way as the normal case, and in particular is done based only on the normal part of the single-particle density matrix. In particular, the natural orbitals are still defined as the eigenvectors of Eq.~\eqref{eq:Dreduced}. However, due to Nambu doubling, the isometry $Q_p$ will in this case be a $2N_k \times 2N_p$ with nonzero matrix elements
\begin{align}
    (Q_p)_{2n-1,2m-1} = [|n)_p]_m, \\
    (Q_p)_{2n,2m} = [|n)_p]_m^*,
\end{align}
where $n = 1,\dots N_k$ and $m=1, \dots, N_p$. In other words, the odd rows and odd columns of $Q_p$ are filled by the eigenvectors of $D_p$, while, to preserve the particle-hole symmetry of the projected Hamiltonian, the even rows and columns are filled by the complex-conjugate of the eigenvectors.


\begin{thebibliography}{60}%
\makeatletter
\providecommand \@ifxundefined [1]{%
 \@ifx{#1\undefined}
}%
\providecommand \@ifnum [1]{%
 \ifnum #1\expandafter \@firstoftwo
 \else \expandafter \@secondoftwo
 \fi
}%
\providecommand \@ifx [1]{%
 \ifx #1\expandafter \@firstoftwo
 \else \expandafter \@secondoftwo
 \fi
}%
\providecommand \natexlab [1]{#1}%
\providecommand \enquote  [1]{``#1''}%
\providecommand \bibnamefont  [1]{#1}%
\providecommand \bibfnamefont [1]{#1}%
\providecommand \citenamefont [1]{#1}%
\providecommand \href@noop [0]{\@secondoftwo}%
\providecommand \href [0]{\begingroup \@sanitize@url \@href}%
\providecommand \@href[1]{\@@startlink{#1}\@@href}%
\providecommand \@@href[1]{\endgroup#1\@@endlink}%
\providecommand \@sanitize@url [0]{\catcode `\\12\catcode `\$12\catcode
  `\&12\catcode `\#12\catcode `\^12\catcode `\_12\catcode `\%12\relax}%
\providecommand \@@startlink[1]{}%
\providecommand \@@endlink[0]{}%
\providecommand \url  [0]{\begingroup\@sanitize@url \@url }%
\providecommand \@url [1]{\endgroup\@href {#1}{\urlprefix }}%
\providecommand \urlprefix  [0]{URL }%
\providecommand \Eprint [0]{\href }%
\providecommand \doibase [0]{https://doi.org/}%
\providecommand \selectlanguage [0]{\@gobble}%
\providecommand \bibinfo  [0]{\@secondoftwo}%
\providecommand \bibfield  [0]{\@secondoftwo}%
\providecommand \translation [1]{[#1]}%
\providecommand \BibitemOpen [0]{}%
\providecommand \bibitemStop [0]{}%
\providecommand \bibitemNoStop [0]{.\EOS\space}%
\providecommand \EOS [0]{\spacefactor3000\relax}%
\providecommand \BibitemShut  [1]{\csname bibitem#1\endcsname}%
\let\auto@bib@innerbib\@empty
%</preamble>
\bibitem [{\citenamefont {Aghaee}\ \emph {et~al.}(2023)\citenamefont {Aghaee}
  \emph {et~al.}}]{Aghaee23}%
  \BibitemOpen
  \bibfield  {author} {\bibinfo {author} {\bibfnamefont {M.}~\bibnamefont
  {Aghaee}} \emph {et~al.} (\bibinfo {collaboration} {Microsoft Quantum}),\
  }\bibfield  {title} {\bibinfo {title} {{InAs-Al hybrid devices passing the
  topological gap protocol}},\ }\href
  {https://doi.org/10.1103/PhysRevB.107.245423} {\bibfield  {journal} {\bibinfo
   {journal} {Phys. Rev. B}\ }\textbf {\bibinfo {volume} {107}},\ \bibinfo
  {pages} {245423} (\bibinfo {year} {2023})}\BibitemShut {NoStop}%
\bibitem [{\citenamefont {Aghaee}\ \emph {et~al.}(2024)\citenamefont {Aghaee}
  \emph {et~al.}}]{Aghaee24}%
  \BibitemOpen
  \bibfield  {author} {\bibinfo {author} {\bibfnamefont {M.}~\bibnamefont
  {Aghaee}} \emph {et~al.} (\bibinfo {collaboration} {Microsoft Quantum}),\
  }\bibfield  {title} {\bibinfo {title} {{Interferometric Single-Shot Parity
  Measurement in an InAs-Al Hybrid Device}},\ }\href
  {https://doi.org/10.48550/arXiv.2401.09549} {\bibfield  {journal} {\bibinfo
  {journal} {arXiv:2401.09549}\ } (\bibinfo {year} {2024})}\BibitemShut
  {NoStop}%
\bibitem [{\citenamefont {Datta}(1997)}]{datta_electronic_1997}%
  \BibitemOpen
  \bibfield  {author} {\bibinfo {author} {\bibfnamefont {S.}~\bibnamefont
  {Datta}},\ }\href@noop {} {\emph {\bibinfo {title} {Electronic transport in
  mesoscopic systems}}}\ (\bibinfo  {publisher} {Cambridge University Press},\
  \bibinfo {year} {1997})\BibitemShut {NoStop}%
\bibitem [{\citenamefont {Nazarov}\ and\ \citenamefont
  {Blanter}(2009)}]{nazarov_quantum_2009}%
  \BibitemOpen
  \bibfield  {author} {\bibinfo {author} {\bibfnamefont {Y.~V.}\ \bibnamefont
  {Nazarov}}\ and\ \bibinfo {author} {\bibfnamefont {Y.~M.}\ \bibnamefont
  {Blanter}},\ }\href@noop {} {\emph {\bibinfo {title} {Quantum {Transport}:
  {Introduction} to {Nanoscience}}}}\ (\bibinfo  {publisher} {Cambridge
  University Press},\ \bibinfo {year} {2009})\BibitemShut {NoStop}%
\bibitem [{\citenamefont {Vigneau}\ \emph {et~al.}(2023)\citenamefont
  {Vigneau}, \citenamefont {Fedele}, \citenamefont {Chatterjee}, \citenamefont
  {Reilly}, \citenamefont {Kuemmeth}, \citenamefont {Gonzalez-Zalba},
  \citenamefont {Laird},\ and\ \citenamefont {Ares}}]{vigneau_probing_2023}%
  \BibitemOpen
  \bibfield  {author} {\bibinfo {author} {\bibfnamefont {F.}~\bibnamefont
  {Vigneau}}, \bibinfo {author} {\bibfnamefont {F.}~\bibnamefont {Fedele}},
  \bibinfo {author} {\bibfnamefont {A.}~\bibnamefont {Chatterjee}}, \bibinfo
  {author} {\bibfnamefont {D.}~\bibnamefont {Reilly}}, \bibinfo {author}
  {\bibfnamefont {F.}~\bibnamefont {Kuemmeth}}, \bibinfo {author}
  {\bibfnamefont {M.~F.}\ \bibnamefont {Gonzalez-Zalba}}, \bibinfo {author}
  {\bibfnamefont {E.}~\bibnamefont {Laird}},\ and\ \bibinfo {author}
  {\bibfnamefont {N.}~\bibnamefont {Ares}},\ }\bibfield  {title} {\bibinfo
  {title} {Probing quantum devices with radio-frequency reflectometry},\ }\href
  {https://doi.org/10.1063/5.0088229} {\bibfield  {journal} {\bibinfo
  {journal} {Appl. Phys. Rev.}\ }\textbf {\bibinfo {volume} {10}},\ \bibinfo
  {pages} {021305} (\bibinfo {year} {2023})}\BibitemShut {NoStop}%
\bibitem [{\citenamefont {Burkard}\ \emph {et~al.}(2023)\citenamefont
  {Burkard}, \citenamefont {Ladd}, \citenamefont {Pan}, \citenamefont
  {Nichol},\ and\ \citenamefont {Petta}}]{burkard_semiconductor_2023}%
  \BibitemOpen
  \bibfield  {author} {\bibinfo {author} {\bibfnamefont {G.}~\bibnamefont
  {Burkard}}, \bibinfo {author} {\bibfnamefont {T.~D.}\ \bibnamefont {Ladd}},
  \bibinfo {author} {\bibfnamefont {A.}~\bibnamefont {Pan}}, \bibinfo {author}
  {\bibfnamefont {J.~M.}\ \bibnamefont {Nichol}},\ and\ \bibinfo {author}
  {\bibfnamefont {J.~R.}\ \bibnamefont {Petta}},\ }\bibfield  {title} {\bibinfo
  {title} {Semiconductor spin qubits},\ }\href
  {https://doi.org/10.1103/RevModPhys.95.025003} {\bibfield  {journal}
  {\bibinfo  {journal} {Rev. Mod. Phys.}\ }\textbf {\bibinfo {volume} {95}},\
  \bibinfo {pages} {025003} (\bibinfo {year} {2023})}\BibitemShut {NoStop}%
\bibitem [{\citenamefont {Karzig}\ \emph {et~al.}(2017)\citenamefont {Karzig},
  \citenamefont {Knapp}, \citenamefont {Lutchyn}, \citenamefont {Bonderson},
  \citenamefont {Hastings}, \citenamefont {Nayak}, \citenamefont {Alicea},
  \citenamefont {Flensberg}, \citenamefont {Plugge}, \citenamefont {Oreg},
  \citenamefont {Marcus},\ and\ \citenamefont {Freedman}}]{karzig2017scalable}%
  \BibitemOpen
  \bibfield  {author} {\bibinfo {author} {\bibfnamefont {T.}~\bibnamefont
  {Karzig}}, \bibinfo {author} {\bibfnamefont {C.}~\bibnamefont {Knapp}},
  \bibinfo {author} {\bibfnamefont {R.~M.}\ \bibnamefont {Lutchyn}}, \bibinfo
  {author} {\bibfnamefont {P.}~\bibnamefont {Bonderson}}, \bibinfo {author}
  {\bibfnamefont {M.~B.}\ \bibnamefont {Hastings}}, \bibinfo {author}
  {\bibfnamefont {C.}~\bibnamefont {Nayak}}, \bibinfo {author} {\bibfnamefont
  {J.}~\bibnamefont {Alicea}}, \bibinfo {author} {\bibfnamefont
  {K.}~\bibnamefont {Flensberg}}, \bibinfo {author} {\bibfnamefont
  {S.}~\bibnamefont {Plugge}}, \bibinfo {author} {\bibfnamefont
  {Y.}~\bibnamefont {Oreg}}, \bibinfo {author} {\bibfnamefont {C.~M.}\
  \bibnamefont {Marcus}},\ and\ \bibinfo {author} {\bibfnamefont {M.~H.}\
  \bibnamefont {Freedman}},\ }\bibfield  {title} {\bibinfo {title} {Scalable
  designs for quasiparticle-poisoning-protected topological quantum computation
  with {Majorana} zero modes},\ }\href
  {https://doi.org/10.1103/PhysRevB.95.235305} {\bibfield  {journal} {\bibinfo
  {journal} {Phys. Rev. B}\ }\textbf {\bibinfo {volume} {95}},\ \bibinfo
  {pages} {235305} (\bibinfo {year} {2017})}\BibitemShut {NoStop}%
\bibitem [{\citenamefont {Derakhshan~Maman}\ \emph {et~al.}(2020)\citenamefont
  {Derakhshan~Maman}, \citenamefont {Gonzalez-Zalba},\ and\ \citenamefont
  {P\'alyi}}]{maman_2020}%
  \BibitemOpen
  \bibfield  {author} {\bibinfo {author} {\bibfnamefont {V.}~\bibnamefont
  {Derakhshan~Maman}}, \bibinfo {author} {\bibfnamefont {M.}~\bibnamefont
  {Gonzalez-Zalba}},\ and\ \bibinfo {author} {\bibfnamefont {A.}~\bibnamefont
  {P\'alyi}},\ }\bibfield  {title} {\bibinfo {title} {Charge noise and
  overdrive errors in dispersive readout of charge, spin, and {Majorana}
  qubits},\ }\href {https://doi.org/10.1103/PhysRevApplied.14.064024}
  {\bibfield  {journal} {\bibinfo  {journal} {Phys. Rev. Appl.}\ }\textbf
  {\bibinfo {volume} {14}},\ \bibinfo {pages} {064024} (\bibinfo {year}
  {2020})}\BibitemShut {NoStop}%
\bibitem [{\citenamefont {Peri}\ \emph {et~al.}(2024)\citenamefont {Peri},
  \citenamefont {Oakes}, \citenamefont {Cochrane}, \citenamefont {Ford},\ and\
  \citenamefont {Gonzalez-Zalba}}]{peri_2024}%
  \BibitemOpen
  \bibfield  {author} {\bibinfo {author} {\bibfnamefont {L.}~\bibnamefont
  {Peri}}, \bibinfo {author} {\bibfnamefont {G.~A.}\ \bibnamefont {Oakes}},
  \bibinfo {author} {\bibfnamefont {L.}~\bibnamefont {Cochrane}}, \bibinfo
  {author} {\bibfnamefont {C.~J.~B.}\ \bibnamefont {Ford}},\ and\ \bibinfo
  {author} {\bibfnamefont {M.~F.}\ \bibnamefont {Gonzalez-Zalba}},\ }\bibfield
  {title} {\bibinfo {title} {Beyond-adiabatic {Quantum} {Admittance} of a
  {Semiconductor} {Quantum} {Dot} at {High} {Frequencies}: {Rethinking}
  {Reflectometry} as {Polaron} {Dynamics}},\ }\href
  {https://doi.org/10.22331/q-2024-03-21-1294} {\bibfield  {journal} {\bibinfo
  {journal} {Quantum}\ }\textbf {\bibinfo {volume} {8}},\ \bibinfo {pages}
  {1294} (\bibinfo {year} {2024})}\BibitemShut {NoStop}%
\bibitem [{\citenamefont {Nathan}\ and\ \citenamefont
  {Rudner}(2020)}]{Nathan20}%
  \BibitemOpen
  \bibfield  {author} {\bibinfo {author} {\bibfnamefont {F.}~\bibnamefont
  {Nathan}}\ and\ \bibinfo {author} {\bibfnamefont {M.~S.}\ \bibnamefont
  {Rudner}},\ }\bibfield  {title} {\bibinfo {title} {Universal {{Lindblad}}
  equation for open quantum systems},\ }\href
  {https://doi.org/10.1103/PhysRevB.102.115109} {\bibfield  {journal} {\bibinfo
   {journal} {Phys. Rev. B}\ }\textbf {\bibinfo {volume} {102}},\ \bibinfo
  {pages} {115109} (\bibinfo {year} {2020})}\BibitemShut {NoStop}%
\bibitem [{\citenamefont {Nielsen}\ and\ \citenamefont
  {Muller}(2010)}]{nielsen2010configuration}%
  \BibitemOpen
  \bibfield  {author} {\bibinfo {author} {\bibfnamefont {E.}~\bibnamefont
  {Nielsen}}\ and\ \bibinfo {author} {\bibfnamefont {R.~P.}\ \bibnamefont
  {Muller}},\ }\bibfield  {title} {\bibinfo {title} {A configuration
  interaction analysis of exchange in double quantum dots},\ }\href
  {https://doi.org/10.48550/arXiv.1006.2735} {\bibfield  {journal} {\bibinfo
  {journal} {arXiv:1006.2735}\ } (\bibinfo {year} {2010})}\BibitemShut
  {NoStop}%
\bibitem [{\citenamefont {Shehata}\ \emph {et~al.}(2023)\citenamefont
  {Shehata}, \citenamefont {Simion}, \citenamefont {Li}, \citenamefont
  {Mohiyaddin}, \citenamefont {Wan}, \citenamefont {Mongillo}, \citenamefont
  {Govoreanu}, \citenamefont {Radu}, \citenamefont {De~Greve},\ and\
  \citenamefont {Van~Dorpe}}]{shehata2023modeling}%
  \BibitemOpen
  \bibfield  {author} {\bibinfo {author} {\bibfnamefont {M.~M. E.~K.}\
  \bibnamefont {Shehata}}, \bibinfo {author} {\bibfnamefont {G.}~\bibnamefont
  {Simion}}, \bibinfo {author} {\bibfnamefont {R.}~\bibnamefont {Li}}, \bibinfo
  {author} {\bibfnamefont {F.~A.}\ \bibnamefont {Mohiyaddin}}, \bibinfo
  {author} {\bibfnamefont {D.}~\bibnamefont {Wan}}, \bibinfo {author}
  {\bibfnamefont {M.}~\bibnamefont {Mongillo}}, \bibinfo {author}
  {\bibfnamefont {B.}~\bibnamefont {Govoreanu}}, \bibinfo {author}
  {\bibfnamefont {I.}~\bibnamefont {Radu}}, \bibinfo {author} {\bibfnamefont
  {K.}~\bibnamefont {De~Greve}},\ and\ \bibinfo {author} {\bibfnamefont
  {P.}~\bibnamefont {Van~Dorpe}},\ }\bibfield  {title} {\bibinfo {title}
  {Modeling semiconductor spin qubits and their charge noise environment for
  quantum gate fidelity estimation},\ }\href
  {https://doi.org/10.1103/PhysRevB.108.045305} {\bibfield  {journal} {\bibinfo
   {journal} {Phys. Rev. B}\ }\textbf {\bibinfo {volume} {108}},\ \bibinfo
  {pages} {045305} (\bibinfo {year} {2023})}\BibitemShut {NoStop}%
\bibitem [{\citenamefont {Bach}\ \emph {et~al.}(1994)\citenamefont {Bach},
  \citenamefont {Lieb},\ and\ \citenamefont {Solovej}}]{bach1994generalized}%
  \BibitemOpen
  \bibfield  {author} {\bibinfo {author} {\bibfnamefont {V.}~\bibnamefont
  {Bach}}, \bibinfo {author} {\bibfnamefont {E.~H.}\ \bibnamefont {Lieb}},\
  and\ \bibinfo {author} {\bibfnamefont {J.~P.}\ \bibnamefont {Solovej}},\
  }\bibfield  {title} {\bibinfo {title} {Generalized {H}artree-{F}ock theory
  and the {H}ubbard model},\ }\href {https://doi.org/10.1007/BF02188656}
  {\bibfield  {journal} {\bibinfo  {journal} {J. Stat. Phys.}\ }\textbf
  {\bibinfo {volume} {76}},\ \bibinfo {pages} {3} (\bibinfo {year}
  {1994})}\BibitemShut {NoStop}%
\bibitem [{\citenamefont {White}(1992)}]{white1992density}%
  \BibitemOpen
  \bibfield  {author} {\bibinfo {author} {\bibfnamefont {S.~R.}\ \bibnamefont
  {White}},\ }\bibfield  {title} {\bibinfo {title} {Density matrix formulation
  for quantum renormalization groups},\ }\href
  {https://doi.org/10.1103/PhysRevLett.69.2863} {\bibfield  {journal} {\bibinfo
   {journal} {Phys. Rev. Lett.}\ }\textbf {\bibinfo {volume} {69}},\ \bibinfo
  {pages} {2863} (\bibinfo {year} {1992})}\BibitemShut {NoStop}%
\bibitem [{\citenamefont {Munk}\ \emph {et~al.}(2020)\citenamefont {Munk},
  \citenamefont {Schulenborg}, \citenamefont {Egger},\ and\ \citenamefont
  {Flensberg}}]{munk2020parity}%
  \BibitemOpen
  \bibfield  {author} {\bibinfo {author} {\bibfnamefont {M.~I.}\ \bibnamefont
  {Munk}}, \bibinfo {author} {\bibfnamefont {J.}~\bibnamefont {Schulenborg}},
  \bibinfo {author} {\bibfnamefont {R.}~\bibnamefont {Egger}},\ and\ \bibinfo
  {author} {\bibfnamefont {K.}~\bibnamefont {Flensberg}},\ }\bibfield  {title}
  {\bibinfo {title} {Parity-to-charge conversion in {M}ajorana qubit readout},\
  }\href {https://doi.org/10.1103/PhysRevResearch.2.033254} {\bibfield
  {journal} {\bibinfo  {journal} {Phys. Rev. Res.}\ }\textbf {\bibinfo {volume}
  {2}},\ \bibinfo {pages} {033254} (\bibinfo {year} {2020})}\BibitemShut
  {NoStop}%
\bibitem [{\citenamefont {Steiner}\ and\ \citenamefont {von
  Oppen}(2020)}]{steiner2020readout}%
  \BibitemOpen
  \bibfield  {author} {\bibinfo {author} {\bibfnamefont {J.~F.}\ \bibnamefont
  {Steiner}}\ and\ \bibinfo {author} {\bibfnamefont {F.}~\bibnamefont {von
  Oppen}},\ }\bibfield  {title} {\bibinfo {title} {Readout of {M}ajorana
  qubits},\ }\href {https://doi.org/10.1103/PhysRevResearch.2.033255}
  {\bibfield  {journal} {\bibinfo  {journal} {Phys. Rev. Res.}\ }\textbf
  {\bibinfo {volume} {2}},\ \bibinfo {pages} {033255} (\bibinfo {year}
  {2020})}\BibitemShut {NoStop}%
\bibitem [{\citenamefont {Sau}\ and\ \citenamefont
  {Sarma}(2024)}]{sau2024capacitance}%
  \BibitemOpen
  \bibfield  {author} {\bibinfo {author} {\bibfnamefont {J.~D.}\ \bibnamefont
  {Sau}}\ and\ \bibinfo {author} {\bibfnamefont {S.~D.}\ \bibnamefont
  {Sarma}},\ }\bibfield  {title} {\bibinfo {title} {Capacitance-based fermion
  parity read-out and predicted {R}abi oscillations in a {M}ajorana nanowire},\
  }\href {https://arxiv.org/abs/2406.18080} {\bibfield  {journal} {\bibinfo
  {journal} {arXiv:2406.18080}\ } (\bibinfo {year} {2024})}\BibitemShut
  {NoStop}%
\bibitem [{\citenamefont {Nakajima}(1958)}]{nakajima_1958}%
  \BibitemOpen
  \bibfield  {author} {\bibinfo {author} {\bibfnamefont {S.}~\bibnamefont
  {Nakajima}},\ }\bibfield  {title} {\bibinfo {title} {On {Quantum} {Theory} of
  {Transport} {Phenomena}: {Steady} {Diffusion}},\ }\href
  {https://doi.org/10.1143/PTP.20.948} {\bibfield  {journal} {\bibinfo
  {journal} {Prog. Theor. Phys.}\ }\textbf {\bibinfo {volume} {20}},\ \bibinfo
  {pages} {948} (\bibinfo {year} {1958})}\BibitemShut {NoStop}%
\bibitem [{\citenamefont {Zwanzig}(1960)}]{zwanzig_1960}%
  \BibitemOpen
  \bibfield  {author} {\bibinfo {author} {\bibfnamefont {R.}~\bibnamefont
  {Zwanzig}},\ }\bibfield  {title} {\bibinfo {title} {Ensemble {Method} in the
  {Theory} of {Irreversibility}},\ }\href {https://doi.org/10.1063/1.1731409}
  {\bibfield  {journal} {\bibinfo  {journal} {J. Chem. Phys.}\ }\textbf
  {\bibinfo {volume} {33}},\ \bibinfo {pages} {1338} (\bibinfo {year}
  {1960})}\BibitemShut {NoStop}%
\bibitem [{\citenamefont {Redfield}(1965)}]{redfield_1965}%
  \BibitemOpen
  \bibfield  {author} {\bibinfo {author} {\bibfnamefont {A.~G.}\ \bibnamefont
  {Redfield}},\ }\bibfield  {title} {\bibinfo {title} {The {Theory} of
  {Relaxation} {Processes}},\ }\href
  {https://doi.org/10.1016/B978-1-4832-3114-3.50007-6} {\bibfield  {journal}
  {\bibinfo  {journal} {Adv. Magn. Opt. Res.}\ }\textbf {\bibinfo {volume}
  {1}},\ \bibinfo {pages} {1} (\bibinfo {year} {1965})}\BibitemShut {NoStop}%
\bibitem [{\citenamefont {Mozgunov}\ and\ \citenamefont
  {Lidar}(2020)}]{Mozgunov20}%
  \BibitemOpen
  \bibfield  {author} {\bibinfo {author} {\bibfnamefont {E.}~\bibnamefont
  {Mozgunov}}\ and\ \bibinfo {author} {\bibfnamefont {D.}~\bibnamefont
  {Lidar}},\ }\bibfield  {title} {\bibinfo {title} {Completely positive master
  equation for arbitrary driving and small level spacing},\ }\href
  {https://doi.org/10.22331/q-2020-02-06-227} {\bibfield  {journal} {\bibinfo
  {journal} {Quantum}\ }\textbf {\bibinfo {volume} {4}},\ \bibinfo {pages}
  {227} (\bibinfo {year} {2020})}\BibitemShut {NoStop}%
\bibitem [{\citenamefont {Clerk}\ \emph {et~al.}(2010)\citenamefont {Clerk},
  \citenamefont {Devoret}, \citenamefont {Girvin}, \citenamefont {Marquardt},\
  and\ \citenamefont {Schoelkopf}}]{clerk2010}%
  \BibitemOpen
  \bibfield  {author} {\bibinfo {author} {\bibfnamefont {A.~A.}\ \bibnamefont
  {Clerk}}, \bibinfo {author} {\bibfnamefont {M.~H.}\ \bibnamefont {Devoret}},
  \bibinfo {author} {\bibfnamefont {S.~M.}\ \bibnamefont {Girvin}}, \bibinfo
  {author} {\bibfnamefont {F.}~\bibnamefont {Marquardt}},\ and\ \bibinfo
  {author} {\bibfnamefont {R.~J.}\ \bibnamefont {Schoelkopf}},\ }\bibfield
  {title} {\bibinfo {title} {Introduction to quantum noise, measurement, and
  amplification},\ }\href {https://doi.org/10.1103/RevModPhys.82.1155}
  {\bibfield  {journal} {\bibinfo  {journal} {Rev. Mod. Phys.}\ }\textbf
  {\bibinfo {volume} {82}},\ \bibinfo {pages} {1155} (\bibinfo {year}
  {2010})}\BibitemShut {NoStop}%
\bibitem [{\citenamefont {Paladino}\ \emph {et~al.}(2014)\citenamefont
  {Paladino}, \citenamefont {Galperin}, \citenamefont {Falci},\ and\
  \citenamefont {Altshuler}}]{paladino2014}%
  \BibitemOpen
  \bibfield  {author} {\bibinfo {author} {\bibfnamefont {E.}~\bibnamefont
  {Paladino}}, \bibinfo {author} {\bibfnamefont {Y.}~\bibnamefont {Galperin}},
  \bibinfo {author} {\bibfnamefont {G.}~\bibnamefont {Falci}},\ and\ \bibinfo
  {author} {\bibfnamefont {B.}~\bibnamefont {Altshuler}},\ }\bibfield  {title}
  {\bibinfo {title} {1/f noise: {Implications} for solid-state quantum
  information},\ }\href {https://doi.org/10.1103/RevModPhys.86.361} {\bibfield
  {journal} {\bibinfo  {journal} {Rev. Mod. Phys.}\ }\textbf {\bibinfo {volume}
  {86}},\ \bibinfo {pages} {361} (\bibinfo {year} {2014})}\BibitemShut
  {NoStop}%
\bibitem [{\citenamefont {Schaller}\ and\ \citenamefont
  {Brandes}(2008)}]{Schaller08}%
  \BibitemOpen
  \bibfield  {author} {\bibinfo {author} {\bibfnamefont {G.}~\bibnamefont
  {Schaller}}\ and\ \bibinfo {author} {\bibfnamefont {T.}~\bibnamefont
  {Brandes}},\ }\bibfield  {title} {\bibinfo {title} {Preservation of
  positivity by dynamical coarse graining},\ }\href
  {https://doi.org/10.1103/PhysRevA.78.022106} {\bibfield  {journal} {\bibinfo
  {journal} {Phys. Rev. A}\ }\textbf {\bibinfo {volume} {78}},\ \bibinfo
  {pages} {022106} (\bibinfo {year} {2008})}\BibitemShut {NoStop}%
\bibitem [{\citenamefont {Kir{\v s}anskas}\ \emph {et~al.}(2018)\citenamefont
  {Kir{\v s}anskas}, \citenamefont {Francki{\'e}},\ and\ \citenamefont
  {Wacker}}]{Kirsanskas18}%
  \BibitemOpen
  \bibfield  {author} {\bibinfo {author} {\bibfnamefont {G.}~\bibnamefont
  {Kir{\v s}anskas}}, \bibinfo {author} {\bibfnamefont {M.}~\bibnamefont
  {Francki{\'e}}},\ and\ \bibinfo {author} {\bibfnamefont {A.}~\bibnamefont
  {Wacker}},\ }\bibfield  {title} {\bibinfo {title} {Phenomenological position
  and energy resolving {{Lindblad}} approach to quantum kinetics},\ }\href
  {https://doi.org/10.1103/PhysRevB.97.035432} {\bibfield  {journal} {\bibinfo
  {journal} {Phys. Rev. B}\ }\textbf {\bibinfo {volume} {97}},\ \bibinfo
  {pages} {035432} (\bibinfo {year} {2018})}\BibitemShut {NoStop}%
\bibitem [{\citenamefont {Mishmash}\ \emph {et~al.}(2020)\citenamefont
  {Mishmash}, \citenamefont {Bauer}, \citenamefont {von Oppen},\ and\
  \citenamefont {Alicea}}]{mishmash2020dephasing}%
  \BibitemOpen
  \bibfield  {author} {\bibinfo {author} {\bibfnamefont {R.~V.}\ \bibnamefont
  {Mishmash}}, \bibinfo {author} {\bibfnamefont {B.}~\bibnamefont {Bauer}},
  \bibinfo {author} {\bibfnamefont {F.}~\bibnamefont {von Oppen}},\ and\
  \bibinfo {author} {\bibfnamefont {J.}~\bibnamefont {Alicea}},\ }\bibfield
  {title} {\bibinfo {title} {Dephasing and leakage dynamics of noisy
  {Majorana}-based qubits: Topological versus {Andreev}},\ }\href
  {https://doi.org/10.1103/PhysRevB.101.075404} {\bibfield  {journal} {\bibinfo
   {journal} {Phys. Rev. B}\ }\textbf {\bibinfo {volume} {101}},\ \bibinfo
  {pages} {075404} (\bibinfo {year} {2020})}\BibitemShut {NoStop}%
\bibitem [{\citenamefont {Wiseman}\ and\ \citenamefont
  {Milburn}(2010)}]{wiseman2010measuerment}%
  \BibitemOpen
  \bibfield  {author} {\bibinfo {author} {\bibfnamefont {H.}~\bibnamefont
  {Wiseman}}\ and\ \bibinfo {author} {\bibfnamefont {G.}~\bibnamefont
  {Milburn}},\ }\href {https://books.google.com/books?id=ZNjvHaH8qA4C} {\emph
  {\bibinfo {title} {Quantum Measurement and Control}}}\ (\bibinfo  {publisher}
  {Cambridge University Press},\ \bibinfo {year} {2010})\BibitemShut {NoStop}%
\bibitem [{\citenamefont {Schulenborg}\ \emph {et~al.}(2023)\citenamefont
  {Schulenborg}, \citenamefont {Kr{\o}jer}, \citenamefont {Burrello},
  \citenamefont {Leijnse},\ and\ \citenamefont
  {Flensberg}}]{schulenborg2023detecting}%
  \BibitemOpen
  \bibfield  {author} {\bibinfo {author} {\bibfnamefont {J.}~\bibnamefont
  {Schulenborg}}, \bibinfo {author} {\bibfnamefont {S.}~\bibnamefont
  {Kr{\o}jer}}, \bibinfo {author} {\bibfnamefont {M.}~\bibnamefont {Burrello}},
  \bibinfo {author} {\bibfnamefont {M.}~\bibnamefont {Leijnse}},\ and\ \bibinfo
  {author} {\bibfnamefont {K.}~\bibnamefont {Flensberg}},\ }\bibfield  {title}
  {\bibinfo {title} {Detecting {Majorana} modes by readout of poisoning-induced
  parity flips},\ }\href {https://doi.org/10.1103/PhysRevB.107.L121401}
  {\bibfield  {journal} {\bibinfo  {journal} {Phys. Rev. B}\ }\textbf {\bibinfo
  {volume} {107}},\ \bibinfo {pages} {L121401} (\bibinfo {year}
  {2023})}\BibitemShut {NoStop}%
\bibitem [{\citenamefont {Vurgaftman}\ \emph {et~al.}(2001)\citenamefont
  {Vurgaftman}, \citenamefont {Meyer},\ and\ \citenamefont
  {Ram-Mohan}}]{vurgaftman01}%
  \BibitemOpen
  \bibfield  {author} {\bibinfo {author} {\bibfnamefont {I.}~\bibnamefont
  {Vurgaftman}}, \bibinfo {author} {\bibfnamefont {J.~R.}\ \bibnamefont
  {Meyer}},\ and\ \bibinfo {author} {\bibfnamefont {L.~R.}\ \bibnamefont
  {Ram-Mohan}},\ }\bibfield  {title} {\bibinfo {title} {Band parameters for
  {III–V} compound semiconductors and their alloys},\ }\href
  {https://doi.org/10.1063/1.1368156} {\bibfield  {journal} {\bibinfo
  {journal} {J. Appl. Phys.}\ }\textbf {\bibinfo {volume} {89}},\ \bibinfo
  {pages} {5815} (\bibinfo {year} {2001})}\BibitemShut {NoStop}%
\bibitem [{\citenamefont {{Madelung}}(2004)}]{madelung04}%
  \BibitemOpen
  \bibfield  {author} {\bibinfo {author} {\bibfnamefont {O.}~\bibnamefont
  {{Madelung}}},\ }\href@noop {} {\emph {\bibinfo {title} {{Semiconductors:
  Data Handbook}}}}\ (\bibinfo  {publisher} {Springer},\ \bibinfo {address}
  {Berlin},\ \bibinfo {year} {2004})\BibitemShut {NoStop}%
\bibitem [{\citenamefont {Shevchenko}\ \emph {et~al.}(2010)\citenamefont
  {Shevchenko}, \citenamefont {Ashhab},\ and\ \citenamefont
  {Nori}}]{shevchenko_2010}%
  \BibitemOpen
  \bibfield  {author} {\bibinfo {author} {\bibfnamefont {S.}~\bibnamefont
  {Shevchenko}}, \bibinfo {author} {\bibfnamefont {S.}~\bibnamefont {Ashhab}},\
  and\ \bibinfo {author} {\bibfnamefont {F.}~\bibnamefont {Nori}},\ }\bibfield
  {title} {\bibinfo {title} {Landau–{Zener}–{Stückelberg}
  interferometry},\ }\href
  {https://doi.org/https://doi.org/10.1016/j.physrep.2010.03.002} {\bibfield
  {journal} {\bibinfo  {journal} {Phys. Rep.}\ }\textbf {\bibinfo {volume}
  {492}},\ \bibinfo {pages} {1} (\bibinfo {year} {2010})}\BibitemShut {NoStop}%
\bibitem [{\citenamefont {Liu}\ \emph {et~al.}(2017)\citenamefont {Liu},
  \citenamefont {Levchenko},\ and\ \citenamefont {Lutchyn}}]{Dong17}%
  \BibitemOpen
  \bibfield  {author} {\bibinfo {author} {\bibfnamefont {D.~E.}\ \bibnamefont
  {Liu}}, \bibinfo {author} {\bibfnamefont {A.}~\bibnamefont {Levchenko}},\
  and\ \bibinfo {author} {\bibfnamefont {R.~M.}\ \bibnamefont {Lutchyn}},\
  }\bibfield  {title} {\bibinfo {title} {Keldysh approach to periodically
  driven systems with a fermionic bath: Nonequilibrium steady state, proximity
  effect, and dissipation},\ }\href
  {https://doi.org/10.1103/PhysRevB.95.115303} {\bibfield  {journal} {\bibinfo
  {journal} {Phys. Rev. B}\ }\textbf {\bibinfo {volume} {95}},\ \bibinfo
  {pages} {115303} (\bibinfo {year} {2017})}\BibitemShut {NoStop}%
\bibitem [{\citenamefont {Bauer}\ \emph {et~al.}(2018)\citenamefont {Bauer},
  \citenamefont {Karzig}, \citenamefont {Mishmash}, \citenamefont {Antipov},\
  and\ \citenamefont {Alicea}}]{bauer_dynamics_2018}%
  \BibitemOpen
  \bibfield  {author} {\bibinfo {author} {\bibfnamefont {B.}~\bibnamefont
  {Bauer}}, \bibinfo {author} {\bibfnamefont {T.}~\bibnamefont {Karzig}},
  \bibinfo {author} {\bibfnamefont {R.}~\bibnamefont {Mishmash}}, \bibinfo
  {author} {\bibfnamefont {A.}~\bibnamefont {Antipov}},\ and\ \bibinfo {author}
  {\bibfnamefont {J.}~\bibnamefont {Alicea}},\ }\bibfield  {title} {\bibinfo
  {title} {Dynamics of {Majorana}-based qubits operated with an array of
  tunable gates},\ }\href {https://doi.org/10.21468/SciPostPhys.5.1.004}
  {\bibfield  {journal} {\bibinfo  {journal} {SciPost Physics}\ }\textbf
  {\bibinfo {volume} {5}},\ \bibinfo {pages} {004} (\bibinfo {year}
  {2018})}\BibitemShut {NoStop}%
\bibitem [{\citenamefont {Majorana}(1932)}]{majorana1932atomi}%
  \BibitemOpen
  \bibfield  {author} {\bibinfo {author} {\bibfnamefont {E.}~\bibnamefont
  {Majorana}},\ }\bibfield  {title} {\bibinfo {title} {Atomi orientati in campo
  magnetico variabile},\ }\href {https://doi.org/10.1007/BF02960953} {\bibfield
   {journal} {\bibinfo  {journal} {Il Nuovo Cimento}\ }\textbf {\bibinfo
  {volume} {9}},\ \bibinfo {pages} {43} (\bibinfo {year} {1932})}\BibitemShut
  {NoStop}%
\bibitem [{\citenamefont {Zener}(1932)}]{zener1932non}%
  \BibitemOpen
  \bibfield  {author} {\bibinfo {author} {\bibfnamefont {C.}~\bibnamefont
  {Zener}},\ }\bibfield  {title} {\bibinfo {title} {Non-adiabatic crossing of
  energy levels},\ }in\ \href {https://doi.org/10.1098/rspa.1932.0165} {\emph
  {\bibinfo {booktitle} {Proceedings of the Royal Society of London A:
  Mathematical, Physical and Engineering Sciences}}},\ Vol.\ \bibinfo {volume}
  {137}\ (\bibinfo {organization} {The Royal Society},\ \bibinfo {year}
  {1932})\ p.\ \bibinfo {pages} {696}\BibitemShut {NoStop}%
\bibitem [{\citenamefont {Landau}(1932)}]{landau1932theorie}%
  \BibitemOpen
  \bibfield  {author} {\bibinfo {author} {\bibfnamefont {L.}~\bibnamefont
  {Landau}},\ }\bibfield  {title} {\bibinfo {title} {{Zur Theorie der
  Energie\"ubertragung. II}},\ }\href@noop {} {\bibfield  {journal} {\bibinfo
  {journal} {Phys. Z. Sowjetunion}\ }\textbf {\bibinfo {volume} {2}},\ \bibinfo
  {pages} {1} (\bibinfo {year} {1932})}\BibitemShut {NoStop}%
\bibitem [{\citenamefont {Pustilnik}\ and\ \citenamefont
  {Glazman}(2004)}]{pustilnik2004kondo}%
  \BibitemOpen
  \bibfield  {author} {\bibinfo {author} {\bibfnamefont {M.}~\bibnamefont
  {Pustilnik}}\ and\ \bibinfo {author} {\bibfnamefont {L.}~\bibnamefont
  {Glazman}},\ }\bibfield  {title} {\bibinfo {title} {Kondo effect in quantum
  dots},\ }\href {https://doi.org/10.1088/0953-8984/16/16/R01} {\bibfield
  {journal} {\bibinfo  {journal} {J. Phys. Condens. Matter}\ }\textbf {\bibinfo
  {volume} {16}},\ \bibinfo {pages} {R513} (\bibinfo {year}
  {2004})}\BibitemShut {NoStop}%
\bibitem [{\citenamefont {Bravyi}\ and\ \citenamefont
  {Gosset}(2017)}]{Bravyi2017}%
  \BibitemOpen
  \bibfield  {author} {\bibinfo {author} {\bibfnamefont {S.}~\bibnamefont
  {Bravyi}}\ and\ \bibinfo {author} {\bibfnamefont {D.}~\bibnamefont
  {Gosset}},\ }\bibfield  {title} {\bibinfo {title} {Complexity of quantum
  impurity problems},\ }\href {https://doi.org/10.1007/s00220-017-2976-9}
  {\bibfield  {journal} {\bibinfo  {journal} {Comm. Math. Phys.}\ }\textbf
  {\bibinfo {volume} {356}},\ \bibinfo {pages} {451} (\bibinfo {year}
  {2017})}\BibitemShut {NoStop}%
\bibitem [{\citenamefont {Boutin}\ and\ \citenamefont
  {Bauer}(2021)}]{boutin2021quantum}%
  \BibitemOpen
  \bibfield  {author} {\bibinfo {author} {\bibfnamefont {S.}~\bibnamefont
  {Boutin}}\ and\ \bibinfo {author} {\bibfnamefont {B.}~\bibnamefont {Bauer}},\
  }\bibfield  {title} {\bibinfo {title} {Quantum impurity models using
  superpositions of fermionic {Gaussian} states: Practical methods and
  applications},\ }\href {https://doi.org/10.1103/PhysRevResearch.3.033188}
  {\bibfield  {journal} {\bibinfo  {journal} {Phys. Rev. Res.}\ }\textbf
  {\bibinfo {volume} {3}},\ \bibinfo {pages} {033188} (\bibinfo {year}
  {2021})}\BibitemShut {NoStop}%
\bibitem [{\citenamefont {L\"owdin}(1955)}]{lowdin1955}%
  \BibitemOpen
  \bibfield  {author} {\bibinfo {author} {\bibfnamefont {P.-O.}\ \bibnamefont
  {L\"owdin}},\ }\bibfield  {title} {\bibinfo {title} {Quantum theory of
  many-particle systems. ii. {S}tudy of the ordinary {H}artree-{F}ock
  approximation},\ }\href {https://doi.org/10.1103/PhysRev.97.1490} {\bibfield
  {journal} {\bibinfo  {journal} {Phys. Rev.}\ }\textbf {\bibinfo {volume}
  {97}},\ \bibinfo {pages} {1490} (\bibinfo {year} {1955})}\BibitemShut
  {NoStop}%
\bibitem [{\citenamefont {Bravyi}(2004)}]{bravyi2004lagrangian}%
  \BibitemOpen
  \bibfield  {author} {\bibinfo {author} {\bibfnamefont {S.}~\bibnamefont
  {Bravyi}},\ }\bibfield  {title} {\bibinfo {title} {Lagrangian representation
  for fermionic linear optics},\ }\href
  {https://doi.org/10.48550/arXiv.quant-ph/0404180} {\bibfield  {journal}
  {\bibinfo  {journal} {arXiv:quant-ph/0404180}\ } (\bibinfo {year}
  {2004})}\BibitemShut {NoStop}%
\bibitem [{\citenamefont {Kraus}\ and\ \citenamefont
  {Cirac}(2010)}]{kraus2010generalized}%
  \BibitemOpen
  \bibfield  {author} {\bibinfo {author} {\bibfnamefont {C.~V.}\ \bibnamefont
  {Kraus}}\ and\ \bibinfo {author} {\bibfnamefont {J.~I.}\ \bibnamefont
  {Cirac}},\ }\bibfield  {title} {\bibinfo {title} {{Generalized Hartree--Fock
  theory for interacting fermions in lattices: numerical methods}},\ }\href
  {https://doi.org/10.1088/1367-2630/12/11/113004} {\bibfield  {journal}
  {\bibinfo  {journal} {New J. Phys.}\ }\textbf {\bibinfo {volume} {12}},\
  \bibinfo {pages} {113004} (\bibinfo {year} {2010})}\BibitemShut {NoStop}%
\bibitem [{\citenamefont {Zgid}\ \emph {et~al.}(2012)\citenamefont {Zgid},
  \citenamefont {Gull},\ and\ \citenamefont {Chan}}]{TCI_dmft_2012}%
  \BibitemOpen
  \bibfield  {author} {\bibinfo {author} {\bibfnamefont {D.}~\bibnamefont
  {Zgid}}, \bibinfo {author} {\bibfnamefont {E.}~\bibnamefont {Gull}},\ and\
  \bibinfo {author} {\bibfnamefont {G.~K.-L.}\ \bibnamefont {Chan}},\
  }\bibfield  {title} {\bibinfo {title} {Truncated configuration interaction
  expansions as solvers for correlated quantum impurity models and dynamical
  mean-field theory},\ }\href {https://doi.org/10.1103/PhysRevB.86.165128}
  {\bibfield  {journal} {\bibinfo  {journal} {Phys. Rev. B}\ }\textbf {\bibinfo
  {volume} {86}},\ \bibinfo {pages} {165128} (\bibinfo {year}
  {2012})}\BibitemShut {NoStop}%
\bibitem [{\citenamefont {Foresman}\ \emph {et~al.}(1992)\citenamefont
  {Foresman}, \citenamefont {Head-Gordon}, \citenamefont {Pople},\ and\
  \citenamefont {Frisch}}]{foresman1992toward}%
  \BibitemOpen
  \bibfield  {author} {\bibinfo {author} {\bibfnamefont {J.~B.}\ \bibnamefont
  {Foresman}}, \bibinfo {author} {\bibfnamefont {M.}~\bibnamefont
  {Head-Gordon}}, \bibinfo {author} {\bibfnamefont {J.~A.}\ \bibnamefont
  {Pople}},\ and\ \bibinfo {author} {\bibfnamefont {M.~J.}\ \bibnamefont
  {Frisch}},\ }\bibfield  {title} {\bibinfo {title} {Toward a systematic
  molecular orbital theory for excited states},\ }\href
  {https://doi.org/10.1021/j100180a030} {\bibfield  {journal} {\bibinfo
  {journal} {J. Phys. Chem.}\ }\textbf {\bibinfo {volume} {96}},\ \bibinfo
  {pages} {135} (\bibinfo {year} {1992})}\BibitemShut {NoStop}%
\bibitem [{\citenamefont {Flensberg}(1993)}]{Flensberg1993}%
  \BibitemOpen
  \bibfield  {author} {\bibinfo {author} {\bibfnamefont {K.}~\bibnamefont
  {Flensberg}},\ }\bibfield  {title} {\bibinfo {title} {Capacitance and
  conductance of mesoscopic systems connected by quantum point contacts},\
  }\href {https://doi.org/10.1103/PhysRevB.48.11156} {\bibfield  {journal}
  {\bibinfo  {journal} {Phys. Rev. B}\ }\textbf {\bibinfo {volume} {48}},\
  \bibinfo {pages} {11156} (\bibinfo {year} {1993})}\BibitemShut {NoStop}%
\bibitem [{\citenamefont {Lutchyn}\ \emph {et~al.}(2016)\citenamefont
  {Lutchyn}, \citenamefont {Flensberg},\ and\ \citenamefont
  {Glazman}}]{Lutchyn16}%
  \BibitemOpen
  \bibfield  {author} {\bibinfo {author} {\bibfnamefont {R.~M.}\ \bibnamefont
  {Lutchyn}}, \bibinfo {author} {\bibfnamefont {K.}~\bibnamefont {Flensberg}},\
  and\ \bibinfo {author} {\bibfnamefont {L.~I.}\ \bibnamefont {Glazman}},\
  }\bibfield  {title} {\bibinfo {title} {Quantum charge fluctuations of a
  proximitized nanowire},\ }\href {https://doi.org/10.1103/PhysRevB.94.125407}
  {\bibfield  {journal} {\bibinfo  {journal} {Phys. Rev. B}\ }\textbf {\bibinfo
  {volume} {94}},\ \bibinfo {pages} {125407} (\bibinfo {year}
  {2016})}\BibitemShut {NoStop}%
\bibitem [{\citenamefont {White}\ and\ \citenamefont
  {Noack}(1992)}]{white1992real}%
  \BibitemOpen
  \bibfield  {author} {\bibinfo {author} {\bibfnamefont {S.~R.}\ \bibnamefont
  {White}}\ and\ \bibinfo {author} {\bibfnamefont {R.~M.}\ \bibnamefont
  {Noack}},\ }\bibfield  {title} {\bibinfo {title} {Real-space quantum
  renormalization groups},\ }\href
  {https://doi.org/10.1103/PhysRevLett.68.3487} {\bibfield  {journal} {\bibinfo
   {journal} {Phys. Rev. Lett.}\ }\textbf {\bibinfo {volume} {68}},\ \bibinfo
  {pages} {3487} (\bibinfo {year} {1992})}\BibitemShut {NoStop}%
\bibitem [{\citenamefont {Schuch}\ \emph {et~al.}(2012)\citenamefont {Schuch},
  \citenamefont {Wolf},\ and\ \citenamefont {Cirac}}]{schuch2012gaussian}%
  \BibitemOpen
  \bibfield  {author} {\bibinfo {author} {\bibfnamefont {N.}~\bibnamefont
  {Schuch}}, \bibinfo {author} {\bibfnamefont {M.~M.}\ \bibnamefont {Wolf}},\
  and\ \bibinfo {author} {\bibfnamefont {J.~I.}\ \bibnamefont {Cirac}},\
  }\bibfield  {title} {\bibinfo {title} {Gaussian matrix product states},\
  }\href {https://doi.org/10.48550/arXiv.1201.3945} {\bibfield  {journal}
  {\bibinfo  {journal} {arXiv:1201.3945}\ } (\bibinfo {year}
  {2012})}\BibitemShut {NoStop}%
\bibitem [{\citenamefont {Fishman}\ and\ \citenamefont
  {White}(2015)}]{fishman2015compression}%
  \BibitemOpen
  \bibfield  {author} {\bibinfo {author} {\bibfnamefont {M.~T.}\ \bibnamefont
  {Fishman}}\ and\ \bibinfo {author} {\bibfnamefont {S.~R.}\ \bibnamefont
  {White}},\ }\bibfield  {title} {\bibinfo {title} {Compression of correlation
  matrices and an efficient method for forming matrix product states of
  fermionic {Gaussian} states},\ }\href
  {https://doi.org/10.1103/PhysRevB.92.075132} {\bibfield  {journal} {\bibinfo
  {journal} {Phys. Rev. B}\ }\textbf {\bibinfo {volume} {92}},\ \bibinfo
  {pages} {075132} (\bibinfo {year} {2015})}\BibitemShut {NoStop}%
\bibitem [{\citenamefont {Schuch}\ and\ \citenamefont
  {Bauer}(2019)}]{schuch2019matrix}%
  \BibitemOpen
  \bibfield  {author} {\bibinfo {author} {\bibfnamefont {N.}~\bibnamefont
  {Schuch}}\ and\ \bibinfo {author} {\bibfnamefont {B.}~\bibnamefont {Bauer}},\
  }\bibfield  {title} {\bibinfo {title} {Matrix product state algorithms for
  {Gaussian} fermionic states},\ }\href
  {https://doi.org/10.1103/PhysRevB.100.245121} {\bibfield  {journal} {\bibinfo
   {journal} {Phys. Rev. B}\ }\textbf {\bibinfo {volume} {100}},\ \bibinfo
  {pages} {245121} (\bibinfo {year} {2019})}\BibitemShut {NoStop}%
\bibitem [{\citenamefont {{Lutchyn}}\ \emph {et~al.}(2010)\citenamefont
  {{Lutchyn}}, \citenamefont {{Sau}},\ and\ \citenamefont {{Das
  Sarma}}}]{Lutchyn10}%
  \BibitemOpen
  \bibfield  {author} {\bibinfo {author} {\bibfnamefont {R.~M.}\ \bibnamefont
  {{Lutchyn}}}, \bibinfo {author} {\bibfnamefont {J.~D.}\ \bibnamefont
  {{Sau}}},\ and\ \bibinfo {author} {\bibfnamefont {S.}~\bibnamefont {{Das
  Sarma}}},\ }\bibfield  {title} {\bibinfo {title} {Majorana {Fermions} and a
  topological phase transition in semiconductor-superconductor
  heterostructures},\ }\href {https://doi.org/10.1103/PhysRevLett.105.077001}
  {\bibfield  {journal} {\bibinfo  {journal} {Phys. Rev. Lett.}\ }\textbf
  {\bibinfo {volume} {105}},\ \bibinfo {pages} {077001} (\bibinfo {year}
  {2010})}\BibitemShut {NoStop}%
\bibitem [{\citenamefont {{Oreg}}\ \emph {et~al.}(2010)\citenamefont {{Oreg}},
  \citenamefont {{Refael}},\ and\ \citenamefont {{von Oppen}}}]{Oreg10}%
  \BibitemOpen
  \bibfield  {author} {\bibinfo {author} {\bibfnamefont {Y.}~\bibnamefont
  {{Oreg}}}, \bibinfo {author} {\bibfnamefont {G.}~\bibnamefont {{Refael}}},\
  and\ \bibinfo {author} {\bibfnamefont {F.}~\bibnamefont {{von Oppen}}},\
  }\bibfield  {title} {\bibinfo {title} {Helical liquids and {Majorana} bound
  states in quantum wires},\ }\href
  {https://doi.org/10.1103/PhysRevLett.105.177002} {\bibfield  {journal}
  {\bibinfo  {journal} {Phys. Rev. Lett.}\ }\textbf {\bibinfo {volume} {105}},\
  \bibinfo {pages} {177002} (\bibinfo {year} {2010})}\BibitemShut {NoStop}%
\bibitem [{\citenamefont {{Stanescu}}\ \emph {et~al.}(2011)\citenamefont
  {{Stanescu}}, \citenamefont {{Lutchyn}},\ and\ \citenamefont {{Das
  Sarma}}}]{Stanescu11}%
  \BibitemOpen
  \bibfield  {author} {\bibinfo {author} {\bibfnamefont {T.~D.}\ \bibnamefont
  {{Stanescu}}}, \bibinfo {author} {\bibfnamefont {R.~M.}\ \bibnamefont
  {{Lutchyn}}},\ and\ \bibinfo {author} {\bibfnamefont {S.}~\bibnamefont {{Das
  Sarma}}},\ }\bibfield  {title} {\bibinfo {title} {{Majorana fermions in
  semiconductor nanowires}},\ }\href
  {https://doi.org/10.1103/PhysRevB.84.144522} {\bibfield  {journal} {\bibinfo
  {journal} {\prb}\ }\textbf {\bibinfo {volume} {84}},\ \bibinfo {eid} {144522}
  (\bibinfo {year} {2011})}\BibitemShut {NoStop}%
\bibitem [{\citenamefont {Hell}\ \emph {et~al.}(2018)\citenamefont {Hell},
  \citenamefont {Flensberg},\ and\ \citenamefont {Leijnse}}]{Hell2018}%
  \BibitemOpen
  \bibfield  {author} {\bibinfo {author} {\bibfnamefont {M.}~\bibnamefont
  {Hell}}, \bibinfo {author} {\bibfnamefont {K.}~\bibnamefont {Flensberg}},\
  and\ \bibinfo {author} {\bibfnamefont {M.}~\bibnamefont {Leijnse}},\
  }\bibfield  {title} {\bibinfo {title} {Distinguishing {Majorana} bound states
  from localized {Andreev} bound states by interferometry},\ }\href
  {https://doi.org/10.1103/PhysRevB.97.161401} {\bibfield  {journal} {\bibinfo
  {journal} {Phys. Rev. B}\ }\textbf {\bibinfo {volume} {97}},\ \bibinfo
  {pages} {161401} (\bibinfo {year} {2018})}\BibitemShut {NoStop}%
\bibitem [{\citenamefont {Kitaev}(2001)}]{Kitaev01}%
  \BibitemOpen
  \bibfield  {author} {\bibinfo {author} {\bibfnamefont {A.~Y.}\ \bibnamefont
  {Kitaev}},\ }\bibfield  {title} {\bibinfo {title} {Unpaired {Majorana}
  fermions in quantum wires},\ }\href
  {https://doi.org/10.1070/1063-7869/44/10S/S29} {\bibfield  {journal}
  {\bibinfo  {journal} {Phys.-Usp.}\ }\textbf {\bibinfo {volume} {44}},\
  \bibinfo {pages} {31} (\bibinfo {year} {2001})}\BibitemShut {NoStop}%
\bibitem [{\citenamefont {Cleland}(2013)}]{cleland2013}%
  \BibitemOpen
  \bibfield  {author} {\bibinfo {author} {\bibfnamefont {A.}~\bibnamefont
  {Cleland}},\ }\href {https://books.google.com/books?id=ofvuCAAAQBAJ} {\emph
  {\bibinfo {title} {Foundations of Nanomechanics: From Solid-State Theory to
  Device Applications}}},\ Advanced Texts in Physics\ (\bibinfo  {publisher}
  {Springer Berlin Heidelberg},\ \bibinfo {year} {2013})\BibitemShut {NoStop}%
\bibitem [{\citenamefont {Fishman}\ \emph {et~al.}(2022)\citenamefont
  {Fishman}, \citenamefont {White},\ and\ \citenamefont
  {Stoudenmire}}]{ITensor}%
  \BibitemOpen
  \bibfield  {author} {\bibinfo {author} {\bibfnamefont {M.}~\bibnamefont
  {Fishman}}, \bibinfo {author} {\bibfnamefont {S.~R.}\ \bibnamefont {White}},\
  and\ \bibinfo {author} {\bibfnamefont {E.~M.}\ \bibnamefont {Stoudenmire}},\
  }\bibfield  {title} {\bibinfo {title} {{The ITensor Software Library for
  Tensor Network Calculations}},\ }\href
  {https://doi.org/10.21468/SciPostPhysCodeb.4} {\bibfield  {journal} {\bibinfo
   {journal} {SciPost Phys. Codebases}\ ,\ \bibinfo {pages} {4}} (\bibinfo
  {year} {2022})}\BibitemShut {NoStop}%
\bibitem [{\citenamefont {Pave\ifmmode \check{s}\else
  \v{s}\fi{}i\ifmmode~\acute{c}\else \'{c}\fi{}}\ and\ \citenamefont
  {\ifmmode~\check{Z}\else \v{Z}\fi{}itko}(2022)}]{Pavesic2022A}%
  \BibitemOpen
  \bibfield  {author} {\bibinfo {author} {\bibfnamefont {L.}~\bibnamefont
  {Pave\ifmmode \check{s}\else \v{s}\fi{}i\ifmmode~\acute{c}\else \'{c}\fi{}}}\
  and\ \bibinfo {author} {\bibfnamefont {R.}~\bibnamefont
  {\ifmmode~\check{Z}\else \v{Z}\fi{}itko}},\ }\bibfield  {title} {\bibinfo
  {title} {Qubit based on spin-singlet yu-shiba-rusinov states},\ }\href
  {https://doi.org/10.1103/PhysRevB.105.075129} {\bibfield  {journal} {\bibinfo
   {journal} {Phys. Rev. B}\ }\textbf {\bibinfo {volume} {105}},\ \bibinfo
  {pages} {075129} (\bibinfo {year} {2022})}\BibitemShut {NoStop}%
\bibitem [{\citenamefont {Stoudenmire}\ and\ \citenamefont
  {White}(2012)}]{Stoudenmire12}%
  \BibitemOpen
  \bibfield  {author} {\bibinfo {author} {\bibfnamefont {E.}~\bibnamefont
  {Stoudenmire}}\ and\ \bibinfo {author} {\bibfnamefont {S.~R.}\ \bibnamefont
  {White}},\ }\bibfield  {title} {\bibinfo {title} {Studying two-dimensional
  systems with the density matrix renormalization group},\ }\href
  {https://doi.org/https://doi.org/10.1146/annurev-conmatphys-020911-125018}
  {\bibfield  {journal} {\bibinfo  {journal} {Annu. Rev. Condens. Matter
  Phys.}\ }\textbf {\bibinfo {volume} {3}},\ \bibinfo {pages} {111} (\bibinfo
  {year} {2012})}\BibitemShut {NoStop}%
\bibitem [{\citenamefont {Wimmer}(2012)}]{Wimmer12}%
  \BibitemOpen
  \bibfield  {author} {\bibinfo {author} {\bibfnamefont {M.}~\bibnamefont
  {Wimmer}},\ }\bibfield  {title} {\bibinfo {title} {Algorithm 923: Efficient
  numerical computation of the {Pfaffian} for dense and banded skew-symmetric
  matrices},\ }\href {https://doi.org/10.1145/2331130.2331138} {\bibfield
  {journal} {\bibinfo  {journal} {ACM Trans. Math. Softw.}\ }\textbf {\bibinfo
  {volume} {38}},\ \bibinfo {pages} {1} (\bibinfo {year} {2012})}\BibitemShut
  {NoStop}%
\end{thebibliography}
\end{document}